\documentclass[aps,rmp,reprint,groupedaddress]{revtex4-2}

\usepackage{graphicx}%
\usepackage{dcolumn}%
\usepackage{bm}%
\usepackage{hyperref}%
\usepackage{orcidlink}
\usepackage{booktabs}
\usepackage{xcolor} 
\usepackage{tabularx}
\usepackage{amsmath}
\usepackage{placeins}

\def\aap{AA}
\def\prl{Phys. Rev. Lett.}
\def\aapr{AA Rev}
\def\apjl{ApJL}

\def\mnras{MNRAS}
\def\apj{ApJ}
\def\apjs{ApJS}
\def\aj{AJ}

\def\pasj{PASJ}
\def\nat{Nat}

\def\prd{PhysRevD}
\def\jcap{JCAP}
\def\physrep{Phys. Rept.}
\def\araa{Annual Review of Astronomy and Astrophysics}
\def\ssr{Space Science Reviews}
    
\newcommand{{\qsub}}{\bf{q}}
\newcommand{{\rsub}}{{\bf{r}}}
\newcommand{{\xlight}}{{\bf{n}}_{\rm{light}}}
\newcommand{{\xlens}}{{\bf{n}}_{\rm{mac}}}
\newcommand{{\data}}{\bf{D}}
\newcommand{{\datan}}{{\bf{d}}_{\rm{n}}}
\newcommand{{\datanprime}}{{{\bf{d}}_{\rm{n}}^{\prime}}}
\newcommand{{\dimg}}{{\bf{d}}_{\mathcal{I}}}
\newcommand{{\dimgprime}}{{\bf{d}}_{\mathcal{I}}^{\prime}}
\newcommand{{\dptsrc}}{{\bf{d}}_{\rm{ptsrc}}}
\newcommand{{\dptsrcprime}}{{\bf{d}}_{\rm{ptsrc}}^{\prime}}
\newcommand{{\dfr}}{{\bf{d}}_{\rm{fr}}}
\newcommand{{\dfrprime}}{{\bf{d}}_{\rm{fr}}^{\prime}}
\newcommand{{\zlens}}{{z_{\rm{d}}}}
\newcommand{{\zsrc}}{{z_{\rm{s}}}}
\newcommand{{\nbody}}{{$N$-body~}}
\newcommand{{\msun}}{{\rm{M}_{\odot}}}

\newcommand{\dt}{\ensuremath{D_{\Delta{\rm t}}}}
\newcommand{\dd}{\ensuremath{D_\mathrm{d}}}    					%

\newcommand{\Ddt}{D_{\Delta{\rm t}}}
\newcommand{\Dd}{D_{\rm d}}

\newcommand{\xx}{\boldsymbol{\theta}}

\newcommand{\lint}{\lambda_{\rm int}}
\newcommand{\ltot}{\lambda_{\rm tot}}

\def\kext{\kappa_{\rm ext}}

\def\physrep{Phys.~Rep.}

\newcommand{\Ho}{H$_0$}
\newcommand{\hunit}{km s$^{-1}$ Mpc$^{-1}$}

\begin{document}

\title{Cosmology with strong gravitational lenses}

\author{Tommaso Treu}
\email{Contact author: tt@astro.ucla.edu}
\affiliation{%
Physics and Astronomy Department\\ University of California, Los Angeles CA 90095
}%

\date{\today}

\begin{abstract}
Three of the most pressing open questions in cosmology -- the nature of dark energy and dark matter, and the growing tension in the value of the Hubble constant H$_0$ -- call for cosmological probes that are independent of, and complementary to, traditional methods. Strong gravitational lensing offers exactly that: a purely gravitational probe of the universe's expansion history and matter distribution, sensitive to dark matter's presence regardless of whether it interacts with light. Time-delay lenses and multiplane lenses can measure H$_0$ and the dark energy equation of state directly, while the distortion of lensed images can reveal the abundance and internal structure of dark matter subhalos. After a brief theoretical and historical introduction, I review recent breakthroughs enabled by JWST and by advances in lens modeling and stellar dynamics. Examples include the recent measurement of H$_0$ with 3\% precision in flat $\Lambda$ cold dark matter cosmology, and a tentative detection of dark matter self-interaction. Strong lensing is entering a pivotal era: wide-field surveys with Euclid, Rubin, and Roman, and radio surveys with the Square Kilometer Array, are poised to increase the number of known strong lenses by orders of magnitude, positioning strong lensing to play a central role in answering these open questions in the decade ahead.
\end{abstract}

\maketitle

\section{Introduction}
\label{sec:intro}

As photons travel from an astronomical source to the observer, their trajectories are affected by the inhomogeneities of the mass distribution of the universe. In general, every source appears weakly lensed to the observer, i.e., slightly distorted and offset on the sky compared to how it would appear in a perfectly homogeneous universe described by the Friedmann-Lemaître-Robertson-Walker (FLRW) metric.  Approximately one in a thousand to one in a hundred sources \citep[e.g.,][]{Taak:2023} appear {\it strongly lensed}: their photons pass close enough to an over-density that distortion is more pronounced, resulting in the formation of multiple images. 

In this review, I describe how strong gravitational lensing can be used as a powerful cosmological probe, independent of and complementary to more commonly used probes. At a fundamental level, the cosmological power of strong lensing comes from two facts. First, strong lensing depends on redshifts and cosmological distances. Therefore, it can be used to measure the expansion history of the universe, and thus quantities such as the Hubble constant \Ho, and the dark energy equation-of-state parameters. Second, strong lensing is sensitive to the distribution of matter on sub-galactic scales, and therefore it can be used to test any cosmological model that makes predictions about the growth of structure on small scales. Importantly, strong lensing does not depend at all on whether the small-scale mass emits light; it is a purely gravitational effect. Therefore, it can be used to probe dark matter models irrespective of their interactions with light or baryons. Examples include sterile neutrinos, primordial black holes, and self-interacting dark matter. 

This is a pivotal moment for strong lensing, as we are about to enter an era of target abundance. Although the past decades have seen enormous progress in the theoretical understanding of strong lensing, practical methodology, and data quality per system, the statistical power has generally been limited by the size of known samples of suitable systems. State-of-the-art strong lensing studies as of 2026 typically rely on samples of tens to hundreds of systems at best.  With the advent of wide-field optical, infrared, and radio surveys -- such as those carried out by  Euclid, Rubin, Roman, and the Square Kilometer Array -- orders-of-magnitude improvements in sample size will be possible. If we can muster the resources to obtain the necessary follow-up for these large samples, and if we can continue to accelerate our data analysis speed using traditional and artificial intelligence methods, strong lensing can play a major role in answering fundamental questions such as ``What is dark energy?" and ``What is dark matter?".

This review is organized as follows. Section~\ref{sec:theory} starts with a brief description of the basics of strong lensing. Section~\ref{sec:history} summarizes the history of strong lensing. Section~\ref{sec:expansion} goes into more detail on how strong lensing time delays can be used to measure the Hubble constant H$_0$. Section~\ref{sec:darkenergy} describes the application of strong lensing to the measurement of the dark energy equation of state parameters. Section~\ref{sec:sys} takes a deep dive into systematic errors and open issues in time-delay cosmography. Section~\ref{sec:darkmatter} focuses on strong lensing as a probe of dark matter. Section~\ref{sec:future} discusses the future of strong lensing cosmology, describing the unprecedented opportunities and the necessary challenges. Section~\ref{sec:summary} provides a short summary of this review.

Owing to space limitations, I cannot cover all aspects of strong lensing or all the relevant literature. Importantly, I am restricting the article to cosmology. I am not discussing any of the many and successful applications of strong lensing to astrophysics. Those include the study of the deflector or source in terms of galaxy formation and evolution, cosmic reionization, black holes, accretion disks, and stellar astrophysics. 
Interested readers can find more extensive treatments in books
\citep{Schneider:1992,Kochanek:2006saasfee,Dodelson2017,Meneghetti:2021} and reviews \citep{B+N92,Treu:2010,TreuMarshall2016,TSM22,Birrer2024,Vegetti2024,T+S24,K+N11,Natarajan:2024,Shajib:2025review,Saha:2024,Suyu:2024,Ellis:2010}.

\begin{figure*}[!t]
\centering\includegraphics[width=0.96\textwidth]{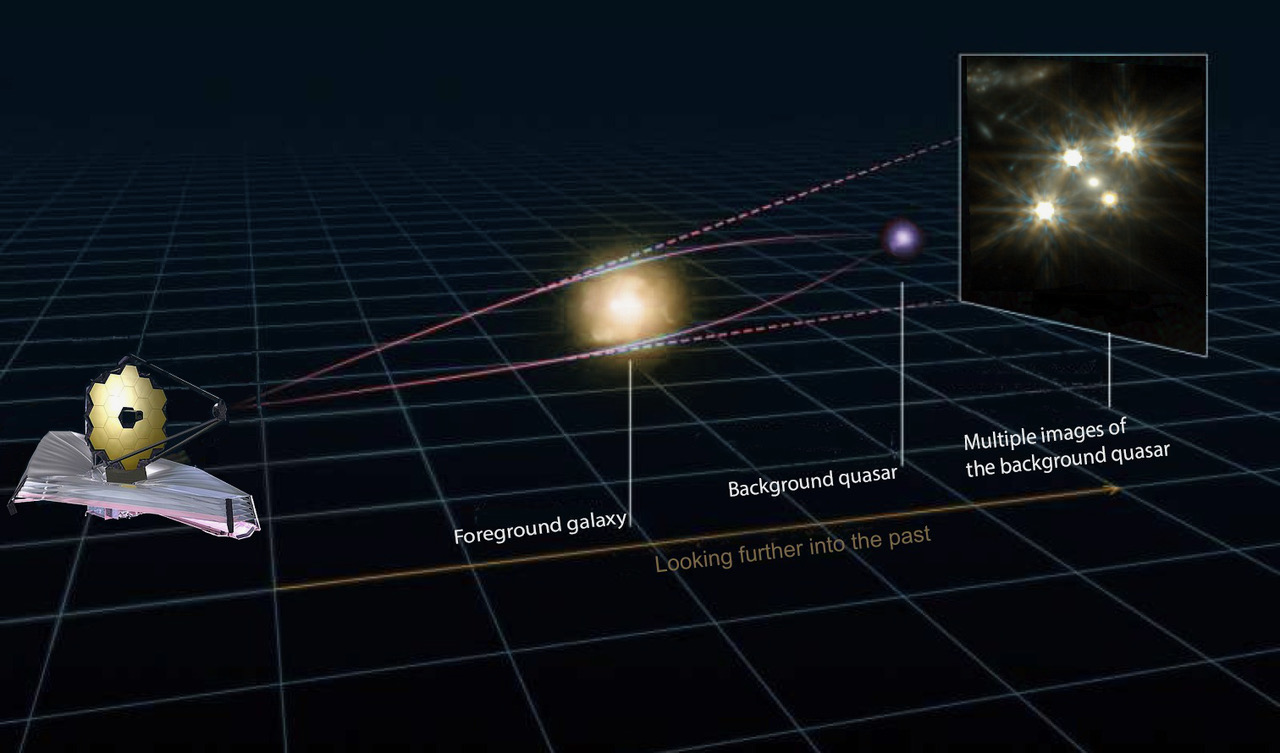}
\caption{Schematic lens diagram, illustrating the deflection of light in the galaxy-scale gravitationally lensed quasar system 2M1134-2103. The deflection is dominated by the deflector, a massive galaxy (graphic credit: Martin Millon). Four images of the background quasar appear in an Einstein Cross configuration. Note the characteristic hexagonal symmetry of the point spread function of JWST used for these observations (JWST-GO-7184; PI Millon; JWST NIRCam).} 
\label{fig:lensdiagram}
\end{figure*}

\section{Strong lensing basics and methods} 
\label{sec:theory}

\subsection{Fermat's principle and the lens equation}

In the case of a single deflector plane\footnote{See, e.g., \citep{Schneider2019}, for multiplane generalization.}, strong lensing can
be  mathematically described as a transformation from the two-dimensional observed coordinates of a photon ${\boldsymbol \theta}$ in the ``image plane" to the two-dimensional coordinates that the photon would be observed at in a homogeneous FLRW universe
${\boldsymbol \beta}$ (i.e., the ``source plane").

As in standard optics, a version of Fermat's principle applies to gravitational optics
\citep[][and references therein]{Schneider:1985,Blanford:1986}. For a
given source position ${\bf \beta}$, the excess time-delay surface in the image plane is given by:

\begin{equation}
 t=\frac{D_d D_s (1+z_d)}{cD_{ds}} \left(\frac{1}{2}|{\boldsymbol
 \theta}-{\boldsymbol \beta}|^2- \psi({\boldsymbol \theta}) \right) \equiv \frac{\Ddt}{c} \Phi,
\label{eq:fermat}
\end{equation}

\noindent
where $D_d$, $D_s$, $D_{ds}$ are the angular diameter
distances between the observer and the deflector (at redshift $z_d$), the observer and the
source, and the deflector and the source. $\Ddt\equiv (1 + z_d)\frac{D_d D_s}{D_{ds}}$ is known as the time-delay distance \citep{Suyu:2010}, and $\Phi$ is known as the Fermat potential. The lensing
geometry is illustrated in Figure~\ref{fig:lensdiagram}.  

The quantity $\psi$ is the two-dimensional gravitational lensing potential, satisfying the
Poisson equation:
\begin{equation}
\nabla^2 \psi = 2 \kappa.
\label{eq:poisson}
\end{equation}
\noindent
Here, the dimensionless quantity $\kappa$ is the convergence, i.e., the surface mass density expressed in 
units of the critical density $\Sigma_c=c^2 D_{s} / (4 \pi
G D_{d} D_{ds})$.

According to Fermat's principle, images form at the extrema of
the time-delay surface given by the condition:
\begin{equation}
\boldsymbol \beta=\boldsymbol \theta-\nabla \psi = \boldsymbol \theta - \boldsymbol \alpha
\label{eq:lensequation}
\end{equation}
which is known as the lens equation. The scaled deflection angle $\boldsymbol \alpha$ is related to the
deflection angle $\hat{\boldsymbol \alpha}$ by
$\boldsymbol \alpha=\frac{D_{ds} \hat{\boldsymbol \alpha}}{D_s}$. Note that the
transformation between source and image planes is achromatic and preserves surface brightness.

Strong lensing occurs when Equation~\ref{eq:lensequation} has multiple solutions corresponding to multiple images.  

\subsection{Magnification and distortion}

The Jacobian of the transformation from the image to the source plane gives the inverse magnification, which can be written as
\begin{equation}
\frac{\partial \boldsymbol \beta}{\partial \boldsymbol \theta}=\delta_{ij}-\frac{\partial^2 \psi}{\partial \theta_i \partial \theta_j}=\left(\begin{array}{cc} 1-\kappa-\gamma_1 & -\gamma_2 \\ 
-\gamma_2 & 1-\kappa+\gamma_1 \end{array} \right)
\label{eq:magnification}
\end{equation}
and describes the distortion of a small source. The convergence $\kappa$ determines the isotropic magnification, while the shear components ($\gamma_1$, and $\gamma_2$) determine its elliptical distortions. Higher-order derivatives of the potential describe more complex distortions. For example, the third-order derivatives are known as flexion \citep{G+N02}.

In the limit of a point source, the local magnification $\mu$ is given by the determinant of the magnification tensor:
\begin{equation}
\mu=\frac{1}{(1-\kappa)^2-\gamma_1^2-\gamma_2^2}.
\label{eq:scalarmag}
\end{equation}
\noindent
For extended sources, the total magnification can be obtained by integration since the surface brightness is conserved. The curves in the image plane that correspond to infinite
magnification are called critical lines. The
corresponding curves in the source plane are called caustics. Magnification factors of several thousand have been observed for compact sources \citep{Kelly:2018}. The total observed flux is always finite for astrophysical sources.

\subsection{The Einstein Radius}

The Einstein radius $\theta_{\rm E}$ is perhaps the single most important property of strong lenses. For a circular deflector, it is the radius of the region inside which the average
surface mass density equals the critical density.  The Einstein radius depends on the enclosed mass $M_{\rm E}$ as well as on the
angular diameter distances to the deflector, to the source, and between the two: 

\begin{equation}
\theta_{\rm E}=\sqrt{\frac{4 G M_{\rm E} D_{ds}}{c^2 D_{d}D_{s}}}.
\label{eq:thetaE}
\end{equation}

The definition of Einstein radius
needs to be modified for non-circular deflectors \citep[e.g.,][]{Kormann:1994}. The Einstein radius summarizes the lensing strength of an object, and can be measured to within 1-2\% precision with good data, including all random and systematic uncertainties.

\subsection{On sources and deflectors}

\begin{figure}[!t]
\centering\includegraphics[width=0.48\textwidth]{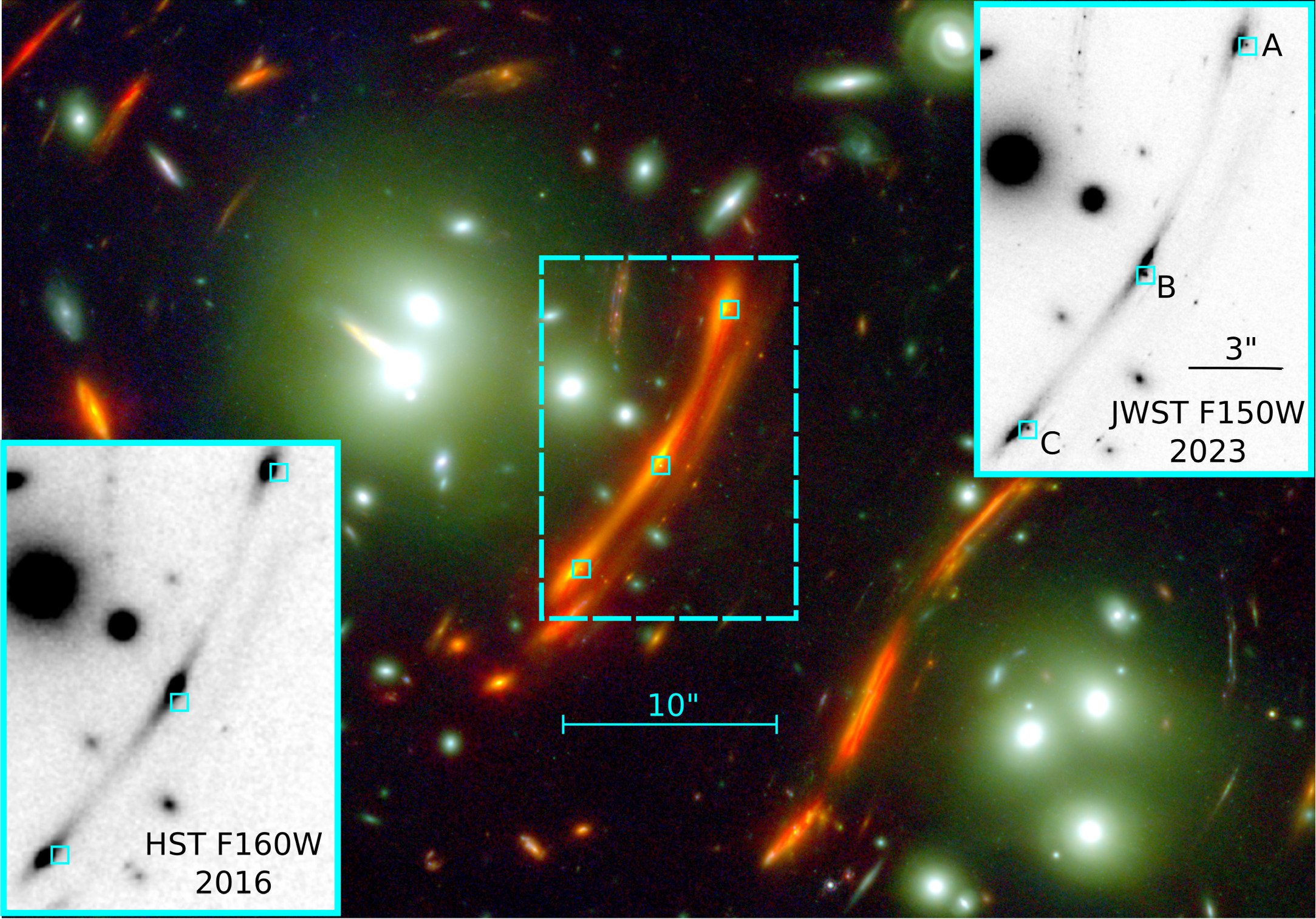}
\caption{The first multiply imaged type Ia supernova with measured time delay, named ``H0pe". The standardizable candle luminosity was used to break the mass-sheet degeneracy and measure \Ho. Figure from~\citet{Pascale:2025}.} 
\label{fig:H0pe}
\end{figure}

The electromagnetic sources of interest for cosmological applications of strong lensing are galaxies, active galactic nuclei, supernovae, and individual stars\footnote{See \citet{Smith2025,Taak:2023} for multi-messenger strong lensing.}. Within this group, galaxies are by far the most abundant, outnumbering active galactic nuclei by 2-3 orders of magnitude. Because the light emission from galaxies does not vary over human timescales, they can be used only for cosmological applications of strong lensing that do not require variability. Active galactic nuclei are typically variable \citep{T+T23} and have been used successfully for time-delay cosmography \citep[e.g.,][]{TreuKoopmans:2002,Suyu:2010,Dux:2025,TDCOSMO25}. AGN appear as point sources, but the host galaxy extended images provide information that is useful to constrain the foreground gravitational potential (Figure~\ref{fig:lensdiagram}). Multiply imaged supernovae were initially suggested by \citet{Refsdal:1964} as targets for time-delay cosmography, but it took 50 years to discover the first example \citep{Kelly:2015}, because they need to be observed at the right time. Several examples of supernovae multiply imaged by clusters \citep[][and example in Figure~\ref{fig:H0pe}]{Pascale:2025,Suyu:2026,Pierel:2026,Bazzanini:2026} and one lensed by a galaxy \citep{Winny5} have now been discovered and used for cosmography. 

The main deflectors in extragalactic strong lensing are typically galaxies, groups, or clusters of galaxies. Of the galaxy-scale lenses, the vast majority are massive elliptical galaxies, which dominate the lensing cross-section \citep{Treu:2010}, although spiral deflectors \citep{Treu:2011} and low mass galaxy deflectors \citep{Shu:2015} can be found via dedicated searches. 

The space of deflectors is continuous and not discrete, as massive galaxies tend to be the centers of groups, and, within groups and clusters, individual galaxy members can act as lenses in their own right. However, it has been customary in the literature to treat galaxy and cluster lenses differently, owing to a few significant differences. First, galaxy-scale lenses have Einstein radii of order arcseconds, while clusters can reach arcminute scale. Thus, the lensing images span 2-3 orders of magnitude more pixels for clusters than for galaxy-scale lenses. Likewise, the area within the caustics is 2-3 orders of magnitude larger for clusters, so that tens of individual sources per lens are multiply imaged at HST/JWST depth and resolution, while typically one or two sources are multiply imaged by galaxies. Thus, cluster-scale lens modelers focused traditionally on using the so-called conjugate points technique, where multiply imaged sources are treated as points (or collection of points) and their positions are used as constraints. Only recently has pixel-based modeling of cluster-scale lenses become feasible \citep{Acebron:2025,Schuldt:2026,Urcelay:2026,Sengul:2025}. In contrast, pixel-based techniques, modeling the full surface brightness of the lensed source, have been practical in galaxy-scale lensing for over two decades \citep{W+D03,TreuKoopmans:2004,Koopmans:2005,B+L06}.

Second, from a deflector standpoint, while the gravitational potential of galaxy-scale lenses is generally dominated by a single galaxy in dynamical equilibrium, the potential of cluster-scale lenses requires the description of multiple components out of dynamical equilibrium.  Thus, for galaxy-scale lenses it has been possible to construct self-consistent models of the lensing and stellar kinematic data based on well-known techniques developed for massive galaxies \citep{TreuKoopmans:2002,Barnabe:2007,Suyu:2010,Yildirim:2020,Shajib:2023,Wang:2025,Sheu:2026}. For cluster-scale lenses, multi-component or free-form models must be used with stellar kinematics used to anchor the individual components \citep{Granata:2025}.

\subsection{The mass-sheet degeneracy}

A key challenge to overcome when using strong lensing time delays for cosmology is the so-called ``Mass-Sheet Degeneracy'' \citep[MSD,][]{Falco:1985,SchneiderSluse:2013,SchneiderSluse:2014}.  The MSD is the most important of the known lensing degeneracies for cosmological applications. 
It will thus be a recurring theme of this article \citep[see, e.g., ][for a pedagogical introduction to strong lensing and its degeneracies]{Saha:2024}. 

In the simplest terms, for any convergence  $\kappa$ that reproduces the observed lensing configuration, the mass-sheet transform (MST) generates a family of solutions that also satisfies the same constraints:

\begin{equation}
\label{eq:mst-kappa}
\kappa_{\ltot}(\xx) = \ltot \kappa(\xx) + (1-\ltot).
\end{equation}

The Fermat potentials are re-scaled by the same mass-sheet parameter $\ltot$, such that $\Delta \Phi_{\ltot} = \ltot \Delta\Phi$ (and therefore, for a given cosmology, $\Delta t_{\ltot} = \ltot \Delta t$). The source magnification and position, and hence luminosity, are also affected by the MST. 
\begin{equation}
\boldsymbol\beta_{\ltot} = \boldsymbol \beta/\ltot ; \quad
\mu_{\ltot}=\mu/{\ltot}^2.
\label{eq:msscaling}
\end{equation}
Under the MST, for a measured time delay, the inferred time-delay distance transforms as
\begin{equation}
\label{eq:Dt_mst}
    D_{\Delta {\rm t},\ltot} = \ltot^{-1} \Ddt,
\end{equation}
showing that the MST must be broken in order to measure absolute distances and obtain cosmological information. The MST can be generalized to multiplane lensing \citep{McCully:2014,Schneider2019,Teodori:2022,Teodori:2026}. Multiplane lensing effects need to be taken into account for precision cosmology. 

Four pieces of information can be  used to break the MST in time-delay cosmography: i) astrophysical information on the mass distribution of galaxies; ii) stellar kinematics; iii) standard candles; iv) weak lensing. Before I discuss these approaches, it is useful to gain some insight by separating $\ltot$ into an internal mass-sheet parameter ($\lint$) and an external mass sheet ($\kext$):
\begin{equation}
\ltot \equiv \lint (1-\kext).
\end{equation}
Physically, the external convergence $\kext$  can be thought of as the first-order approximation of the combined effect of all the matter over- and under-density along the line of sight. It is related to the mass density between the source and deflector by the following equation \citep{Birrer:2020,Johnson:2026}

\begin{equation}
    1-\kappa_{\rm ext} \equiv \frac{(1-\kappa_{\rm d})(1-\kappa_{\rm s})}{1-\kappa_{\rm ds}} \approx 1 - \kappa_{\rm d}-\kappa_{\rm s} + \kappa_{\rm ds}.
    \label{eq:kappa_ext_breakdown}
\end{equation}

For a perfectly homogeneous universe $\kext=0$, overdense lines of sight will have $\kext>0$, and underdense ones $\kext<0$. 
$\kext$ can be estimated statistically using cosmological numerical simulations. It is typically found to be a few percent \citep{Suyu:2010,Fassnacht:2011,Greene:2013,Rusu:2017,Wells:2024,Johnson:2026}. Direct measurements using weak lensing yield consistent results \citep{Tihhonova:2018,Tihhonova:2020}. In this approach, individual mass components along the line of sight that contribute more than a few percent convergence need to be identified and modeled within a multiplane framework \citep[e.g.,][]{Rusu:2020}. 

\begin{figure*}[!t]
\centering\includegraphics[width=\textwidth]{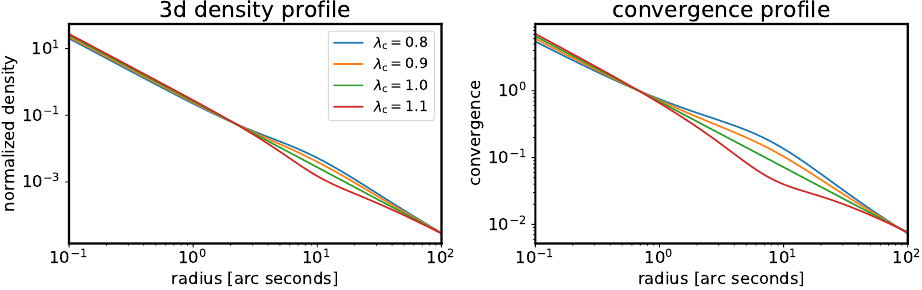}
\caption{Illustration of the implementation of the approximate mass-sheet transformation on a power law mass density profile. All the models shown here are physical (no mass at infinity) and yet reproduce the image positions on arcsecond scales within the observational errors. Figure from \citet{Birrer:2020}.} 
\label{fig:MST}
\end{figure*}

In contrast, the internal MST parameter $\lint$ can be thought of as flexibility in the mass density profile of the main deflector. Galaxies do not have infinite mass sheets. Therefore, for sufficiently large radii, the mass sheet must vanish. Nevertheless, one can produce the same effect with an approximate MST \citep{Birrer:2020,Shajib:2023}, as shown in Figure~\ref{fig:MST}, in which the mass sheet tapers off at sufficiently large radii to leave the lensing configuration unaffected without requiring galaxies to extend to infinity \citep{Blum:2020}. In practice, one can start from a standard mass density profile, for example the elliptical power-law that has been shown to describe galaxy-scale lenses well \citep{Treu:2010}, and then modify it with the MST, thus effectively maximizing the uncertainty on the time-delay distance.

\subsubsection{Breaking the mass-sheet degeneracy}

I now turn to the four ways to break the MST. Approach one consists of using what we know about galaxies to choose a parametrization of the mass density profile of the main deflector that is accurate enough and flexible enough to not require additional freedom in the form of $\lint$.  This was the approach followed by \citet{Wong:2020} and \citet{Millon:2020} who used power-law models inspired by X-ray and stellar kinematic observations of galaxies, and theory-inspired models, consisting of a \citet{NFW} halo and stars, finding consistent results. This is the approach referred to as ``assertive'' by \citet{TSM22}. It has the advantage of yielding high precision, but it could be biased if the assumed parametrization is incorrect.

The other three approaches are more conservative. They break the MSD using non-strong-lensing information and take fully into account the uncertainty associated with the residual MST. Approach number two consists of using stellar kinematics as a lensing-independent measure of the gravitational potential of the main deflector \citep{TreuKoopmans:2002,Shajib:2018,Birrer:2020,B+T20,Yildirim:2023,Wang:2025}, which is a massive early type galaxy in 90\% of galaxy-scale lenses \citep{Treu:2010}. Exploiting stellar kinematics requires additional parameters describing the stellar tracers, the elongation along the line of sight, the anisotropy of the stellar orbits, and the angular diameter distance to the deflector. With high-quality stellar kinematics \citep{Knabel:2025} and using well-established methods in stellar dynamics \citep{Cappellari2016} it is possible to constrain simultaneously the mass-sheet degeneracy, the mass-anisotropy degeneracy \citep{BinneyMamon:1982,Courteau2014}, and obtain cosmological information \citep{Shajib:2023,TDCOSMO25,Sheu:2026}. 

The fundamental equation is that relating the predicted stellar velocity dispersion projected along the line of sight $\sigma_{\rm los}$ to cosmological distances and to a dimensionless quantity $J$ \citep[see e.g.,][]{Birrer:2016,Birrer:2019}
that describes the internal structure of the deflector ($\boldsymbol{\xi}_{\rm lens}$) and the observational conditions (seeing and spectroscopic aperture) of the velocity dispersion measurement \citep[e.g.,][]{BinneyMamon:1982, TreuKoopmans:2004, Suyu:2010}:
\begin{equation}\label{eqn:los_sigma_v}
    \sigma_{\rm los}^2 =  \frac{1-\kappa_{\rm s}}{1 - \kappa_{\rm ds}} \frac{D_{\rm s}}{D_{\rm ds}}c^2 J(\boldsymbol{\xi}_{\rm lens}, \boldsymbol{\beta}_{\rm ani},\lambda_{\rm int}).
\end{equation}
Here $c$ is the speed of light; $\boldsymbol{\beta}_{\rm ani}$ characterizes the anisotropy profile of the stellar orbits; and $\kappa_{\rm s}$ and $\kappa_{\rm ds}$ are the LoS external convergences from the observer to the source and the deflector to the source, respectively.  The equation can be easily generalized to spatially resolved kinematics.

Approach number three exploits the fact that supernovae Ia are standardizable candles to obtain an absolute measurement of magnification in lensed supernovae, thus breaking the MSD \citep{K+B98,Birrer:2022,Pascale:2025}. One potential caveat to this method is that the stellar microlensing \citep{Foxley:2018,Goldstein:2018} or millilensing \citep{Vegetti++24} could impact the flux measurement given the small size of the photosphere \citep[see, also, discussion by][]{Kelly:2023}.

\begin{figure}
\begin{center}
 \includegraphics[width=\linewidth]{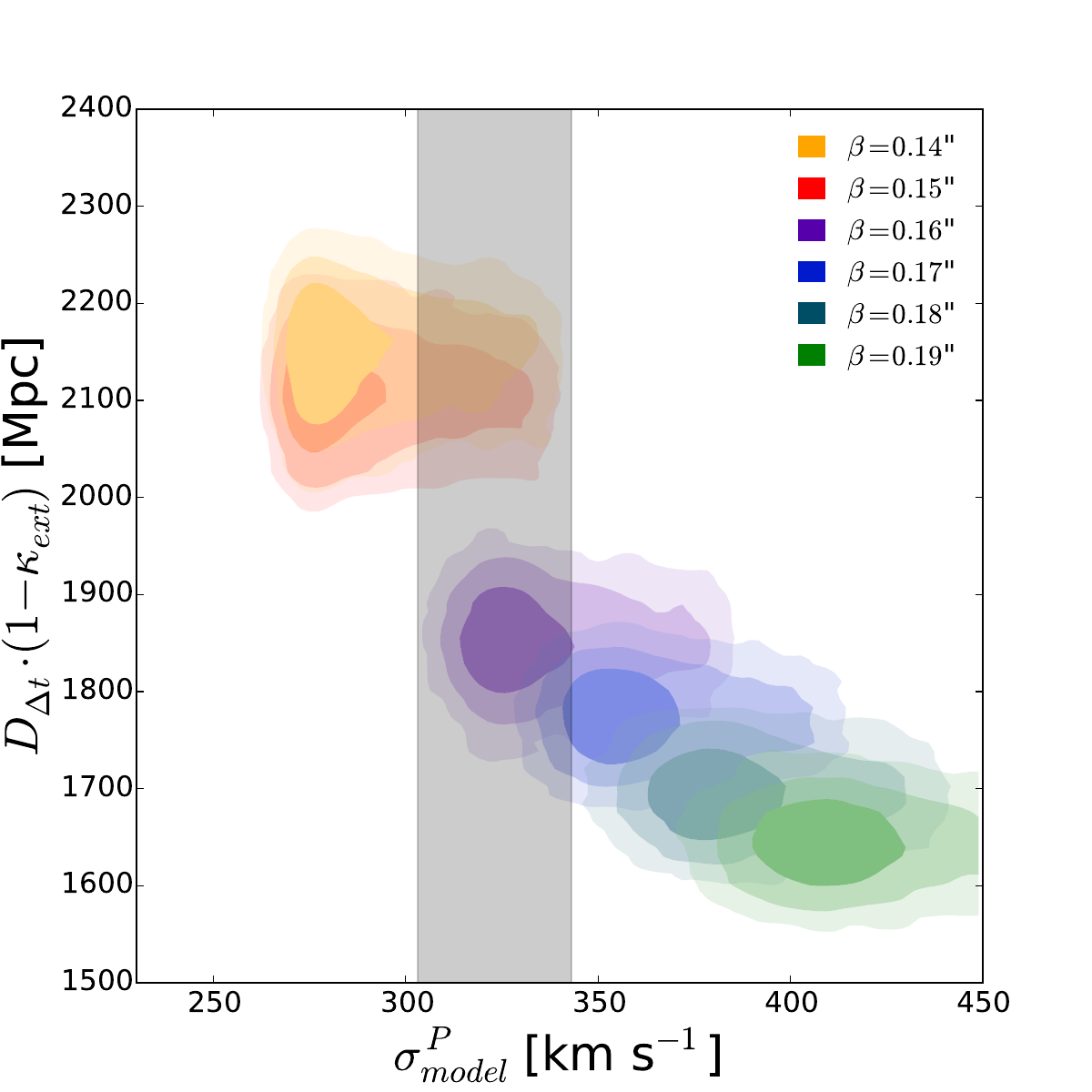}
 \caption{Estimated line of sight central velocity dispersions $\sigma_{\rm p}$ as a function of time-delay distance $\Ddt$ of a sample
of lens models with varying source size $\beta$. Note that varying the source size corresponds to performing a mass-sheet transformation (Eq~\ref{eq:msscaling}). The gray band reflects
the 68\% uncertainty range of the LOS velocity dispersion estimates from the data available when this plot was produced in 2016. This plot illustrates how stellar velocity dispersion helps break the MSD. Figure from \citet{Birrer:2016}.}
 \label{fig:Birrer16}
\end{center}
\end{figure}

\begin{figure}
\begin{center}
\includegraphics[width=\linewidth]{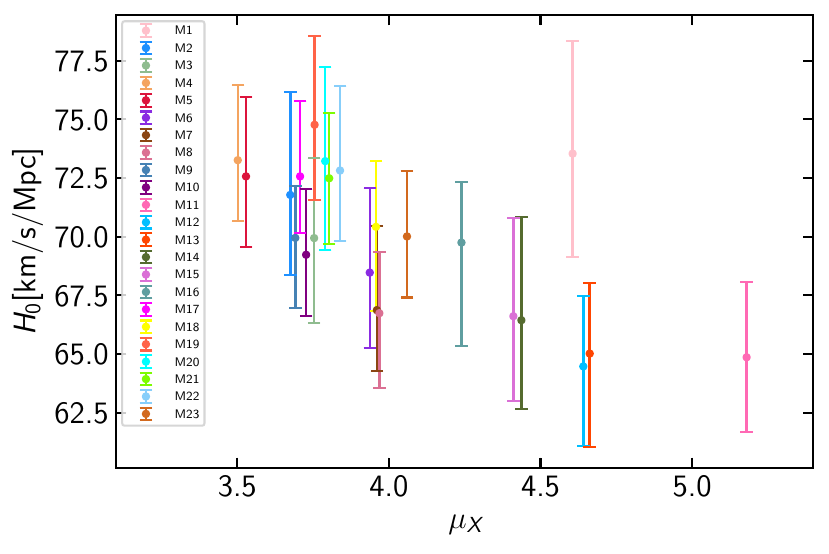}
\includegraphics[width=\linewidth]{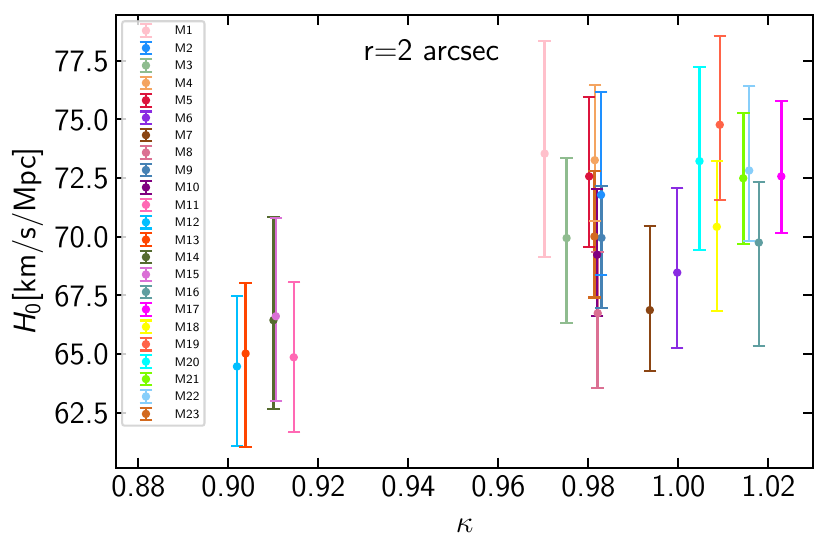}
\includegraphics[width=\linewidth]{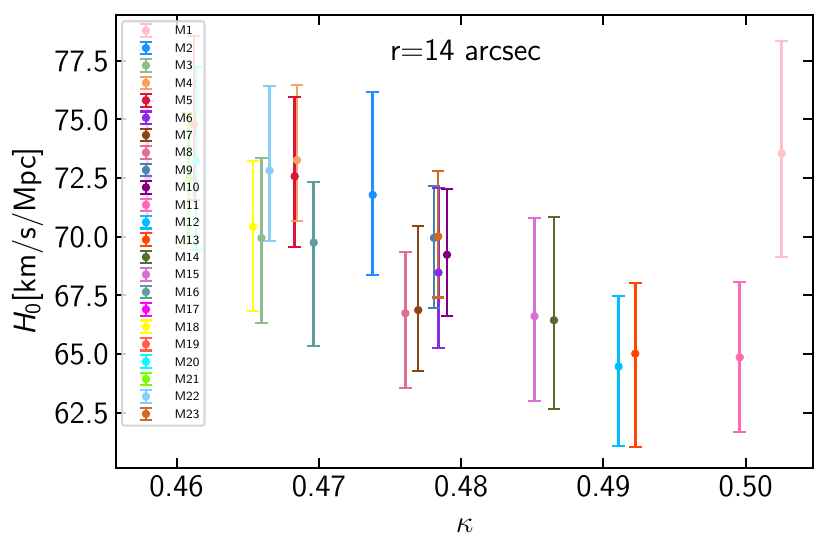}
\caption{Top panel: Correlation between magnification factors of a multiple supernova image and Hubble constant measurements for 23 different lens mass models of cluster MACS1149 lensing supernova Refsdal. The correlation is effectively along the mass-sheet transformation. If the source is a standard candle, magnification is measurable, thus breaking the MSD. Middle and bottom panels: Correlation between the convergence at two different radii and the Hubble constant for the same models. This plot illustrates the modification to the projected mass density profiles, resulting in an effective MSD, similar to the mathematical transformation shown in Figure~\ref{fig:MST}. Figures from \citet{L+O25}.} 
 \label{fig:Liu25}
\end{center}
\end{figure}

Illustrations of how to use approach number two and three to break the MSD are shown in Figures~\ref{fig:Birrer16} and~\ref{fig:Liu25} respectively.
 
Approach number four, proposed by \citet{Khadka:2024}, consists of using weak lensing signal from stacks of large numbers of massive galaxies to constrain the convergence profile from infinity to the Einstein radius, the region most relevant for cosmography \citep{Kochanek:2002}. This approach could become competitive with the other three with the large number of shear measurements that are expected from wide field imaging surveys with Euclid, Rubin, and Roman. The main challenge that will need to be overcome to implement this approach  is that it relies on using large samples of non-lens galaxies well matched to the time-delay lenses, because there are not enough time delay lenses. Therefore, one will need to understand the selection function with residual uncertainties and bias that are smaller than the desired precision.

Approaches two, three, and four are completely independent and can be used as a mutual check of systematics. They can be combined in the case of a supernova Ia lensed by a foreground galaxy. 

\subsection{Strong lensing time-delay cosmography}

The cosmographic power of strong lensing time delays, when using stellar kinematics to break the MSD, is summarized by the following two equations:
\begin{equation}\label{eqn:Ddt_cosmography}
    \Ddt = \frac{1}{(1-\kappa_{\rm ext}) \lambda_{\rm int}}\frac{c \Delta t_{\rm AB}}{\Delta \Phi_{\rm AB}}
\end{equation}
and 
\begin{equation}\label{eqn:Dd_cosmography}
    D_{\rm d} = \frac{1}{1 + z_{\rm d}}
    \frac{c \Delta t_{\rm AB}}{\Delta \Phi_{\rm AB}}  \frac{c^2 J(\boldsymbol{\xi}_{\rm lens}, \boldsymbol{\beta}_{\rm ani}, \lambda_{\rm int})}{\lambda_{\rm int}\sigma^2_{\rm los}},
\end{equation}
where $\Delta t_{\rm AB}$ and $\Delta \Phi_{\rm AB}$ are respectively the measured time-delay difference and the predicted Fermat potential difference between two of the multiple images. If there are more than two images, additional time-delay equations provide additional constraints \citep[e.g.,][]{Dux:2025,Dux:2025b,Sheu:2026,Galan:2026}.

For any given cosmological model, $\Ddt$ and $\Dd$ are related. Therefore, the constraints in the two-dimensional parameter space are effectively reduced to a line in the plane.  Thus, in practice, when using stellar kinematics and accounting for the MSD, the cosmological information comes from a combination of time-delay distance and angular diameter distance to the deflector \citep{Birrer:2016,Jee:2015,Jee:2019}. In this review, I sometimes refer to the measurement just as time-delay distance for simplicity.

\subsection{Multiplane and compound lenses}

\begin{figure*}[!t]
\centering\includegraphics[width=\textwidth]{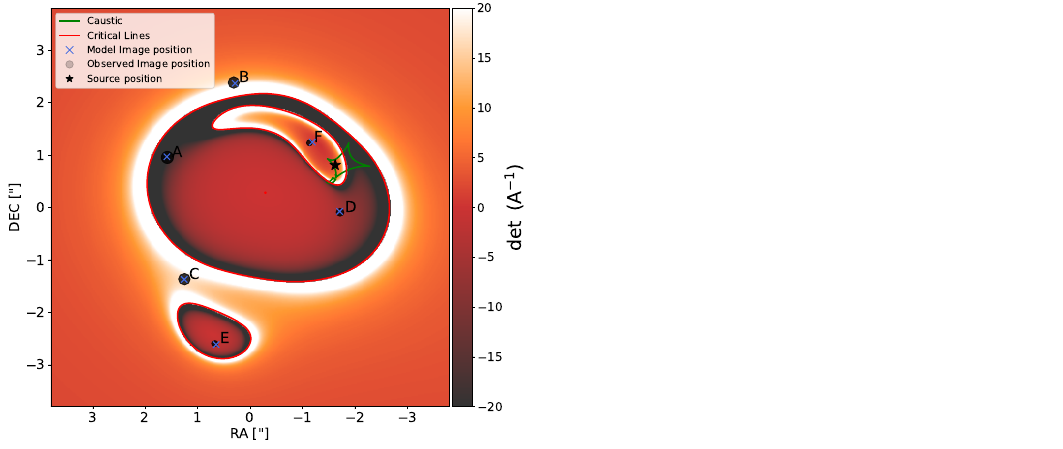}
\caption{Configuration of the first discovered Einstein Zig-Zag lens. The galaxy at $z=1.885$ is multiply imaged by the foreground deflector, but also acts as an additional deflector for the background quasar, producing six images to the observer. Figure from~\citet{Dux:2025b}.} 
\label{fig:zigzag}
\end{figure*}

When more than one source at different redshifts happens to be multiply imaged by a foreground mass, an additional opportunity for cosmography arises. Suppose, for simplicity, that the deflector is a point mass. Then, the ratio of Einstein radii of the two sources is

\begin{equation}
\frac{\theta_{\rm E,1}}{\theta_{\rm E,2}}=\sqrt{\frac{D_{\rm ds,1}D_{\rm s,2}}{D_{\rm s,1}D_{\rm ds,2}}}.
\label{eq:thetaEratio}
\end{equation}
Thus, multiplane lenses provide in principle a way to measure ratios of angular diameter distances and are thus particularly sensitive to quantities such as the matter density $\Omega_{\rm m}$ \citep{Golse:2002} and the dark energy equation of state parameters \citep{Gavazzi:2008,Collett:2014}. In practice, since the deflectors are extended, the Einstein-radius ratio depends on the mass enclosed between the image radii of the different sources, so the radial mass profile must be modeled in each deflector plane. In addition, the nearest multiply imaged source can contribute non-negligible deflection to the trajectory of the light rays from the most distant one \citep{Gavazzi:2008}, hence the name compound lens. In special cases, compound lenses can produce exotic configurations such as the six quasar images produced by the Einstein zig-zag lens \citep{Collett:2016b,Dux:2025b,Schmidt:2025} shown in Figure~\ref{fig:zigzag}. The mass-sheet degeneracy is also relevant for this kind of application \citep{Schneider:2014,JFM26} and needs to be broken with non-lensing information.

When applying this method to real systems, clusters and galaxy-scale lenses offer complementary advantages and disadvantages. For cluster-scale systems, the main advantage is the availability of tens of sets of multiple images, at the cost of  complexity in the gravitational potential of the deflector. For galaxy-scale systems, the main advantage is the simplicity of the deflector, at the cost of a smaller number of multiply imaged sources. Great progress is being made in both regimes \citep[e.g.,][]{DAN:2011,Collett:2014,Sahu:2025,Johnson:2025,Caminha:2022,Urcelay:2026,Bowden:2025}.

\subsection{From distances to cosmology}

\label{ssec:TD2COSMO:dist}

When kinematics is used, each time-delay lens produces a two-dimensional probability distribution function in the two-dimensional parameter space of the two distances $\Ddt$ and $\Dd$ (Equations~\ref{eqn:Ddt_cosmography} and~\ref{eqn:Dd_cosmography}). Compound lenses measure ratios of angular diameter distances. Like every distance in cosmology, $\Ddt$ and $\Dd$ depend on both the cosmological model parameters and the lens/source redshifts, as I summarize below.  

The FLRW metric that describes a homogeneous and isotropic universe is
\begin{equation}
\label{eq:FLRWmetric}
{\rm d}s^2 = c^2 {\rm d}t^2 - a^2(t)\left [{\rm d}\chi^2 +f_{K}^2 (\chi)\left({\rm d}\vartheta^2 + \sin^2\vartheta {\rm d} \varphi^2\right) \right ],
\end{equation}
where $c$ is the speed of light, $t$ is the cosmic time, $a(t)$ is the cosmic scale factor (normalized to 1 today at $t_0$, i.e., $a(t_0)=1$), $\chi$ is the comoving radial coordinate, $(\vartheta, \varphi)$ are the angular coordinates on a unit sphere, and $f_{K} (\chi)$ is the comoving angular diameter distance that depends on the spatial curvature $K$ as
\begin{equation}
\label{eq:fk}
f_{K}(\chi) = \left\{ \begin{array}{ll}
 K^{-1/2} \sin\left(K^{1/2}\chi\right) & \textrm{for $K>0$}\\
 \chi & \textrm{for $K=0$} \\
 (-K)^{-1/2}\sinh\left[(-K)^{1/2}\chi \right] & \textrm{for $K<0$}
  \end{array} \right. .
\end{equation}
The angular diameter distance between two redshifts $z_1$ and $z_2$ is
\begin{equation}
\label{eq:DA}
D_{\rm A}(z_1,z_2) = \frac{1}{1+z_2}f_{K}[\chi(z_1,z_2)].
\end{equation}

In general, $\chi(z_1,z_2)$ is related to cosmological parameters by an integral relation in the cosmological model, $\Lambda$CDM, with density parameters $\Omega_{\rm m}$ for matter, $\Omega_{\rm k}$ for spatial curvature, $\Omega_{\rm DE}$ for dark energy:
\begin{equation}
\label{eq:chi}
\begin{split}
\chi(z_1,z_2) & = \frac{c}{{\rm H}_0}\int_{z_1}^{z_2} {\rm d}z' [\Omega_{\rm r}(1+z')^4 +\Omega_{\rm m}(1+z')^3 \\
& + \Omega_{\rm k} (1+z')^2 + \Omega_{\rm DE}(1+z')^{3(1+w)} ]^{-1/2}. \\
\end{split}
\end{equation}
Note that $K=-\Omega_{\rm k}{\rm H}_0^2/c^2$ is the spatial curvature, and $\Omega_{\rm r}$ is the energy density in relativistic particles (photons and meutrinos).  Consequently, the angular diameter distance $D_{\rm A}$ depends on \{H$_0$, $\Omega_{\rm m}$, $\Omega_{\rm k}$, $\Omega_{\rm DE}$, $w(z)$\} in the $w$CDM cosmological model. If dark energy is the cosmological constant $\Lambda$, then $w=-1$ and $\Omega_{\rm DE}=\Omega_{\Lambda}$. Since $\dt$ is a combination of angular diameter distances between the observer, lens, and source, one can use $P(\dt,\dd)$ to infer cosmological parameters. Similarly, for compound lenses, one can use the ratio of distances to infer cosmological parameters.

\subsection{Probes of dark matter: Flux ratio anomalies, astrometric anomalies, time-delay anomalies}

The sensitivity of strong lensing to the distribution of matter on sub-galactic scales makes it a unique probe of certain classes of dark matter models. For example, in standard cold dark matter (CDM) cosmologies, galaxies are predicted to be surrounded by a hierarchy of subhalos, with specific number density and internal mass distribution \citep[e.g.,][]{Springel:2005}. In warm dark matter (WDM) models, the number of subhalos is suppressed below a certain scale corresponding to the free streaming length, and the concentration of subhalos is generally lower than in CDM \citep{Ludlow++16}. Given current limits on the free streaming length of dark matter \citep{Gilman26,Irsic:2024,Villasenor++23,Nadler++21,NRD26}, the most interesting regime to probe is that of subhalo masses below 10$^8$ M$_\odot$, where most of the subhalos are expected not to contain stars and can therefore only be detected gravitationally. The Einstein radii of these subhalos are typically of the order of milliarcseconds \citep{Treu:2010}, and thus the phenomenon is often referred to as millilensing.

\begin{figure*}
		\centering
		\includegraphics[trim=2cm 0.4cm 1cm
		0cm,width=0.9\textwidth]{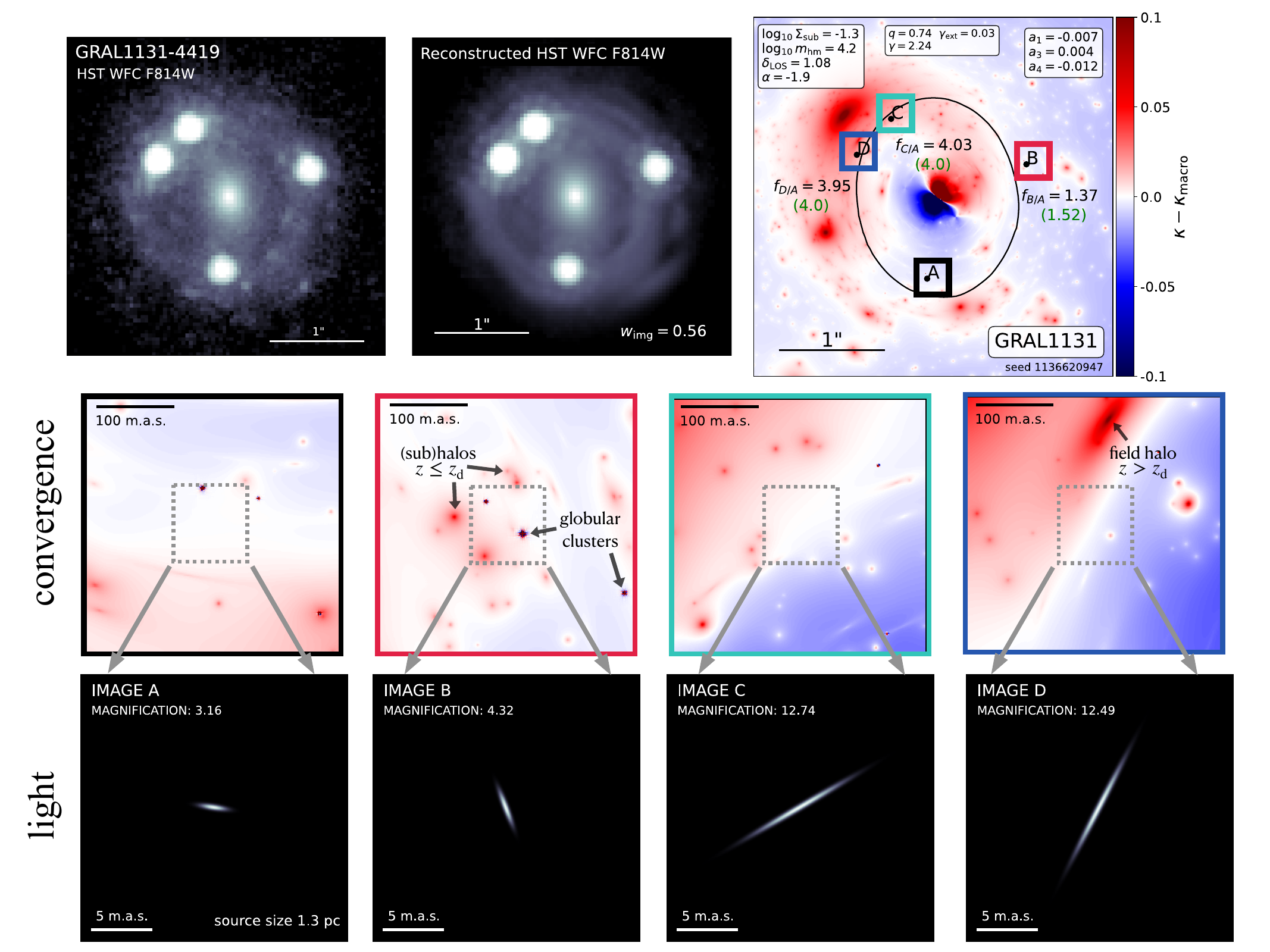}
		\caption{\label{fig:gilman-1131} Illustrations of the calculation of the likelihood function in modern methods that combine flux ratio anomalies with extended source reconstruction to infer dark matter properties \citep{Gilman:2024,Gilman26}. The quadruply imaged quasar GRAL1131+4419 is used as an example.  {\bf{Top row:}} Observed (left) and reconstructed (center) image. The center panel shows the importance weight of the imaging data while the right panel shows the effective convergence. The dark matter hyper-parameters are listed. The black line is the critical curve. {\bf{Center row:}} Zoomed in regions around each lensed image showing the effective convergence near the image. {\bf{Bottom row:}} Model-predicted quasar images obtained by ray tracing through the lens system to a background source with a size 1.3 pc, characteristic of the compact warm dust region responsible for the measured flux. The panels show how a lensed image would appear at infinite resolution. Figure from \citet{Gilman26}.}
	\end{figure*}

There are several ways in which strong lensing can probe dark matter substructure. One is the flux-ratio anomalies technique. It exploits the fact that magnification is sensitive to the second derivatives of the potential (Equation~\ref{eq:scalarmag}), and thus subhalos can modify the observed flux ratios of unresolved images by factors of order unity or more with respect to a smooth mass distribution with the same mass. In order to apply this technique to probe the nature of dark matter by measuring the mass function and density profile of subhalos, one has to be sure that the anomalies are not due to other astrophysical effects, such as microlensing by stars or extinction by dust. Thus, the intrinsic size of the unresolved source must be small enough to be affected by milli-lensing but large enough to be insensitive to microlensing, i.e., between milli- and microarcseconds on the sky, and preferably observable at wavelengths unaffected by dust. Popular choices have been radio emission for radio-loud quasars \citep{Mao:1998,M+M01,Dalal++02,Hsueh++20}, narrow line emission for active galactic nuclei \citep{Moustakas2003,Nierenberg++14,Nierenberg++20,Paugnat:2026},  hot dust emission for active galactic nuclei \citep{Agol:2000,Chiba2005,Nierenberg++24}, and individual stars \citep{Dai:2020}. 
In current state-of-the-art applications of this technique, the extended emission from the host galaxy of the active nucleus is modeled together with the observed flux ratios in order to constrain the distribution of matter on the scale of the lens galaxy and mitigate potential degeneracies between, e.g., the angular structure of the main deflector and the inferred dark matter properties \citep[Figure~\ref{fig:gilman-1131},][]{Gilman:2024,Gilman26}.

\begin{figure}
    \includegraphics[angle=270,width=0.48\textwidth]{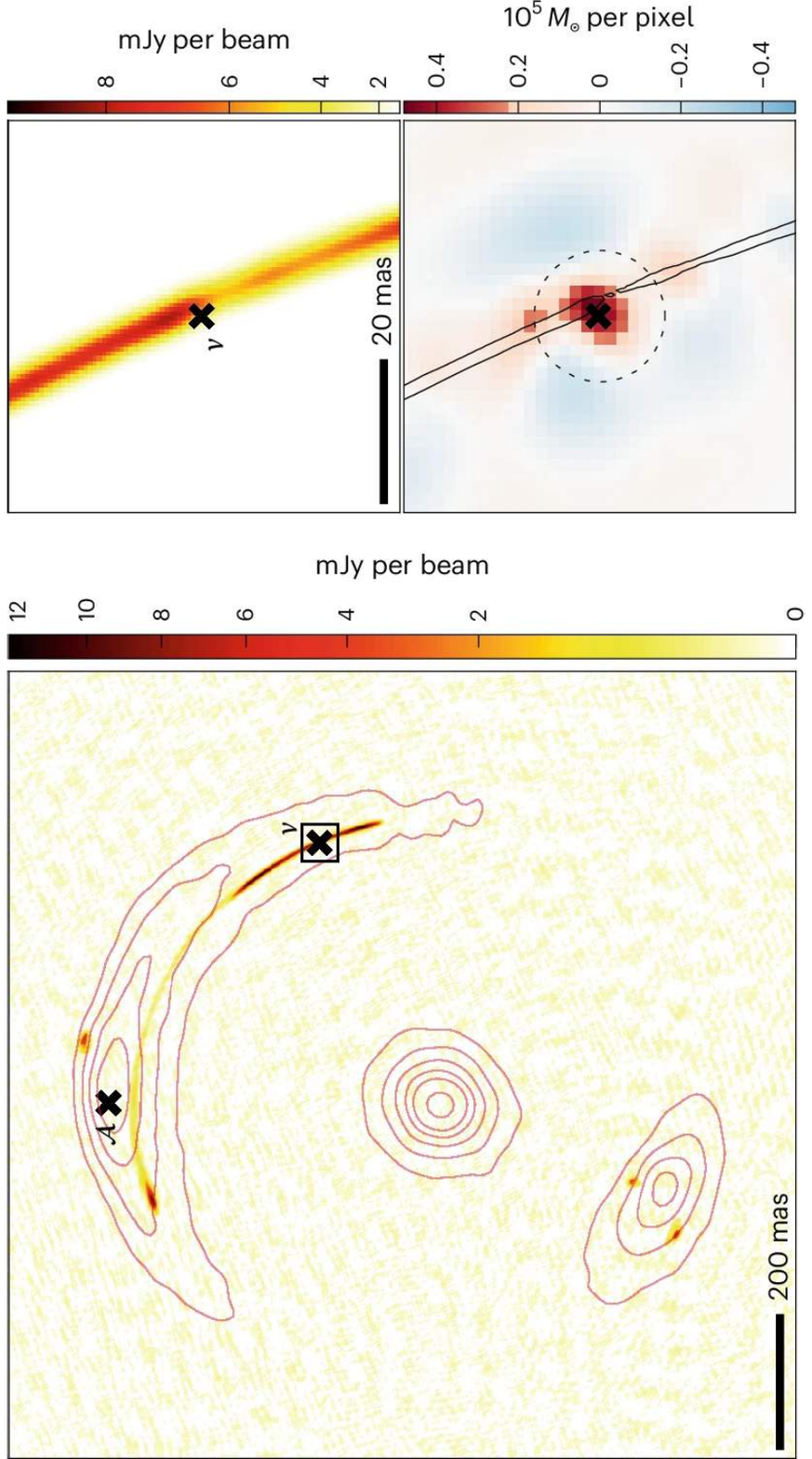}
    \caption{Surface brightness model of the 1.7-GHz global VLBI observation used by \citet{Powell++25}. The near infrared observations with the W.~M.~Keck Observatory adaptive-optics system are shown as red contours. The positions of two low-mass perturbers are each marked with a black X. The $2\times10^8$M$_\odot$ object first detected by \citet{Vegetti++12} is labeled $\mathcal{A}$, and the $10^6$M$_\odot$ detection reported by \citet{Powell++25} is labeled $\nu$. The zoomed-in region shown in the right-hand two panels is indicated by the black square, which has a side length of 60mas. Top right, detail of the bright arc around $\nu$, with the color scale modified to emphasize the gap in the arc produced by the gravitational perturbation of $\nu$. Bottom right, corrections to the lensing convergence, showing a compact, positive feature whose position and mass are consistent with the independent parametric modeling results for $\nu$. The dashed black circle has a radius of 80pc and the lensed emission is indicated by the black contours. Figure and caption from~\citet{Powell++25}.\label{fig:powell}}
\end{figure}

A second technique is the so-called astrometric anomaly technique. It relies on the fact that subhalos perturb the first derivative of the potential and thus the deflection angles and the observed positions of point-like and extended sources. In the case of extended sources, it is often called gravitational imaging \citep{Koopmans:2005}, while for point sources it is usually called astrometric perturbations \citep{Chen:2007}. Angular resolution is key to detecting the effect. Individual satellites have been detected with masses of the order of 10$^{8-9}$ M$_\odot$ using Hubble Space Telescope imaging \citep{Vegetti:2010,Cao:2026} or ground-based adaptive optics in the near infrared \citep{Vegetti++12}. Radio observations \citep{Hezaveh:2013,Hezaveh++16,Amvrosiadis:2026} have been pushed to even smaller masses \citep[Figure~\ref{fig:powell},][]{Powell++25}. It should be noted here that the masses quoted by gravitational imaging papers are the observed ones, which can be orders of magnitude smaller than the mass of the subhalo prior to infall into the host halo and tidal stripping. This is an important distinction to keep in mind when comparing between works.

A third technique is the so-called time-delay anomaly technique. In this approach, one relies on the perturbation introduced by the subhalos on the potential, and thus on the arrival time of the images \citep{Keeton:2009}. The effect is subtle and requires extraordinary precision in time-delay measurements to be detectable. To date, no detection has been performed using this approach.

\section{A brief history of strong lensing cosmology}
\label{sec:history}

Owing to space limitations, I give a brief summary of the history of strong lensing. For a more extensive treatment of the history of strong lensing, see, e.g., \citet{Treu:2010,T+E15,VallsGabaud:2006}. 
For simplicity, I divide the history of strong lensing into three major epochs, summarized in Figure~\ref{fig:timeline}. The first epoch covers the period from the initial theoretical developments to the first discoveries of strong lensing systems (Section~\ref{ssec:early}). The second epoch covers the period from 1990-2020. This was a period of great methodological development and stunning first applications of the new methods to increasingly larger samples of lenses, with better and better data quality. It is summarized in Section~\ref{ssec:firstapp}. I call the period starting around 2020 the era of precision strong lensing cosmology. Major milestones in this period include the launch of the JWST, Euclid,  and Roman satellites, and first light of the Rubin Observatory. This period is briefly introduced in Section~\ref{ssec:precision} and then expanded in the remainder of this article. 

\subsection{Early days: from theory to the discovery of the first lenses}
\label{ssec:early}

\citet{Einstein:1936} famously predicted strong lensing by stars (the phenomenon we now call stellar microlensing), but underestimated how  technology would advance, stating “of course there is no hope of observing this phenomenon directly.”
\footnote{Perhaps less known is the fact that the phenomenon had been predicted nine months earlier by \citet{Link:1936}, in more optimistic terms \citep{VallsGabaud:2006}.}
The field was mostly dormant for decades -- with the exception of key theoretical breakthroughs such as that by \citet{Refsdal:1964} on the use of time delays for measuring distances -- until the first multiply imaged quasars were discovered in the late 1970s \citep{Walsh:1979}, followed by giant arcs lensed by clusters in the 1980s \citep{Soucail:1988} and radio rings \citep{Hewitt:1988}. The observation of the phenomenon generated a wave of interest, leading to great progress in both theory and observation. This early phase of progress is well captured by the foundational books and articles of that time \citep[e.g.,][]{Schneider:1985,BlandfordNarayan86,Falco:1985,Schneider:1992}. Several applications of strong lensing from cosmology to the use of lenses as natural telescopes to magnify background sources were successfully demonstrated. Progress, however, remained slow for two reasons. First, so few strong lenses were known that practitioners of the field knew each one by coordinates and by an endearing nickname. Second, until the refurbishment of the Hubble Space Telescope and the installation of the Wide Field and Planetary Camera 2 in 1993, only radio telescopes had sufficient angular resolution to characterize extended lensed features. This further limited the field. 

\subsection{1990-2020: methodological developments and first applications}
\label{ssec:firstapp}

The pace picked up in the 1990s with dedicated radio  and optical lens searches \citep[e.g.,][]{Myers:2023},  dedicated monitoring campaigns \citep{Schechter:1997}, and Hubble Space Telescope imaging of galaxy-scale \citep{Lehar:2000} and cluster-scale lenses \citep{Smail:1996}.

In the 2000s, progress further accelerated. First, the sample size of known lenses expanded considerably, owing to large-scale optical surveys and the Gaia satellite \citep[e.g.,][]{Bolton:2008,Inada:2012,Lemon:2020,Lemon:2023,Ducourant:2026}, and systematic follow-up of clusters with the Hubble Space Telescope \citep[e.g.,][]{Smith:2001,Postman:2012,Lotz:2017}. Key breakthroughs were achieved in modeling galaxy- and cluster-scale lenses \citep{W+D03,Suyu:2006,Kneib:2011,GLAFIC,N+D15,Nightingale++18,Birrer_lenstronomy,Lombardi:2024,GIGALENS}. High-cadence monitoring of multiply imaged quasars started to provide high-precision time delays at scale \citep[e.g.,][]{Fassnacht:2002,Eigenbrod:2005,Millon2020b,Dux:2025,Paraficz:2010,Hjorth:2002,Schechter:1997,Kundic:1997}. Evidence supporting the existence of dark substructure accumulated \citep{Vegetti:2010,Nierenberg++14,Dalal++02,M+M01,Mao:1998}. Stellar kinematics became a key tool for constraining the gravitational potential of the deflector and breaking the mass-sheet degeneracy \citep{TreuKoopmans:2002,Koopmans:2003,Suyu:2010,Birrer:2020,Shajib:2023,Shajib:2026,Sheu:2026,Wang:2025}. To prevent experimenter bias, blinding was implemented in time-delay cosmography measurements \citep{Suyu:2013}. The first multiply imaged supernova was finally discovered and named ``Refsdal" on the 50th anniversary of Refsdal's paper \citep{Kelly:2015}.  The timing, brightness, and position of the appearance of an additional image were successfully predicted \citep{Treu:2016,Kelly:2016,Grillo:2016,Oguri:2015}.

\subsection{2020-present: Strong lensing precision cosmology}
\label{ssec:precision}

The launch of JWST in 2021 with its unprecedented imaging and spectroscopic capabilities further accelerated progress, marking the era of precision strong lensing cosmology.  The most recent results since 2020, including measurements of H$_0$ from a lensed supernova \citep{Kelly:2023,Grillo:2024} as predicted by \citet{Refsdal:1964},  will be discussed in the next sections. At the time of writing, many of the applications are still limited in their precision by the small number of suitable lenses. This is expected to change dramatically in the next decade through wide-area surveys carried out by the Euclid, Roman, and Rubin observatories. Strong lensing cosmology in the era of target abundance is described in Section~\ref{sec:future}.

\begin{figure*}
\begin{center}
 \includegraphics[width=\linewidth]{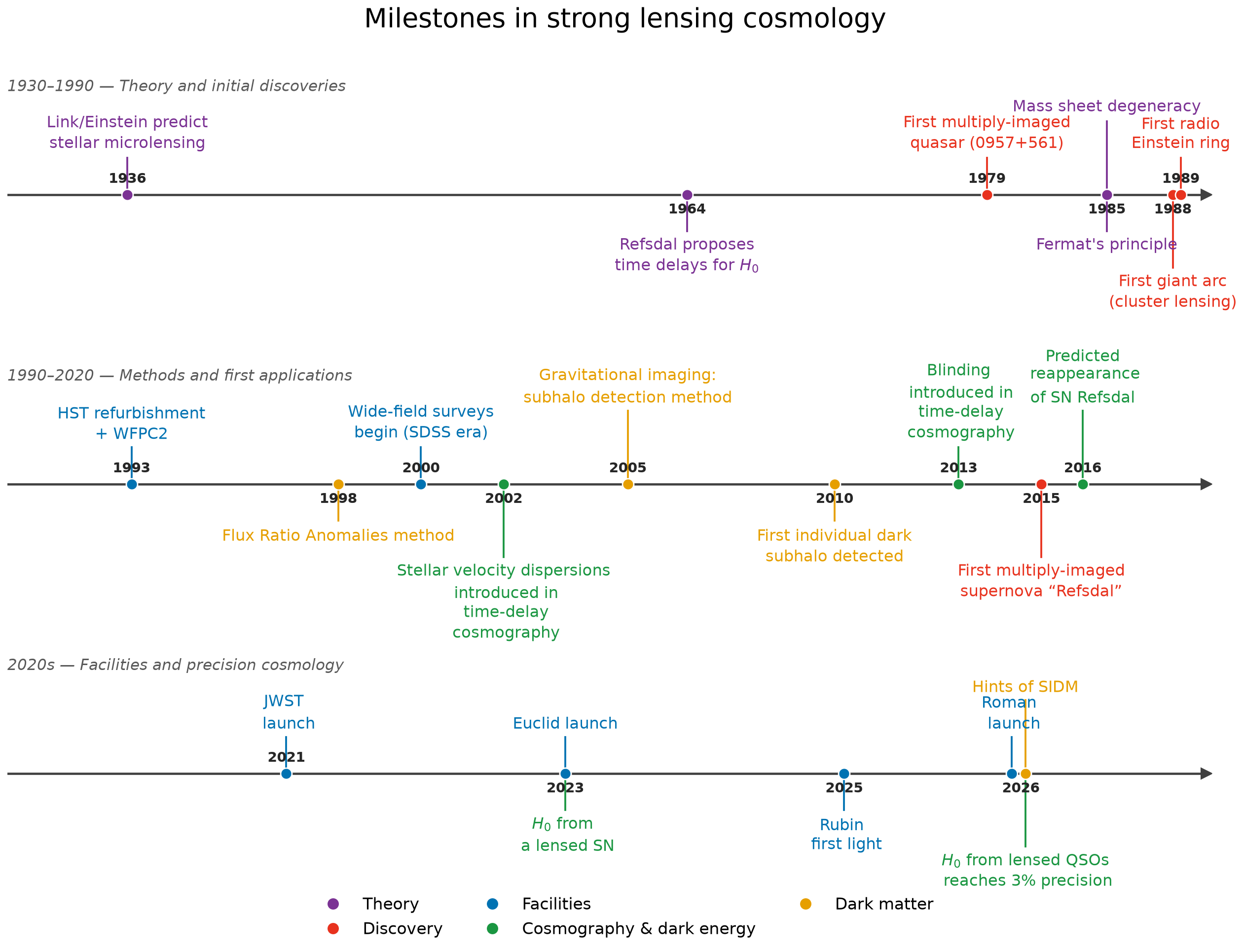}
 \caption{Strong lensing cosmology milestones, divided into three epochs. Highlights of the first epoch include mostly theory and initial discoveries. The second epoch is characterized by methodological developments and applications to larger and higher quality datasets. The third epoch of precision cosmology is the main topic of this review.} 
 \label{fig:timeline}
\end{center}
\end{figure*}

\section{Time delay cosmography, H$_0$, and the ``Hubble Tension"}
\label{sec:expansion}

If there were a prize for the most controversial number in cosmology, the Hubble constant H$_0$ would surely be a strong contender. After decades of debate with measurements ranging between 50 and 100, the question appeared to be settled by the turn of the millennium when the Hubble Key Project delivered a measurement with 10\% precision, settling on H$_0 = 72\pm8$ \hunit~\citep{Freedman:2001}. 

\begin{figure*}
\begin{center}
 \includegraphics[width=\linewidth]{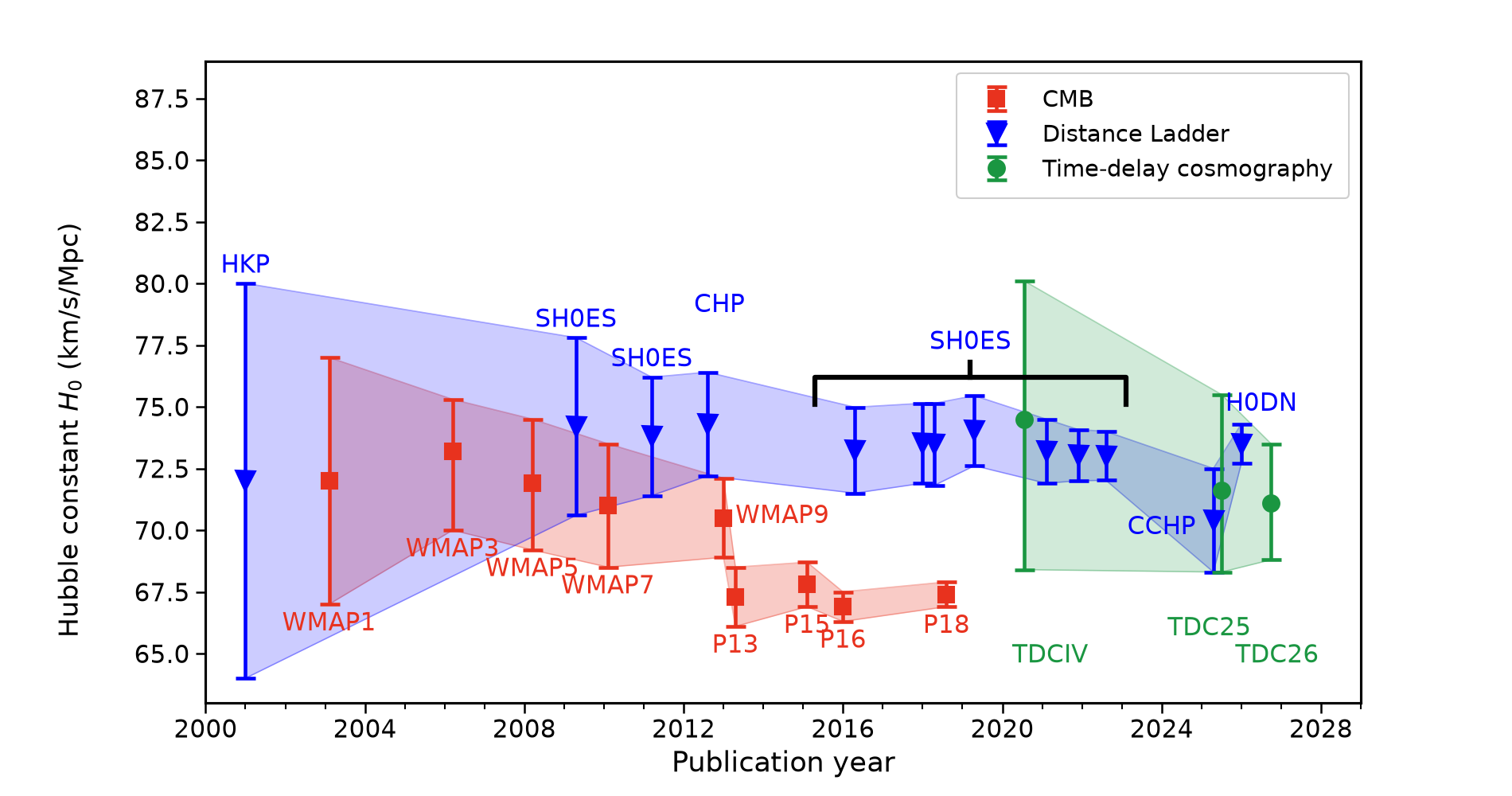}
 \caption{History of the measurement of the Hubble constant H$_0$ in flat $\Lambda$CDM using the local distance ladder \citep[blue; see][and references therein]{Freedman:2025,H0DN:2026}, the cosmic microwave background \citep[red; see][and references therein]{WMAP9,Planck++20}, and time delay cosmography \citep[green; see][and references therein]{TDCOSMO26}. Note that rapid improvement in precision by the TDCOSMO collaboration.} 
 \label{fig:tension_history}
\end{center}
\end{figure*}

In the past ten years, however, a potentially much more interesting problem has emerged, known as the ``Hubble tension". The most precise measurements of H$_0$ based on modern versions of the classical Hubble distance ladder method \citep[][hereafter H0DN]{Riess:2022,Riess21,H0DN:2026} have reached  1\% precision: H$_0 = 73.50 \pm 0.81$ \hunit. While the central value has not changed significantly, the 10-fold increase in precision has created a significant tension with the predicted value of  H$_0$ based on early-universe probes in standard flat $\Lambda$CDM. For example, the cosmic microwave background (CMB) data measured by the Planck satellite \citep{Planck++20} yield H$_0$ = $67.4 \pm 0.5$ \hunit, in flat $\Lambda$CDM. The tension is not limited to the H0DN and Planck measurements. Most early-universe probes tend to favor values around 67 \hunit, while most late-universe probes tend to favor values above 70, as  summarized by \citet{COSMOVERSE25,Abdalla:2022,Freedman:2021,Freedman:2025}.  A plot of the history of the tension is shown in Figure~\ref{fig:tension_history}.

If this ``Hubble tension" is real, and not due to unknown systematic errors in multiple measurements, it implies that flat $\Lambda$CDM is not the correct cosmological model and new physics is required. A number of possible solutions have been proposed, but at the moment there is no clear-cut contender  \citep[e.g.,][]{KnoxMillea2020,Vagnozzi2023,COSMOVERSE25,Turner26,Schonberg:2022,Schonberg:2026a,Schonberg:2026b}. 

Strong lensing gravitational time delays can play a pivotal role in settling the debate on the Hubble tension, by measuring the late-universe H$_0$ in a way that is completely independent of the traditional distance-ladder method. Time-delay cosmography is a stand-alone, independent method with several unique advantages: i) distances are measured over several Gpc, thus insensitive to local inhomogeneities of the universe; ii) distances do not rely on standard candles (except when the source is a lensed SN Ia), and therefore extinction and photometric zero points do not matter; iii) it is based on well-understood gravity and does not require modeling complex baryonic physics; iv) it is primarily sensitive to H$_0$ with mild dependence on other cosmological parameters and the background cosmological model.

In practice, four ingredients are needed to measure H$_0$ using gravitational time delays: i) the time delay itself; ii) a lens model that accurately predicts the difference in Fermat potential between the image positions; iii) characterization of the effect of the environment and line of sight; iv)  information to break the mass-sheet degeneracy. I review the state-of-the-art of each ingredient, before summarizing current results on H$_0$. 

As I will show in this section, and anticipated in Figure~\ref{fig:tension_history}, strong lensing has come a long way since its origin. Progress has been rapid and the measurement of H$_0$ based on 11 lensed quasars by the TDCOSMO collaboration \citep{TDCOSMO26,TDCOSMO26L} has reached 3\%, second only to the local distance ladder amongst late universe probes.

\subsection{Time delay}

Measuring a time delay requires monitoring a variable signal. While the nature of the source does not matter in principle, 
there are important differences between the phenomenology of lensed quasars and supernovae, each with its own pros and cons.

\begin{figure}
        \includegraphics[width=0.45\textwidth]{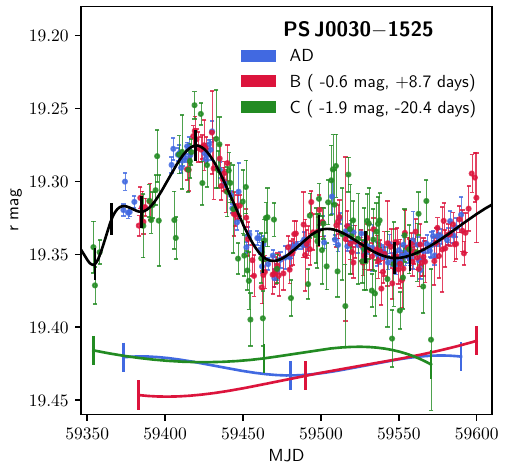}
		\caption{\label{fig:dux} Lensed quasar lightcurves matched between different images with splines. Due to microlensing, which adds independent lower-frequency variations to each curve, the curves could not be matched in this way with time and magnitude shifts only. Thus, an extra modulation is applied to each separately. These modulations are represented by lower order splines, displayed at the bottom. Physically, the black spline models the intrinsic variations of the source, while the colored splines undo the effects of microlensing. Figure and caption from \citet{Dux:2025}.}
\end{figure}

Quasars are thought to be switched on for timescales that are significantly longer than a human life \citep[see, e.g.,][for discussion]{Brewer:2026}, so they can be monitored for extended periods of time. However, their light curves (see example in Figure~\ref{fig:dux}) are stochastic and the amplitude of variability is usually of the order of 0.1-0.2 mags \citep{T+T23} over timescales of a year or so, and it increases with light-curve duration \citep[with notable exceptions, e.g.,][]{Dahle:2015}. 

An additional complication is stellar microlensing that can affect the light curves, especially for galaxy-scale lenses where the density of stars in front of the images of the quasar is high. The relevant timescale for microlensing is given by the time it takes for the source to cross a caustic \citep{SaasFee2006}, usually of order months to years. 

In practice, the observing strategy depends on photometric precision. If a few $10^{-3}$ relative precision can be achieved on the individual light curves, then one can take advantage of small-amplitude variations on timescales much shorter than those associated with microlensing and measure a time delay in a single season \citep{Courbin:2018}. Instead, if precision is only a few percent, multi-year campaigns are needed to catch peaks and valleys of sufficient amplitude to not be confused with the microlensing signal \citep{Tewes:2013}. 

Most of the lensed quasar monitoring to date has been done at ground-based imaging resolution \citep{Dux:2025}. Even using higher-resolution space-based template images and sophisticated deblending techniques, confusion sets a floor to the precision that can be achieved on individual epochs. The noise floor effectively limits successful monitoring to the widest-separation images, imposes long-duration campaigns, and makes it hard to achieve percent-level precision delays.  As a result, after decades of intense efforts of ground-based monitoring \cite[e.g.,][]{Tewes:2013,Millon2020b,Dux:2025}, only 12 time delays of lensed quasars have been measured with 2\% precision or better. In principle, one can measure time delays from unresolved light curves \citep{Bag:2022}, but it has never been demonstrated in practice that one can reach sufficient precision to go from identification to cosmology.

\begin{figure}
        \includegraphics[width=0.45\textwidth]{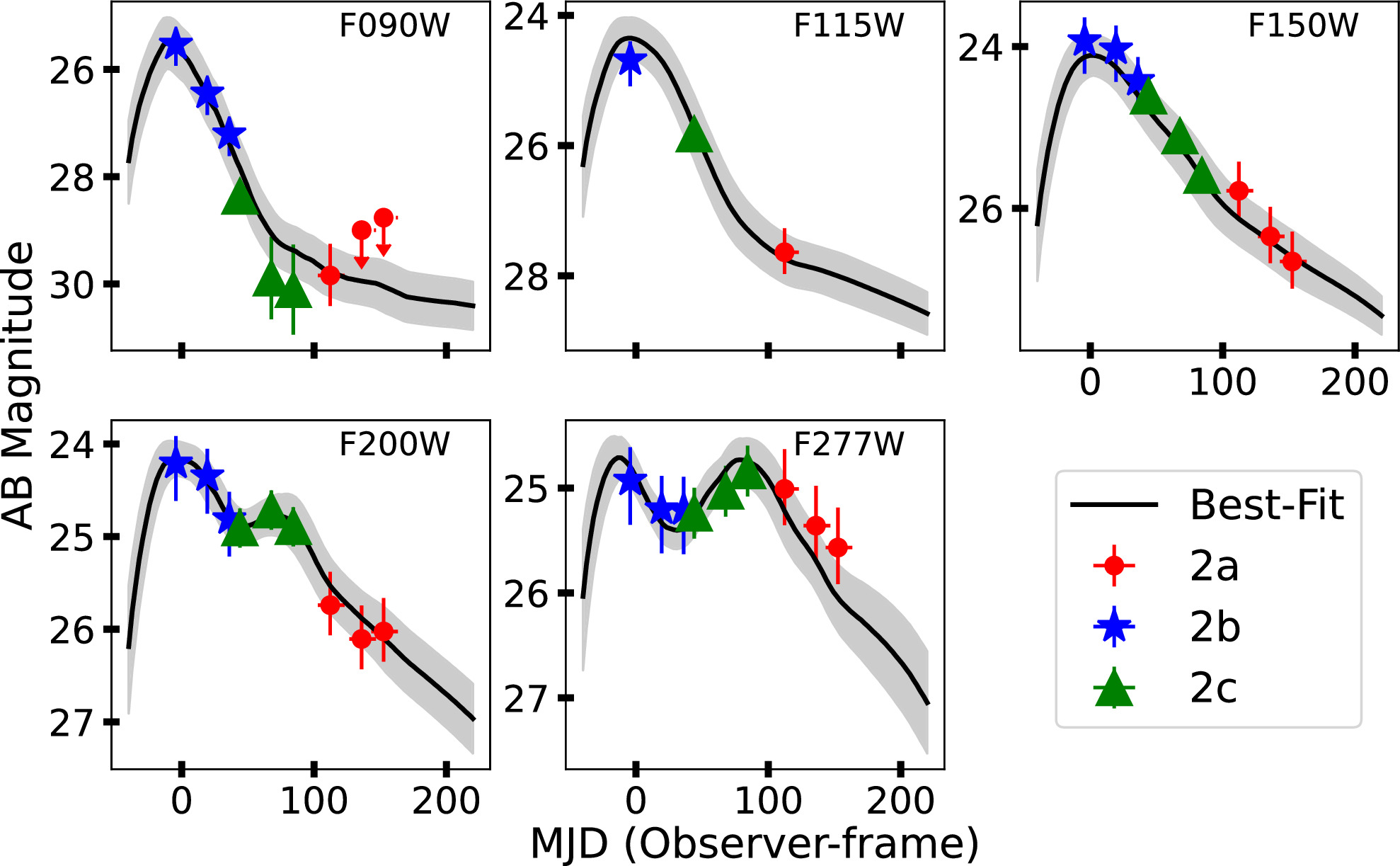}
		\caption{\label{fig:pierel} Reconstructed light curve for SN H0pe after applying the best-fit time delays and magnifications relative to image 2b. The vertical error bars are the photometric precision, while the horizontal error bars are the 16th and 84th percentiles on the time delay error. The gray shaded region is the best-fit model from the SNTD Color method. Figure and caption from \citet{Pierel:2024}.}
\end{figure}

The pros and cons of supernovae for time delays are in some sense the opposite of those of quasars. Their light curves are well known and well behaved with ample variability and clear peaks. However, they are only visible for timescales of weeks to months and there are few intrinsic peaks and valleys. This is illustrated in Figure~\ref{fig:pierel}. The supernova light curve fades after the initial peak. If one had only the F090W, F115W, and F150W light curves, it would be impossible to measure a delay because the light curve is virtually monotonic in those bands. Fortunately, for Ias, the secondary peak at longer wavelengths provides sufficient information in just a few epochs. In contrast to the case of quasars, 3/4 time delays for lensed supernovae measured so far \citep{Rodney:2016,Kelly:2023,Pierel:2024,Pierel:2026} have relied on space-based imaging from HST and JWST for their unique ability to deliver precision photometry of faint and blended images. The only exception is Winny \citep{Winny4}, but the photometric time delay suffers from a 20\% uncertainty. Only one time delay has been measured to better than 2\% precision for a lensed Type II supernova, using extensive HST monitoring \citep{Rodney:2016,Kelly:2023}. No lensed supernova Ia has had its time delay measured to 2\% precision or better \citep{Pascale:2025}. In the future, LSST will measure some time delays for lensed supernovae \citep{Bag:2021,Bag:2024,Bag:2026,Erickson26}, but it seems clear that space-based imaging will be needed to take full advantage of this cosmological probe.

\subsection{Lens model}

In the context of time-delay cosmography, the goal of lens models is to provide accurate and precise estimates of the Fermat potential differences (Equation~\ref{eq:fermat}) and their associated uncertainty, starting from the observational data. In practice, the problem of lens modelers is well summarized by a sentence attributed to Einstein ``Everything should be made as simple as possible, but not simpler." Build too simple a model, and the result will possibly be biased and the uncertainties underestimated. Build too complex a model, and --- if it is calculable --- it will likely result in overfitting, overestimated errors, uninformative, and possibly biased results. 

Historically, two approaches have been taken to solve this problem. ``simply parametrized models" represent the mass distribution by a set of physically justified and computationally tractable functional forms, described by a handful to a few dozen parameters. ``Free-form" models use a much larger number of free parameters, usually grids of the two-dimensional mass or potential or some basis set of functions.  In order to avoid overfitting some form of regularization is usually imposed. There are also hybrid approaches combining parametric mass models with free-form perturbations, or parametric mass models with free-form reconstruction of the lensed source.

The problem of finding the appropriate degree of complexity is exacerbated by statistical and computational challenges. In modern datasets, the information content of the images or datacubes \citep{2008ApJ...686..851R} is enormous, 10$^{3-6}$ pixels with high signal-to-noise ratio or hundreds of multiple images in conjugate point analysis. Modeling these data down to the noise level would often require an absurd amount of complexity, which would not necessarily improve the inference of the Fermat potential. Furthermore, exploring the parameter space is challenging because of its high dimensionality, the presence of local minima, and inherent degeneracies. One of the most infamous degeneracies is that between flattening of the main deflector galaxy and ``external shear". External shear is a term that was introduced in the early days of strong lensing to summarize to first order the distortion attributed to large scale structure and nearby mass structures not included in the main deflector. It is easy to find lens models with external shear of the order of 10-15\%, much larger than expected from large scale structures, and likely describing some missing ingredient in the main deflector \citep{KKS97,Etherington:2024}. At cluster scales, external shear can mimic the effect of real secondary peaks in the mass distribution. The degeneracy between ellipticity and shear can be mitigated with sufficient quality data and informative priors \citep{Schmidt:2023,Williams:2025}. However, it needs to be handled with care: if the effect were due to external and unmodeled mass components, it would also have an associated convergence, which is not observable but affects cosmology through the mass-sheet degeneracy.

Overcoming these difficulties and producing cosmography-grade lens models requires exquisite data, many CPU hours (easily $10^5-10^6$) and extensive exploration of modeling choices. In the simply parametrized approach, a good practice is to increase the complexity of the model until a satisfactory fit is obtained, while checking that the Fermat potential is stable. In the free-form approach, it is important to vary and possibly optimize the strength and form of regularization. It is also important to generate models sampling from the prior, in order to understand what effective prior is being imposed on the parameters of interest by the prior volume. When the prior volume has dimensions $\gg100$, and the parameters are not easily interpretable in terms of physical properties of the mass distribution, it is easy to swamp the information content of the likelihood and one needs to understand what is being imposed. Unfortunately, all these approaches are computationally intensive and time-consuming for the investigator. Advances in computing techniques \citep{Lombardi:2024,GIGALENS,Galan:2022,krizic2026}, machine learning \citep{Erickson2025,Schuldt:2021,Hezaveh:2017}, and GPU-based hardware \citep{Wang:2025} are helpful, but reducing compute and investigator time is a critical need going forward.

The good news is that lens models have also come a long way in the past two decades, when only four image (quasar or galaxy) positions with ground-based astrometry were used as input \citep[e.g.,][]{Schechter:1997}. I now summarize the current state of the art for galaxy- and cluster-scale lens models, leaving the discussion of limitations and systematic errors to Section~\ref{sec:sys} and future work to Section~\ref{ssec:future-analysis}. The mass-sheet degeneracy is discussed separately in Section~\ref{ssec:msd}.

\begin{figure}
        \includegraphics[width=0.45\textwidth]{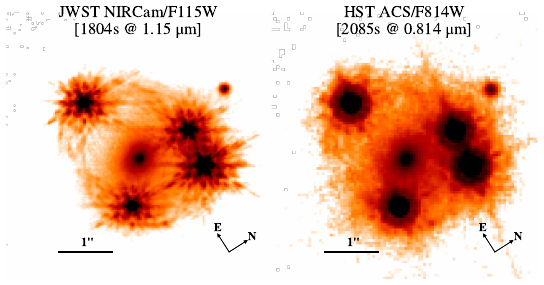}
		\caption{\label{fig:williams0} Comparison of a JWST NIRCam image of the lens WFI2033-4723 through filter F115W, to the HST imaging of the same system using ACS through filter F814W. The ring structure is maximized in each visualization, as it provides the tightest constraints on the deflector’s radial mass profile. Figure and caption from \citet{Williams:2025}.}
\end{figure}

\begin{figure*}
        \includegraphics[width=\textwidth]{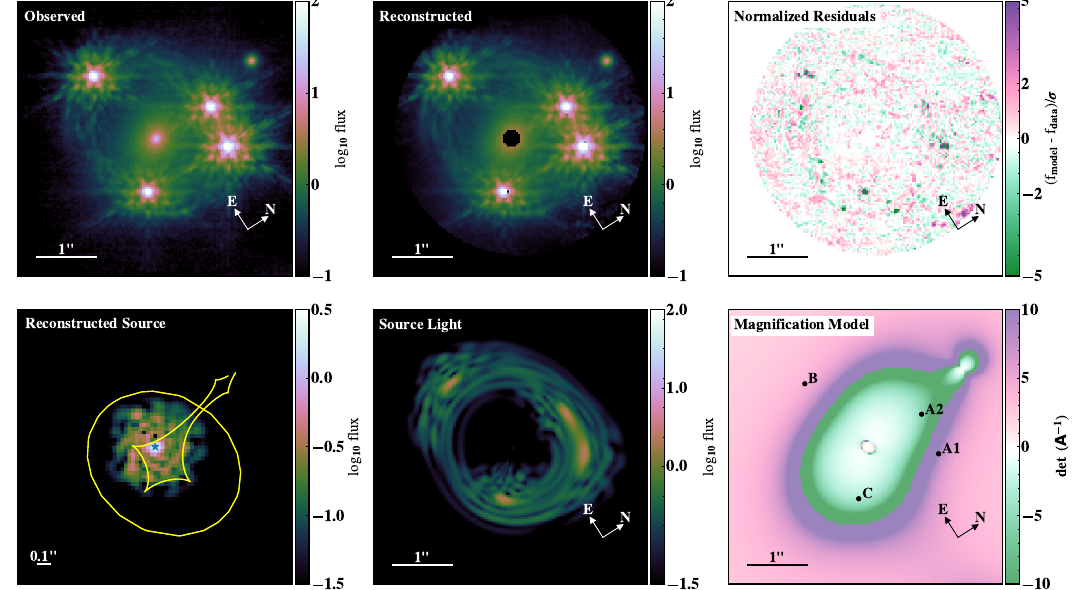}
		\caption{\label{fig:williams} Lens model of a quadruply imaged quasar lensed by a foreground galaxy and imaged by JWST-NIRCam. The top-left panel shows the observed image, the top-middle panel shows the model-predicted reconstruction of the image, and top-right panel shows the fit’s residuals normalized by the estimated uncertainty of each pixel. At the bottom left is the reconstructed source plot, featuring more detail than any parametric source modeled for time-delay cosmography thus far. The star symbol denotes the location of the quasar host galaxy’s centroid. The bottom-middle panel shows the lensed unconvolved extended-source light, and the bottom-right panel shows the magnification map. Figure and caption from \citet{Williams:2025}.}
\end{figure*}

As far as galaxy-scale deflectors are concerned, state-of-the-art methods applied to space-based \citep{Suyu:2010,Bonvin:2018,Williams:2025}, or adaptive-optics assisted \citep{Chen:2019,Chen:2022} imaging, yield Fermat potential differences that are precise to within 2-3\%, modulo the MSD.
The necessary ingredients to achieve this precision are the following. First, detection of extended images with sufficient resolution and signal-to-noise ratio to constrain the radial dependency of the deflection angle, which is in turn related to the slope of the mass density profile and thus the difference in Fermat potential between images. Second, identification and modeling of all the perturbers along the line of sight that can cause flexion shifts $\gtrsim 10^{-4}$ \citep{McCully:2014}, using multiplane lensing if needed. Third, sufficient flexibility in the mass model and in the reconstruction of the source to take advantage of the available information. In principle, features in the mass model and in the source surface brightness are distinguishable. However, if there is not enough flexibility in either one, the other may overcompensate. Fourth, accurate models of the point spread function are needed to take advantage of features on scales down to the angular resolution of the images, and achieve the required milliarcsecond-level astrometric precision \citep{BirrerTreu:2019}. Recent advances in PSF reconstruction techniques \citep{Birrer:2021,Millon++24,krizic2026,Chen:2016} enable cosmography-grade models even in the presence of bright quasar images. These advances take advantage of the superior angular resolution and sensitivity of JWST-NIRCam compared to HST, as shown in Figures~\ref{fig:williams0}~and~\ref{fig:williams}. It is reassuring that the reconstructed source looks like a realistic spiral galaxy even though no such symmetry was imposed.

\begin{figure*}
        \includegraphics[width=\textwidth]{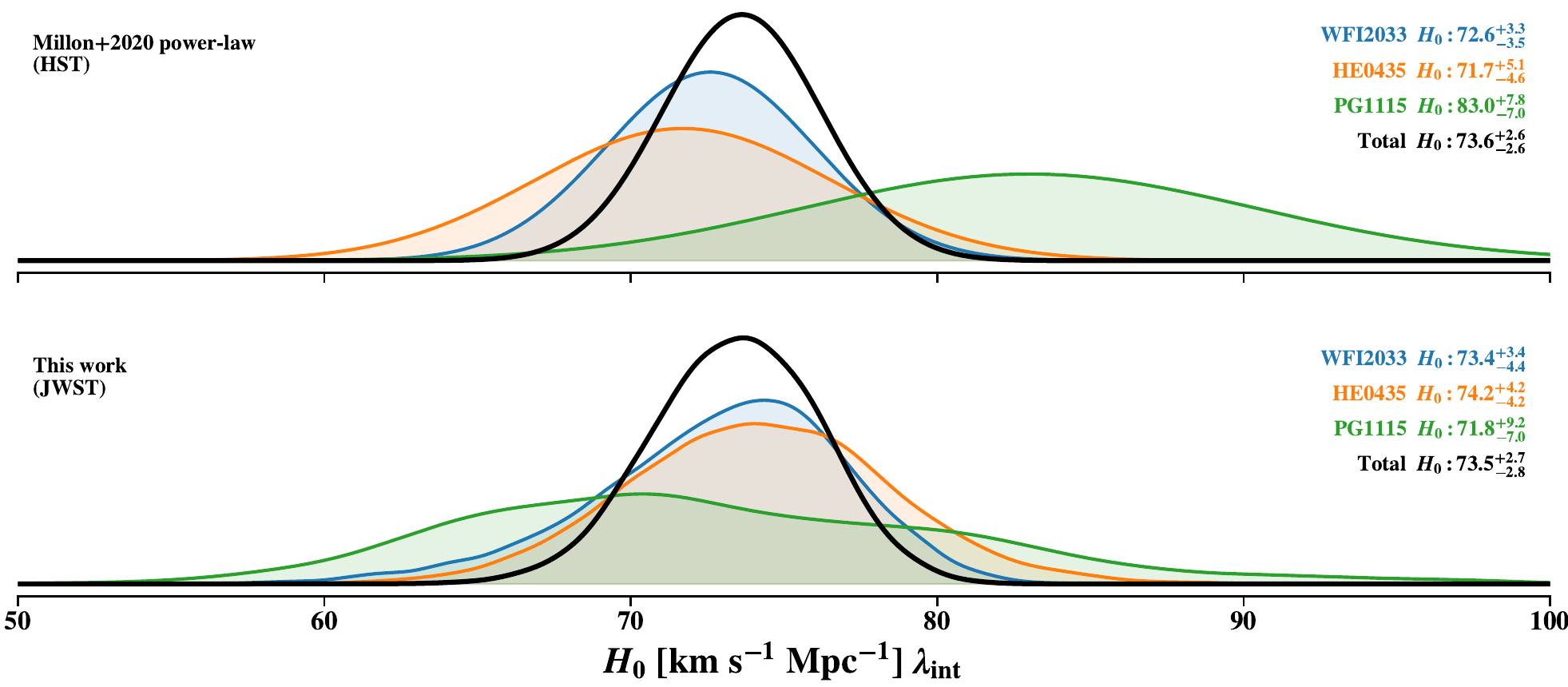}
		\caption{\label{fig:williams2} Comparison of JWST-based H$_0$ results with previous HST-based power-law results \citep{Millon:2020}. Top row: H$_0$ HST-based posteriors for each system from TDCOSMO~I, with results shown as Gaussian approximations to its quoted median and $16/84\%$ uncertainties and the black curve giving the combined constraint (the product of the individual posteriors). Bottom row: same as the second row, but results from \citet{Williams:2026} based on JWST-NIRCam imaging. Quoted values in the lower panels list the median $^{84\%}_{16\%}$ for each system and their combination. Note the improved agreement between the three systems. Note also that this inference assumes $\lint=1$, and does not account for the MSD. Figure and caption adapted from \citet{Williams:2026}.}
\end{figure*}

These advances are exemplified in the recent work by \citet{Williams:2026}. They carried out a blind analysis of two lenses, bringing the total number of published systems with JWST-NIRCam imaging to three together with the one analyzed by \citet{Williams:2025}. As summarized in Figure~\ref{fig:williams2}, they found H$_0$ measurements consistent with previous ones based on different data (JWST vs HST imaging; JWST-NIRSpec vs ground spectroscopy), different lensing softwares \citep[Lenstronomy vs GLEE;][]{Birrer:2021,SuyuHalkola:2010,Suyu:2012}, and different line-of-sight reconstruction techniques \citep{Wells:2024,Johnson:2026}. Importantly, the scatter between the individual systems appears reduced, even though analysis of larger samples is needed to determine whether this is significant. Note that the results shown in Figure~\ref{fig:williams2} are for power-law models and do not include the effects of the MSD, which would increase the uncertainty and potentially shift the mean. An independent analysis of the NIRCam images of WFI2033-4723 by a different team with independent software \citep{Li:2026} yields results consistent with \citet{Williams:2025}.
 
In principle, lens models could be further improved in the case of galaxy-scale lensing of supernovae \citep{Winny5}, because late time imaging will not suffer from the contamination from the bright lensed point source that affects quasar hosts \citep{Ding:2021}.

For cluster-scale deflectors, state-of-the-art models have been able to take advantage of the improved data to make great progress even while facing higher complexity than in galaxies. In the best-studied clusters \citep[e.g.,][]{Treu:2016,Grillo:2024,Pascale:2025,Suyu:2026}, the uncertainty of the Fermat potential difference between multiple images is estimated to be of the order of 5-10\%, depending on the number of spectroscopically confirmed families of multiply imaged sources and on the availability of spectroscopic redshifts and stellar velocity dispersion of cluster member galaxies \citep{J+S16}.

\begin{figure}
        \includegraphics[width=0.45\textwidth]{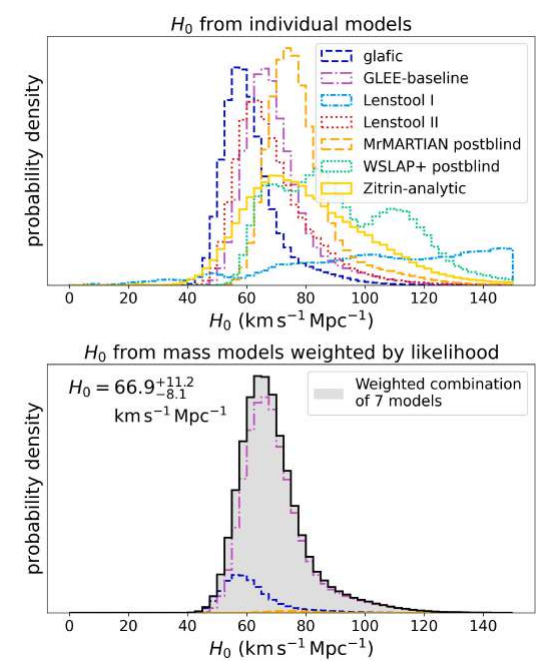}
		\caption{\label{fig:suyu} H$_0$ inference from SN Encore using the time-delay measurement $\Delta$t$_{1b,1a}$ by \citet{Pierel:2026}. Top: Inferred H$_0$ distribution from the mass models of the seven modeling teams via blind analysis, where the mass models and the time delay were kept blind to each other throughout and combined after unblinding without modifications. Bottom: H$_0$ from the combined mass model, weighted by each model’s likelihood. The weights were determined before unblinding. The inferred H$_0 = 66.9_{-8.1}^{+11.2}$\hunit\ has most of its uncertainty stemming from the current time-delay measurement, which will be significantly reduced thanks to approved HST and JWST observations of this unique lens system. Figure and caption from~\citet{Suyu:2026}.}
\end{figure}

A good snapshot of current work is given in the recent analysis by \citet{Suyu:2026}. The authors consider seven models, including five simply parametrized ones, and two free-form ones. The inferred H$_0$ for every model is shown in the upper panel of Figure~\ref{fig:suyu}. The bottom panel shows the combination of the posteriors weighted based on the relative likelihood given the observables that are independent of the Hubble constant (image positions). The result is consistent with that of previous similar studies \citep{Treu:2016,Kelly:2023,Pascale:2025}. 
The simply-parametrized codes are in broad agreement. Those that were designed specifically for time-delay cosmography -- GLEE \citep{SuyuHalkola:2010,Suyu:2012} and GLAFIC \citep{GLAFIC} -- provide the best agreement with the data. However, Lenstool-1 and Lenstool-2 show markedly distinct posteriors despite employing the same software. This is a reminder that lens modeling requires many difficult choices, not just accurate software and good data. The free-form models have much more flexibility, resulting in much broader posteriors. In this case, the weighting on observables is effective in selecting a subset of models.  

\citet{Bazzanini:2026} take the analysis one step further. Using the speed of {\sc gravity.jl} \citep{Lombardi:2024}, they model the lensing cluster under a variety of parametrizations and calculate the Bayesian evidence. They construct the evidence-weighted mean of the models and find H$_0  = 69.2^{ + 9.6}_{-8.0}$ \hunit. This value is consistent with the one reported by \citet{Suyu:2026} but includes the uncertainty from the choice of parametrization in the error budget. The additional mass-sheet degeneracy and line-of-sight contribution remain to be quantified.

The recent paper by \citet{Schuldt:2026b} is an instructive example of the work that remains to be done to estimate realistic uncertainties with cluster-scale lens models. \citet{Schuldt:2026b} increase the number of constraints by modeling the surface brightness of the lensed arcs for a fixed lens model. This reduces the formal uncertainties to 3\%. However, their formal statistical uncertainty does not include the contributions of the line of sight, of the mass-sheet degeneracy, and of the choice of parametrization of the mass components \citep{Bazzanini:2026}. Each of those contributions is of the order of a few percent. Thus, the true uncertainty is larger than the formal reported one. More work is needed to quantify the full error budget, in supernova Refsdal and in other cluster-scale lenses.

Work is also underway to exploit cluster-scale lensing of quasars \citep{Napier:2023,Liu:2023,Acebron:2022a,Acebron:2022b}. From a modeling perspective, the challenges are similar to those encountered for supernovae lensed by clusters. 

\subsection{Line of sight}

In addition to the plane of the main deflector, and perhaps a few individual perturbers or additional deflectors along the line of sight, one needs to also consider the cumulative effect of all other structures between the observer and the most distant source. For typical lines of sight, this effect is of the order of a few percent \citep{Wells:2024}, and it needs to be accounted for if one wants to do precision cosmology. In the context of galaxy-scale lenses, it is generally estimated by selecting simulated light cones that have similar number counts to the ones observed in the field of the deflector and producing a probability distribution function of $\kext$. In a recent study, \citet{Johnson:2026} use Euclid Flagship simulations to estimate $\kext$ for a sample of time-delay lenses, taking into account the contributions of mass between the observer and deflector, and between the deflector and the source. The residual uncertainties per system are of the order of 2-3\%, and the results are in good agreement with older estimates based on Millennium simulations and the approximation $\kext\approx\kappa_{\rm s}$ \citep{Wells:2024}. Importantly, whereas some studies in the past used external shear to select simulated lines of sight in addition to number counts \citep{Rusu:2017}, this has now been deprecated, owing to the difficulties of interpreting external shear discussed above.

In the case of cluster-scale deflectors, the effect of the line of sight remains an area of active research \citep{Roche:2026}. Only a few studies have allowed for the contribution of a sheet of external convergence \citep{Ertl:2025,Grillo:2024}. No comprehensive study of the line-of-sight effects has been performed to date. In principle, the presence of families of multiple images at different redshifts provides a way to do some sort of tomography, modulo the multi plane mass-sheet degeneracy \citep{Schneider:2014}. One important question is the selection function of clusters, and whether, for example, optically selected clusters vs X-ray-selected clusters tend to reside in statistically different lines of sight, and may thus need different corrections. As the precision of cluster-based time-delay cosmology improves, the urgency of characterizing the line of sight will increase.  

\subsection{mass-sheet degeneracy}
\label{ssec:msd}

I review here the most recent implementation of methods to break the MSD, leaving the discussion of residual errors and systematics to Section~\ref{sec:sys}. For time-delay systems dominated by a single deflector, one can apply the methodology of stellar kinematics to infer a measurement of mass and slope of the mass density profile, independent of lensing. A recent example is given by \citet{Sheu:2026}, who model Keck-KCWI stellar kinematics with axisymmetric Jeans modeling and combine it with HST-based lens models. 

Two results of this paper are worth commenting on. The first is the overall precision on H$_0$ (6.5\%), which is significantly better than what had been achieved in the past \citep[e.g., 9\%][]{Shajib:2023} using similar techniques. The reduced uncertainties are due to: i) improvements in stellar velocity dispersion techniques \citep{Knabel:2025} that reduce the covariance between stellar kinematic measurements across different spatial elements; ii) the availability of three time delays, the ratios of which help break degeneracies in the lens model; iii) the clear detection of a kinematic axis aligned with the
galaxy major axis that constrains the intrinsic axis
ratio of light, the inclination, and oblateness (versus prolateness) probability; iv) the use of informative priors on the anisotropy informed by local studies \citep{Cappellari2026}. The authors also allowed for the presence of a possible  central supermassive black hole, showing that in this case it does not affect the results.

\begin{figure}
\begin{center}
 \includegraphics[width=\linewidth]{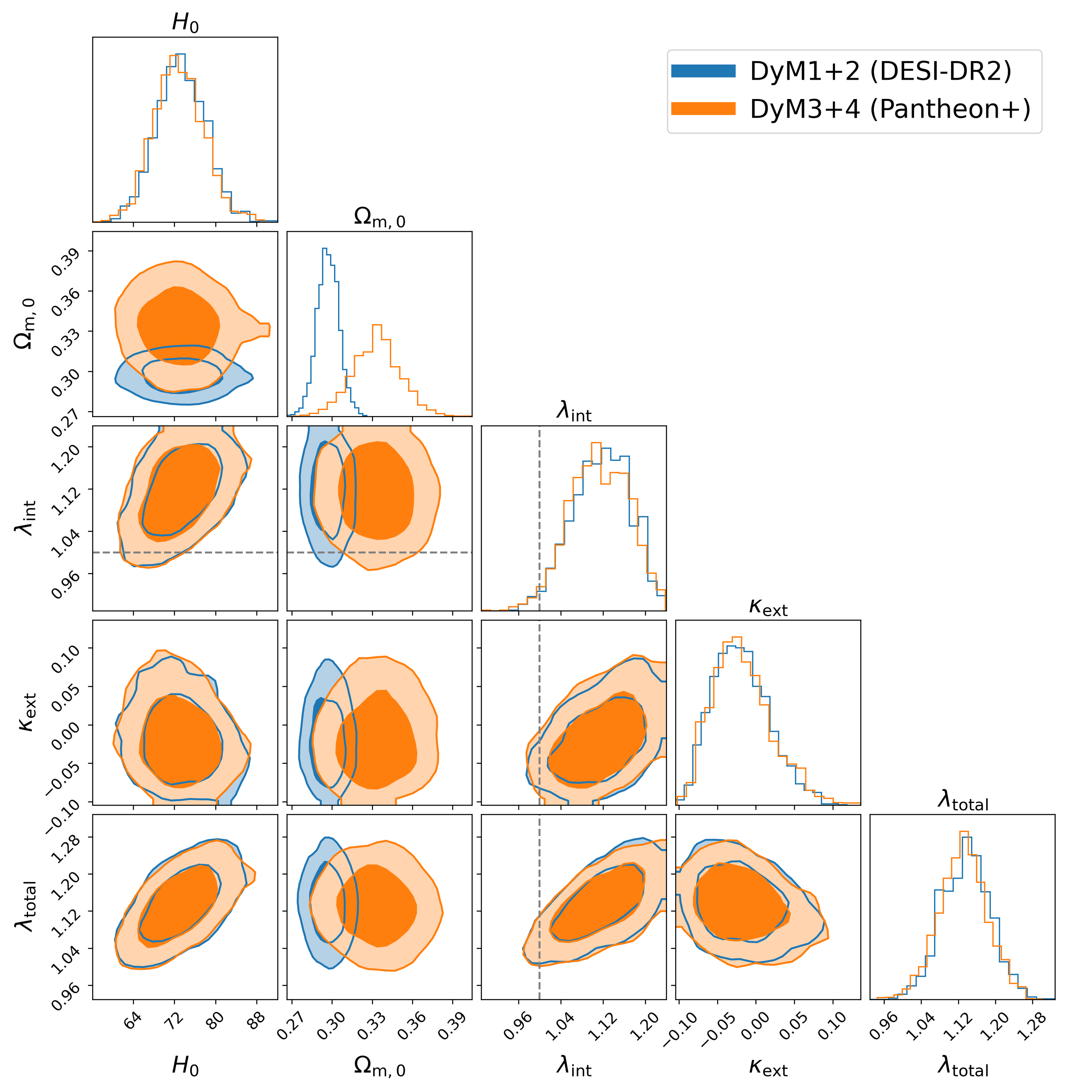}
 \caption{Cosmological constraints in flat $\Lambda$CDM based on lensing and dynamical models of the lens SDSSJ1433+6007 from \citet{Sheu:2026}.  The blue and orange contours are based on different priors on $\Omega_{\rm m, 0}$ from Baryonic Acoustic Oscillations and Supernovae, respectively. Note that H$_0$ is robust w.r.t. the choice of $\Omega_{\rm m}$ prior. The dotted line indicates $\lambda_{\rm int} = 1$.  The inferred values are H$_0=73.2^{+4.8}_{-4.7}$\hunit (a 6.5\% precision), and $\lint=1.12^{+0.05}_{-0.06}$. This is the first galaxy-scale time-delay lens for which evidence of  $\lambda_{\rm int} \neq 1$ is found. This is interpreted to be due to the small Einstein radius compared to the effective radius. Thus, the strong lensing effect is dominated by the stellar mass, which does not follow exactly a power-law profile, similar to some of the systems studied by \citet{Knabel:2026}. Figure from \citet{Sheu:2026}.} 
 \label{fig:final_res_corner}
\end{center}
\end{figure}

The second important result is that \citet{Sheu:2026} found $\lint$ to be larger than unity at 2-$\sigma$ for the lens SDSSJ1433+6007. This is the first galaxy-scale time-delay lens for which evidence of departure from a power-law is found, at variance with the other ten systems previously studied by the TDCOSMO collaboration. This exception underscores the importance of allowing for this extra flexibility. Had $\lint$ been fixed to unity, H$_0$ would have been significantly underestimated (see bottom left corner panel of Figure~\ref{fig:final_res_corner}). For comparison, \citet{Knabel:2026} analyze a sample of 14 SLACS lenses and find that although $\lint=1.01\pm0.03$ on average, individual systems can have significantly different values with an intrinsic scatter of $0.11\pm0.03$. In contrast, no evidence for intrinsic scatter is found for the time-delay lenses in the TDCOSMO-2025 sample \citep{TDCOSMO25}. 

A possible explanation is that the ratio between Einstein and effective radius is smaller than one for SDSSJ1433+6007 and the SLACS lenses, while it is generally larger for the other time-delay lenses. The multiple images of the time-delay lenses happen to be in a region of the deflectors for which the power-law description of the total mass density profile is an excellent approximation resulting from the bulge-halo conspiracy \citep{D+T14,TreuKoopmans:2004,Sheu:2025}. At radii much smaller than the effective radius, the power-law approximation must break down, because the total mass is dominated by the stars, which do not follow a power law. For the SLACS lenses and SDSSJ1433+6007, $\lint\neq1$ may be the effect of this breakdown. Incidentally, a power-law with slope close to minus two cannot be the correct description at larger radii, because the mass profile would not converge. 
These considerations have motivated the past use of composite profiles \citep[][and references therein]{Millon:2020}. Note that, in order to account for the radial dependencies of $\lint$, \citet{TDCOSMO25} and \citet{TDCOSMO26,TDCOSMO26L} allow for the population parameter $\lint$ to depend on the ratio between the Einstein and effective radius of each lens.

To conclude, it seems that for galaxy-scale deflectors, accounting for the MSD through the $\lint$ parameter gives the models sufficient flexibility to describe the data, while retaining generality and computational speed. However, the residual intrinsic scatter suggests that one could gain  by determining $\lint$ at an individual galaxy level if the data quality allows. Alternatively, a more sophisticated model could be found that describes the population better and thus reduces the scatter. As the data improve, it may become necessary or beneficial to consider altogether different frameworks. 

\begin{figure}
\begin{center}
 \includegraphics[width=\linewidth]{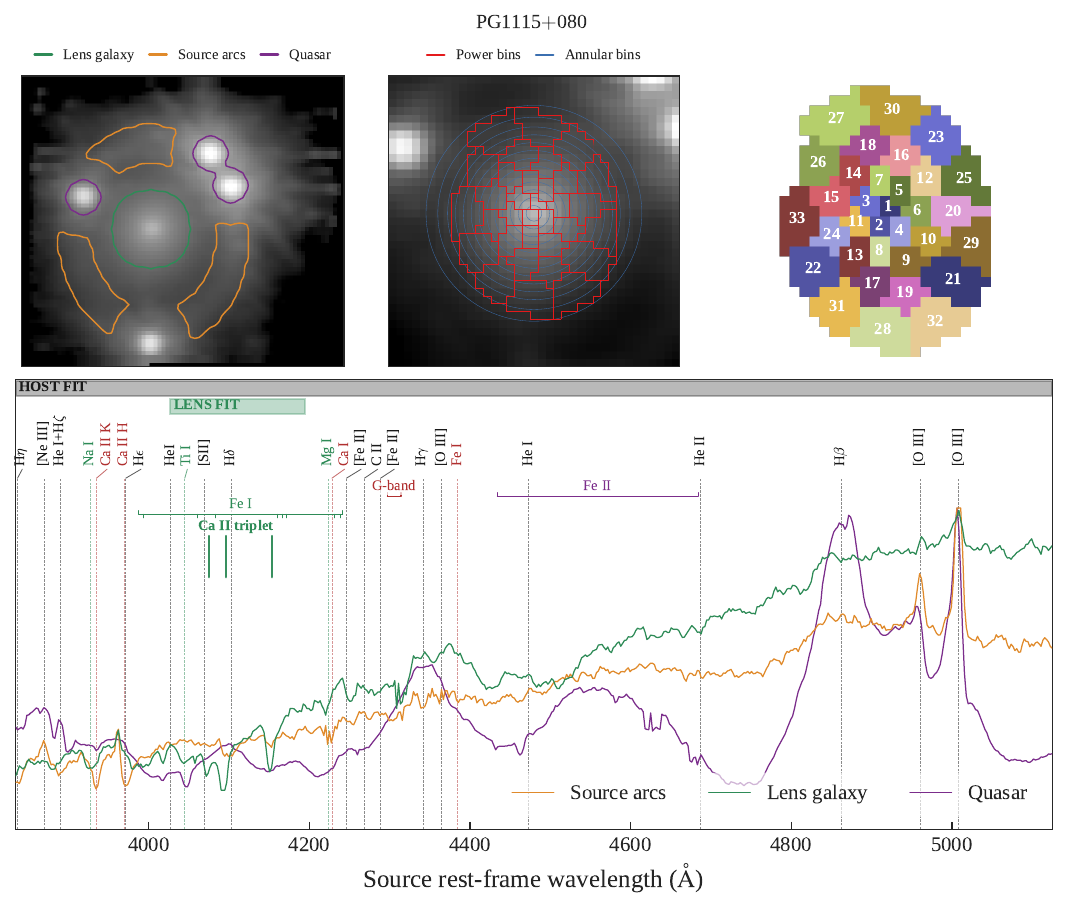}
 \caption{Spectral extraction from the JWST-NIRSpec datacube for PG1115+080. \textit{Upper left}: White light image integrated over wavelength. Outlined regions show the aperture from which the corresponding spectra in the lower panel were extracted. \textit{Upper middle}: White light zoomed to $0.9\times0.9$ arcsec$^2$. Outlined regions show the Power bins and annular bins. \textit{Upper right}: Power bin map. \textit{Lower}: Spectra integrated from apertures shown in upper left panel, in the rest frame of the background source galaxy/quasar. The full wavelength range, highlighted with the gray band, is the range used for the fit to the host galaxy stellar component and emission lines. The green band shows the range fitted for the lens deflector stellar kinematics. Spectral lines are shown with dotted vertical lines, and band features and line complexes are shown with horizontal brackets. Ticks under the horizontal line are centers of lines in the complex. Stellar absorption features for the background host galaxy are marked with red dotted vertical lines and labels. Typical emission lines associated with the host galaxy (and those shared with the quasar spectrum) are labeled in black. Lines and bands associated with the quasar only are shown in purple. Lines and bands associated with the deflector galaxy are shown in green, with the CaT lines displayed with short solid lines for visibility of the line shapes. Figure from \citet{Knabel:2026b}.} 
 \label{fig:PG1115_Knabel}
\end{center}
\end{figure}

The next improvement will come from observing  a large sample of lenses with the JWST NIRSpec integral field spectrograph (see example in Figure~\ref{fig:PG1115_Knabel}) \citep{Shajib:2026,Knabel:2026b,Mozumdar:2026}. NIRSpec IFS enables higher angular resolution kinematic maps, which improves the ability to break the MSD and the mass-anisotropy degeneracy and obtain more accurate and precise measurements of H$_0$ as a result \citep{Shajib:2018,Yildirim:2020,Shajib:2023,Shajib:2026}. 

\begin{figure}
\begin{center}
 \includegraphics[width=\linewidth]{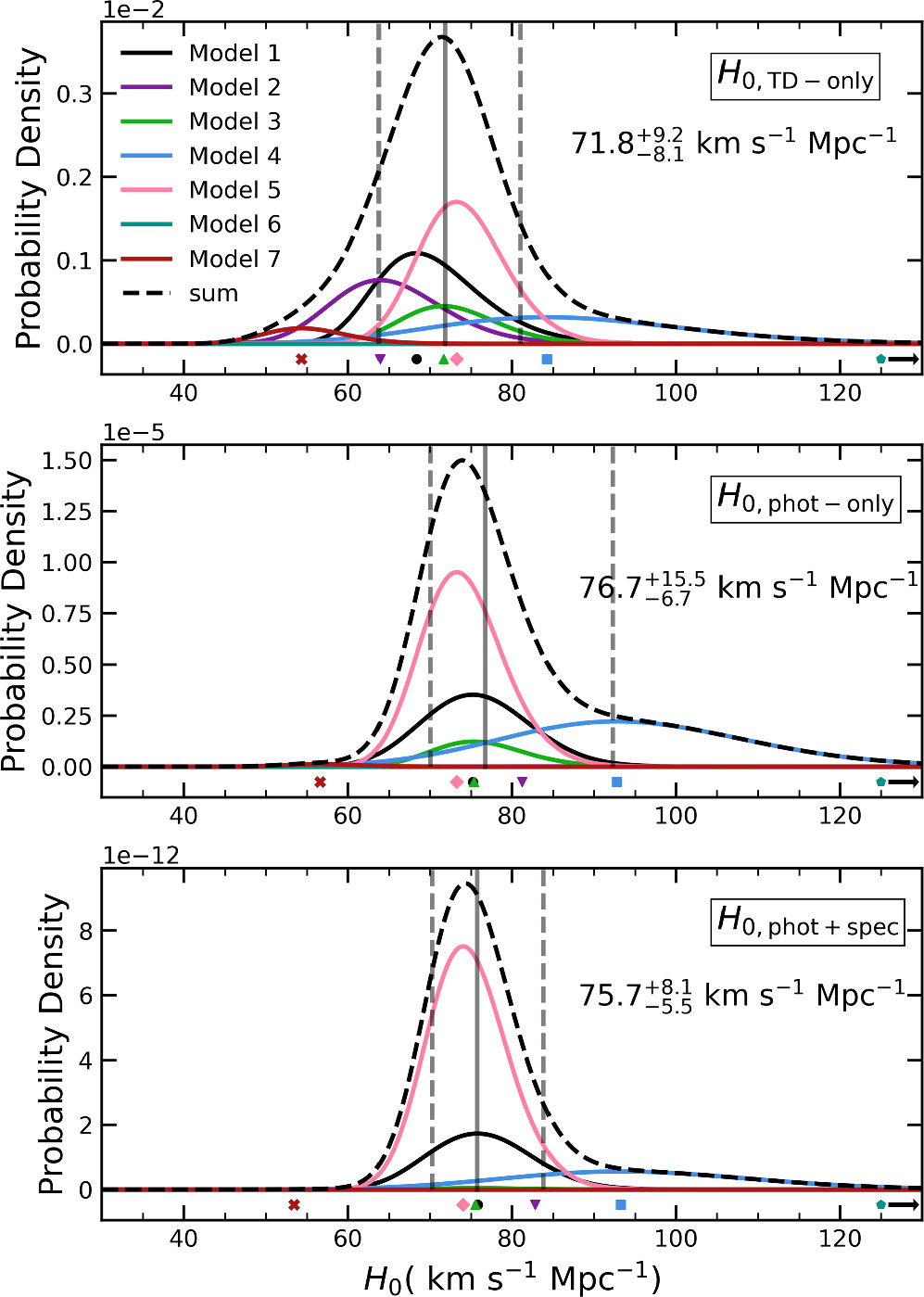}
 \caption{Inferred H$_0$ from comparing the lens-model-predicted time delays and
absolute magnifications to those measured from (in order of increasing
constraints) only time delays (top), only photometry (middle), and combined
photometry and spectroscopy (bottom). Individual lens model inferences (solid curves) are shown with normalization
equal to their weighting, and the median prediction of each model is plotted as
markers below the curves. The median of model 6 pushes up to the H$_0>130$\hunit
prior in all cases, and is instead shown with an arrow. Figure and caption from \citet{Pascale:2025}.} 
 \label{fig:Pascale}
\end{center}
\end{figure}

A different approach is needed to address the MSD for cluster-scale lenses, owing to the increased complexity compared to galaxy-scale lenses. Most studies so far have explored the MSD indirectly, for example by adding a sheet of external convergence \citep{Grillo:2024}, or by comparing different models that appear to be related by some form of MST \citep{L+O25}. A good example of this technique is the study by \citet{Pascale:2025} who used the luminosity of a lensed supernova Ia to weight a set of models. They show that selecting the models that best fit the luminosity - and thus are most likely doing the best to mitigate the MSD - significantly improves the precision of H$_0$ (Figure~\ref{fig:Pascale}). However, this is not a complete exploration of the MST. Developing and applying a formalism to explicitly account for the residual mass-sheet degeneracy appears to be a high priority for future work.

In terms of information available to break the MSD, multiple options are available. In addition to the use of lensed standard candles discussed above \citep{Pascale:2025}, one can take advantage of the presence of multiple families of sources at different redshifts, at the cost of increased complexity \citep{Schneider2019,Teodori:2026}.  Another approach is to use stellar or satellite kinematics. Unfortunately, clusters are not systems in dynamical equilibrium. Therefore, the interpretation of the motion of galaxies is not as straightforward as the interpretation of the motion of stars in galaxies. Methods exist \citep{Diaferio:2005,Biviano:2006}, but it is not clear whether they can reach the precision required for cosmography. A more promising route could be the use of the stellar kinematics of the brightest cluster galaxy \citep{Newman:2013,Odonnell:2026}, and individual cluster members \citep{Bianchetti:2026}.

\subsection{A time delay cosmography view of the Hubble tension}
\label{ssec:tension26}

The recent breakthroughs in time delay cosmography for quasars lensed by galaxies have been combined to produce the most recent estimate of H$_0$ from the TDCOSMO collaboration \citep{TDCOSMO26,TDCOSMO26L}. Their baseline result is H$_0=71.1^{+2.4}_{-2.3}$ \hunit\ in flat $\Lambda$CDM from 11 multiply imaged quasars with measured time delays. The lens models are based on HST and JWST-NIRcam images. The $\kext$ estimates are based on Euclid Flagship simulations. They break the mass-sheet degeneracy using spatially resolved kinematics from JWST-NIRSpec observations for six galaxies, ground-based spatially resolved kinematics from MUSE on VLT and KCWI on Keck for four (one overlaps with JWST), and integrated velocity dispersions from VLT-MUSE for two galaxies. The dynamical analysis is based on axisymmetric Jeans modeling. They combine lensing and dynamical information in a hierarchical analysis, using a prior on $\Omega_{\rm m}$ from the recent DES-Dovekie supernova sample \citep{DESDovekie}. Extensive tests show that the result is robust to the choice of priors and modeling details \citep{TDCOSMO26,TDCOSMO26L}. 

Adding information from two samples of non-time delays lenses, SLACS \citep{Bolton:2008} and SL2S \citep{Gavazzi12}, slightly improves the precision to H$_0=71.3^{+2.2}_{-2.1}$ \hunit. The measurements reach 3\% precision, a considerable improvement from the $\sim$5\% precision reached just a year earlier \citep{TDCOSMO25}. Both values are consistent with previous measurements from the TDCOSMO collaboration and with the local distance ladder values of \citet{H0DN:2026} and \citet{Freedman:2025}. They remain in mild tension with Planck, although the precision is not (yet) sufficient to make any definitive statements. 

\begin{figure}
\centering
 \includegraphics[width=\linewidth]{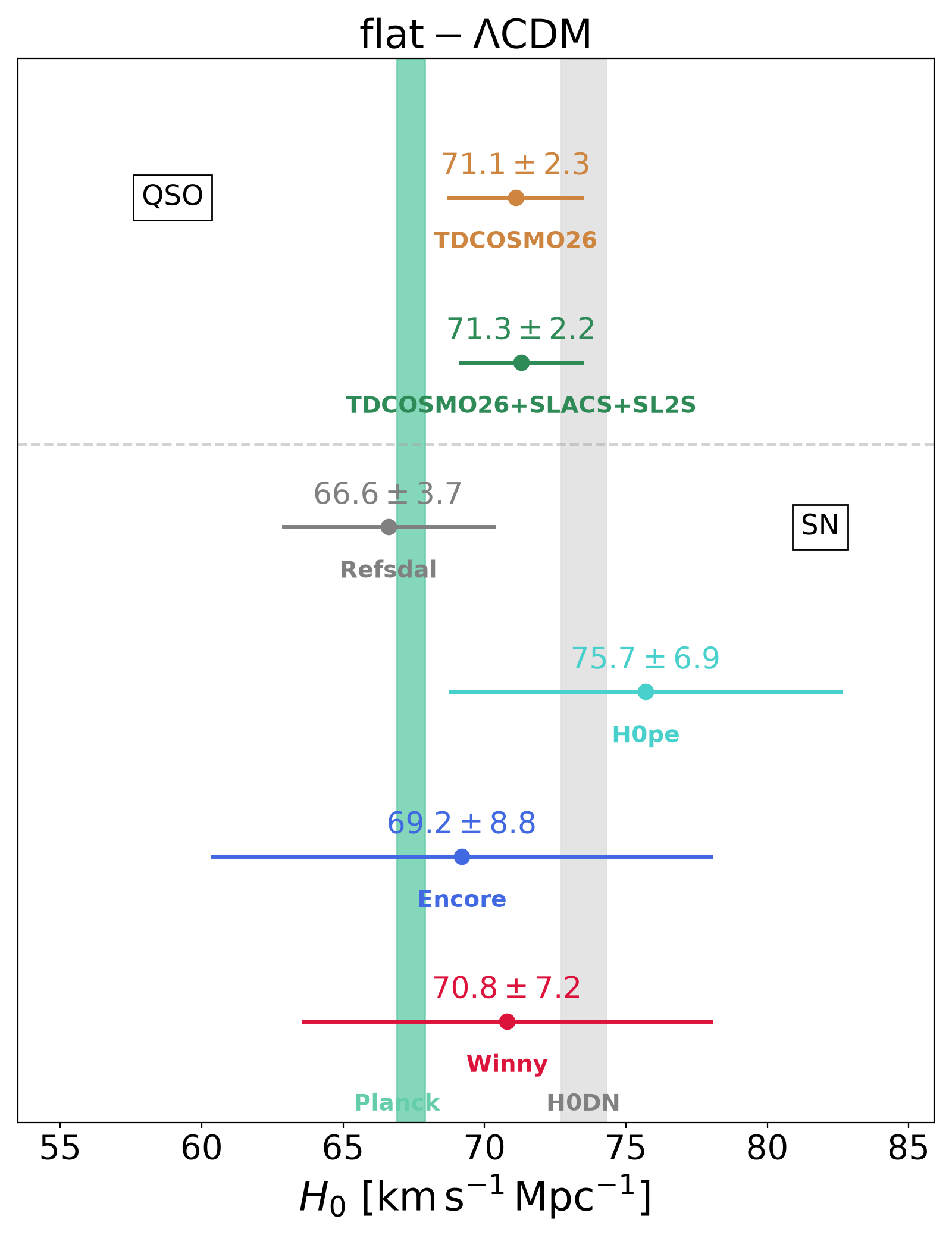}
 \caption{Time delay cosmography perspective on the Hubble tension in flat $\Lambda$CDM as of 2026. The Planck \citep{Planck++20} and H0DN \citep{H0DN:2026} measurements are shown as vertical bands for context. The top panel shows the recent TDCOSMO26 measurements based on 11 lensed quasars with time delays \citep{TDCOSMO26,TDCOSMO26L}. The measurement in combination with non-time delay lenses is also shown as TDCOSMO26+SLACS+SL2S. The bottom panel shows the three published measurements based on individual supernovae multiply imaged by clusters \citep{Kelly:2023,Pascale:2025,Bazzanini:2026}, and the one based on a supernova multiply imaged by a galaxy \citep{Winny5}. For simplicity, I have symmetrized the error bars.  Note that the uncertainties are not directly comparable: the uncertainties of TDCOSMO include the effects of the line of sight and the residual mass-sheet degeneracy, while those of the lensed supernovae do not. The precision of time delay cosmography is not yet sufficient to settle the tension. However, the rapid progress in the number of lensed QSOs and SN with measured time delays and state-of-the-art models means that time delay cosmography is on track to reach a conclusion in 1-2 years.} 
 \label{fig:tension26}
\end{figure}

The TDCOSMO-2026 results are summarized in the top panel of~Figure~\ref{fig:tension26}. In the bottom panel of the same figure, I collect the three published measurements from supernovae lensed by clusters \citep[using the values from][]{Kelly:2023,Pascale:2025,Bazzanini:2026}, and the one from a supernova lensed by a galaxy \citep{Winny5}. The uncertainties are not directly comparable because the lensed supernovae studies do not fully account for the mass-sheet degeneracy and the effects of the line of sight. However, it is encouraging that the lensed supernovae results are statistically consistent with the TDCOSMO ones, because many of the systematic uncertainties of the two methods are  different. The rapid pace of discovery of lensed supernovae, as discussed in Section~\ref{sec:future}, means that the statistical uncertainties of lensed supernovae are about to plummet. 

\section{Time delay cosmography and dark Energy}
\label{sec:darkenergy}

Ever since the discovery of the acceleration of the universe \citep{Riess1998,Perlmutter1999}, determining $w(z)$, and thus achieving some understanding of the nature of dark energy, has been a major goal of cosmology. 

Until recently, measurements of $w(z)$ have been consistent with $w=-1$, with dark energy being Einstein's cosmological $\Lambda$ \citep[e.g.,][]{Planck++20,Scolnic:2018}. In contrast, Baryonic Acoustic Oscillation measurements by the DESI Collaboration \citep{DESI-DR2}, in combination with other probes \citep[e.g.,][]{DES2026}, prefer, at a $\sim$3-$\sigma$ level, a model with $w\neq-1$, known as dynamical dark energy. The connection between dynamical dark energy and Hubble tension needs further investigation \citep{SCL26,Turner2026}.

Given the high stakes, a lot of effort is spent understanding whether dynamical dark energy is real and not due to systematic errors \citep{Efstathiou2025a,Efstathiou2025b}. Having multiple independent measurements of distance is a powerful way to make progress in this crucial debate. 

\begin{figure}
\begin{center}
 \includegraphics[width=\linewidth]{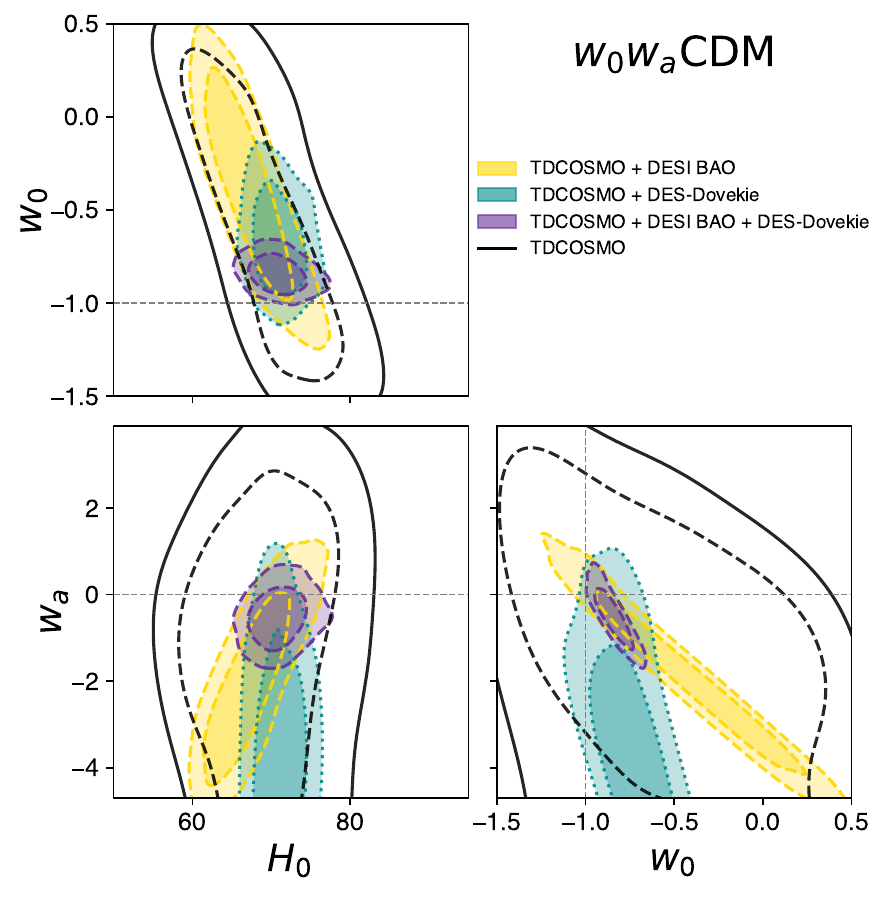}
 \caption{Posterior distribution of H$_0$, $w_0$ and $w_a$ in flat $w_0w_a$CDM cosmology. The combination of TDCOSMO (including the TDCOSMO-2026, SLACS, and SL2S lensing datasets) with the DESI-DR2 BAO and DES-Dovekie, and the combination of both external datasets, is shown with colored contours described in the figure legend. The contour levels represent the 1-$\sigma$ and 2-$\sigma$ constraints. Figure and caption from \citet{TDCOSMO26}.} 
 \label{fig:w0waTDCOSMO}
\end{center}
\end{figure}

Strong lensing is well positioned to be a major contributor to the determination of $w(z)$. Time-delay and multiple-plane lenses offer complementary approaches. Importantly, both are independent of other dark energy probes, independent of dust extinction and photometric calibration, and sensitive primarily to the evolution of distances in the redshift range $z=0-1$, where a departure from $w=-1$ has been suggested. In the case of time delays, there is a slight degeneracy between $w_0$ and H$_0$ (Figure~\ref{fig:w0waTDCOSMO}). In combination with DESI-DR2 and DES-Dovekie, TDCOSMO is inconsistent with $w=-1$ at the 2.3-2.5-$\sigma$ level. This is the case also in other parametrizations of dark energy \citep{TDCOSMO26}.

\begin{figure}
\begin{center}
 \includegraphics[width=\linewidth]{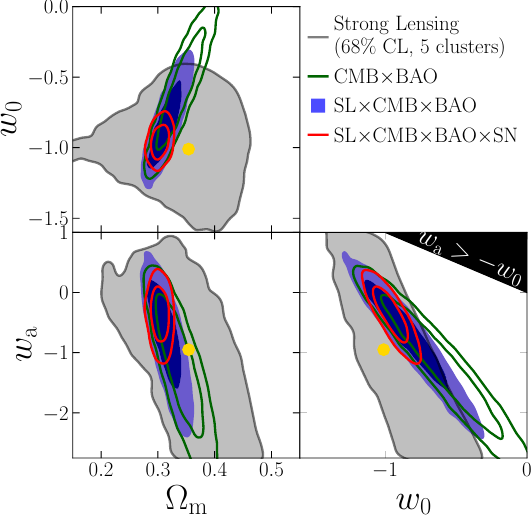}
 \caption{Confidence levels for the $w_0$--$w_a$ model. Here we show in grey the 68\% confidence level for the strong-lensing-only analyses for better visualization. The 68\% and 95\% combined constraints of different probes are shown with colored contours. The region where $w_a>-w_0$ is excluded because it produces a non-physical cosmology. In this cosmology with one additional free parameter, the strong lensing models are still capable of improving the constraints when combined with classical probes, especially on the parameters $w_0$ and $w_a$ of the dark energy equation of state. Figure and caption from~\citet{Caminha:2022}.} 
 \label{fig:Caminha}
\end{center}
\end{figure}

With multiple plane lensing, H$_0$ cancels out (Figure~\ref{fig:Caminha}). The encouraging results shown in the figures are based on just a handful of systems. As I discuss in Section~\ref{sec:future}, the samples will increase by orders of magnitude in the next few years, making these probes a powerful complement to more traditional methods.

\section{Systematics and open issues in time delay cosmography}
\label{sec:sys}

In this section, I discuss the sources of random error and potential systematic bias that can affect the determination of H$_0$ and $w$ from strong gravitational lensing. For clarity, I define the terms as follows. A measurement with 5\% precision -- repeated many times -- falls within 5\% of the truth 68\% of the times. A measurement with 3\%  systematic bias -- averaged many times -- will converge to a value that is 3\% \ higher or lower than the truth. I also refer to these two properties of measurements as precision and accuracy, respectively.

In the following, I review the main sources of random error and potential biases, assessing the current levels for galaxy- and cluster-scale deflectors, and what is needed to take advantage of future improvements in statistical precision.   In Table~\ref{tab:bias}, I summarize my own estimates of each individual source of random error and bias expressed in terms of time-delay distance per lens. The summary is based on my personal experience and reading of the literature. I assume that the errors and bias terms arising from different sources are uncorrelated. This seems a reasonable assumption, considering that the terms are related to independent data and techniques. I set a future goal of $<0.3$\% bias for each term somewhat arbitrarily, motivated by the need to measure percent-level differences in distances with at least 3-$\sigma$ significance, including potential bias. 

\begin{table}[]
    \centering
    \begin{tabular}{l|c|c|c|c}
    Source & Random & Bias & Random    & Bias  \\ 
     & (current) & (current) & (goal)   & (goal)  \\ 
    \hline 
    Time Delay & 1-6\% & $<$1\% & 2\%   & $<$0.3\%  \\ 
    Lens Model (gal) & 3\% & $<$1\% &  3\%  & $<$0.3\%  \\
    Lens Model (clu) & 5\% & ?  &  3\%  &  $<$0.3\% \\
    Kinematics/Dynamics & 3-5\% & $<$1\% &  3\%  &  $<$0.3\% \\
    SN Luminosity     & 5\% & $<$3\% &  5\%  &  $<$0.3\% \\
    L.O.S. (gal) & 3\% & $<$1\% &  3\%  & $<$0.3\%  \\
    L.O.S. (clu) & 5\% & ? &  3\%  & $<$0.3\%  \\
    Peculiar Vel & $<$1\% & $<$0.24\% & 1\%   &  $<$0.3\% \\
    \hline
    \end{tabular}
    \caption{For each contributor to the error budget listed in the first column, columns two and three list the estimated random errors and bias per lens based on current state-of-the-art data and techniques (current). Columns four and five (goal) list the precision and accuracy that is desirable per system in order to take full advantage of the large samples that will be available in the next decade. All errors are quoted in terms of corresponding error in absolute distance (and thus H$_0$).}
    \label{tab:bias}
\end{table}

\begin{itemize}
    \item Time delay. It has been shown via simulations and data challenges that they can be measured with precision as good as 2\% and subpercent accuracy given light curves with sufficiently high SNR per epoch and sufficient number of epochs, both for lensed quasars and supernovae \citep{Dobler:2015,Liao:2015,Courbin:2018,Dux:2025,Kelly:2023,Rodney:2016,Pierel:2024}. A potential source of bias is the microlensing time delay \citep{Tie2017}, but it has been shown to be subpercent in practice \citep{Chen2018}. Simulations (Venkatraman, 2026, priv. comm.) show that 2\% precision and $<0.3$\% bias is attainable with sufficiently good data, without any known limiting factor.
    \item Galaxy lens model. The precision and accuracy of Fermat potential differences, neglecting the MSD, has been tested primarily through multiple independent analyses of the same systems using different data, codes, and modeling assumptions \citep{Shajib++22,Williams:2025,Williams:2026,Galan++24,Galan:2026,Li:2026}. These comparisons show that multiple independent analyses yield the same results within the uncertainties, thus suggesting that the precision of a few percent is correctly estimated. Similarly, statistical analysis of multiple systems shows no evidence of unmodeled sources of scatter \citep{Millon:2020,TDCOSMO25,Williams:2026}, again suggesting that the estimated precision is correct. In a different test, \citet{TDCOSMO3} simulated lenses with dark matter substructure and modeled them neglecting it. They found that substructure introduces a negligible $\sim1$\% scatter and no bias. \citet{TDCOSMO7} found that realistic amounts of boxiness and diskiness affect the inferred H$_0$ by less than 1\%. \citet{krizic2026} developed new techniques to speed up modeling of the PSF and source parameters, showing on simulated data that subpercent accuracy can be achieved.
    Perhaps the strongest argument suggesting that any bias is smaller than the estimated precision is the good agreement between simply parametrized and free-form models \citep{Suyu:2010,Coles:2008}. A lens modeling challenge carried out a few years ago turned out to be inconclusive \citep{Ding_tdlmc2018,Ding:2020} because the simulations of massive galaxies had insufficient resolution and thus unphysical mass density profiles. Owing to the recent improvements in precision and sample size, it seems timely to carry out a new modeling challenge based on higher-resolution models of the deflectors and larger numbers of systems.
    \item Cluster lens model. Setting aside the MSD as for galaxy-scale lenses, comparison between  state-of-the-art models provides an estimate for the precision at around 6\% \citep{Kelly:2023,Pascale:2025,Suyu:2026,Bazzanini:2026}. Initial explorations of the uncertainties associated with the choice of parametrization show that quantifying more accurately this term of the error budget is becoming possible \citep{Bazzanini:2026}. An encouraging result in terms of accuracy was the blind prediction of the position, timing, and brightness of the reappearance of Supernova Refsdal \citep{Treu:2016,Kelly:2016}. However, the degeneracy between timing and H$_0$ means that this experiment tested accuracy only at the level of several percent. The predicted re-appearance of SN Requiem should soon provide another similar test with a longer baseline and better precision \citep{PierelRequiem:2025,Suyu:2026}. Tests of accuracy based on simulations are even harder than at galaxy scales, owing to the cost of numerical simulations, and the challenge of making them realistic in terms of lensing properties \citep{Meneghetti:2023}.
    \item Stellar kinematics/dynamics. This is the primary means used to break the MSD for galaxy-scale deflectors. Potential sources of error and bias can be introduced by  kinematic measurements and dynamical models. For kinematic measurements, \citet{Knabel:2025} showed that subpercent precision and accuracy can be achieved with sufficiently good data and sufficiently good stellar templates. It is crucial to have a template library that includes stars spanning a range of type, metallicity, and elemental abundances, with signal-to-noise ratio and spectral resolution much superior to those of the spectra that are being analyzed. While existing libraries are sufficient, it needs to be evaluated whether further improvements require dedicated ones. It is necessary to know the line spread function of the spectrograph that is being used \citep{Shajib:2025}. 
    For dynamical modeling, standard Jeans techniques require assumptions about symmetry and orbital anisotropy. \citet{Huang:2025} showed that axisymmetric models are sufficient to model intrinsically triaxial galaxies with subpercent precision and accuracy. \citet{V+M26} showed that adopting constant anisotropy is sufficient to obtain percent-level precision and accuracy, and that generalized Osipkov-Merritt parametrization \citep{Osipkov:1979,Merritt:1985} can do even better. As data improve, it will also be helpful to consider higher-order moments of the absorption lines, beyond Gaussian approximation \citep{Cappellari2016,Fores:2026}. Finally, it is important to properly reconstruct the spectroscopic PSF and adopt as tracer of the stellar population the surface brightness profile at the wavelength of the features used  for the kinematic measurement \citep{Fores:2026}. Both issues can be addressed using integral field spectroscopy \citep{Shajib:2026}. Recent work has started to include a supermassive black hole at the center of the deflector. The sphere of influence is marginally resolved in the nearest deflectors at JWST angular resolution \citep{Wang:2025}, but in practice it has a negligible impact in terms of cosmological parameters \citep{Sheu:2026}.
    \item Supernova luminosity. When supernovae Ia are used as standard candles to break the MSD, there are three potential sources of bias and uncertainty. One is dust, for which I refer to discussions in the supernova literature \citep{BR:2024,Dhawan:2025}. The only difference with respect to standard supernovae Ia is that the lensed ones are observed behind a foreground galaxy or cluster and therefore they may be affected by more dust than not-lensed ones, on average. Uncorrected dust extinction could in principle make the lensed supernovae appear fainter, and thus more distant, than they are, leading to biased H$_0$. The second potential source of bias and uncertainty is microlensing by stars in the foreground deflector. This is discussed in theory by \citet{Foxley:2018,Goldstein:2018,Huber:2021} and in practice by \citet{Pierel:2025,Suyu:2026}. A third source of potential bias is millilensing by dark subhalos. These effects are smaller than the current statistical uncertainties, but additional work will be needed as the sample sizes increase.  
    \item Galaxy line of sight. \citet{Johnson:2026} recently computed over- and under-densities for the lines of sight of TDCOSMO lenses by matching observed galaxy counts to lines of sight in the Euclid Flagship simulations. The comparison with earlier work based on Millennium Simulations \citep{Hilbert:2009,Greene:2013,Rusu:2017,Wells:2024} shows that the results are consistent within 0.4\%, and with comparable error bars. From this comparison, I can conclude that the precision is 2-3\% for most cases and the accuracy is $<$1\%. 
    \item Cluster line of sight. This is an area of open inquiry. \citet{DAN:2011} studied the effects of substructure and line of sight on multiplane cluster cosmology, finding that they account for $0.2"-0.3"$ scatter in image positions \citep[see, also,][]{Chirivi:2018}. \citet{Grillo:2024} included a $\kext$ term in their analysis of supernova Refsdal and showed that it is constrained by the families of multiple images to be zero within a few percent. Work along these lines needs to be carried out systematically for all clusters \cite[see, e.g.,][]{Suyu:2026} in order to quantify the precision and accuracy. One aspect that needs investigating is whether the selection method of clusters (e.g., X-ray vs optical) may introduce systematic differences between the lines of sight. With current uncertainties of order 10\%, this is probably negligible. However, the issue will rise in priority as the sample size increases. 
    \item Selection function in hierarchical analysis. \citet{TDCOSMO8} tested the ability of the hierarchical framework to infer an unbiased H$_0$ by using mock lenses with a composite (dark matter + baryons) mass distribution that they modeled with a power-law mass profile. The framework successfully recovered H$_0$ within 1.5$\sigma$ of the fiducial value when using time-delay lenses alone. The precision was improved to $\sim$ 2\% and accuracy to 0.7\% median offset when combining time-delay lenses with an external sample of non-time-delay systems at lower $z$, similar to the SLACS sample. As the field moves from the analysis of individual systems to the hierarchical analysis of large numbers of systems with varying degrees of information content \citep{B+T20,Erickson26}, characterizing the selection function will become an important goal. This will be essential, for example, for using information from much larger samples of non time-delay lenses or non-lensing galaxies, or for modeling simultaneously time-delay lenses with different selection functions (e.g., quads and doubles, or X-ray vs optically selected, or deflector vs source selected). Some work has been done on this topic \citep{Sonnenfeld:2024,Sonnenfeld++23}, but more remains to be done. 
    \item Peculiar velocities. Motion relative to the CMB reference frame has a minimal impact on time-delay distances.  \citet{Dalang2023}  computed the effect on redshift and lensing aberration and estimated the amplitude to be $0.24$ \% for the \citet{Millon:2020} sample. This is a small effect and can be corrected to high precision. It is therefore not a concern.
    \item Astrometry. The requirements on astrometry are non-trivial, in some cases reaching milliarcseconds for the position of multiple images \citep{BirrerTreu:2019}. Meeting these requirements requires excellent distortion corrections and models of the PSF \citep[e.g.,][]{Millon++24}.
\end{itemize}

In conclusion, in the case of galaxy-scale deflectors, systematic biases are at the sub-percent level when using the best available data and state-of-the-art techniques, thus well below statistical errors. For cluster-scale deflectors, systematic biases are at the level of a few percent, and therefore also smaller than the statistical errors when using appropriate data and techniques. For both kinds of deflectors, further work is needed to ensure that systematic bias remains below statistical errors as overall precision improves with increasing sample size. 

\section{The nature of dark matter}
\label{sec:darkmatter}

Although the evidence for non-baryonic dark matter appears to be incontrovertible within standard gravity, the goal of understanding its fundamental nature remains elusive \citep[see, e.g.,][for a historical perspective]{B+H18}. Earth-based direct detection and accelerator searches have ruled out vast swaths of parameter space but have not achieved any detection \citep[see, however,][for a tantalizing event]{LZ:2026}. 

Similarly, astrophysical direct and indirect probes have made great progress but have not yet delivered any smoking gun evidence. As experiments continue to push the boundaries, the search is becoming more and more difficult. Earth based experiments and astrophysical direct searches are predicated on dark matter having some non-gravitational interaction with standard model particles. Astrophysical indirect searches based on electromagnetic tracers, such as those based on the Milky Way satellites and lyman-$\alpha$ forest \citep[e.g.,][]{Villasenor++23}, are inherently limited to structures large enough to produce and retain stars or gas, i.e., subhalo masses larger than $\sim10^7$M$_\odot$ \citep{Nadler++25}. 

The only tracers that can probe dark matter halos on scales smaller than this limit are the purely gravitational ones, i.e., stellar streams \citep{Banik++21} and gravitational lensing \citep{Vegetti++24,Nierenberg2026}. Of those two methods, gravitational lensing is the only one that can probe substructure independent of baryons outside the Local Group, and determine its characteristics as a function of cosmic time. I focus here on strong lensing as a probe of dark matter, referring to the recent excellent review by \citet{NRD26} for a comprehensive summary of astrophysical dark matter models and results based on other methods.

In the following subsections, I review recent lensing results, organized by broad dark matter phenomenology. I start by reviewing lensing tests of the predictions of the standard cold dark matter model (CDM, \ref{ssec:CDM}). I then discuss warm dark matter (WDM, \ref{ssec:WDM}) and self-interacting dark matter (SIDM, \ref{ssec:SIDM}). I briefly review results on other dark matter models, including ``fuzzy", ``mixed", and primordial black hole dark matter in Section~\ref{ssec:ODM}.

\subsection{A fundamental test of cold dark matter (CDM)}
\label{ssec:CDM}

In the standard cold dark matter model, based on massive non-relativistic and non-interacting particle(s), dark matter clusters in a hierarchy of halos and subhalos. The number density of halos, known as the halo mass function, is predicted to depend on the halo mass as $M^{-1.9}$, extending to planet-like scales \citep{Springel:2005,Wang:2020}. Dark-matter-only cosmological numerical simulations also predict the mass density profile of the halos. Baryons and electromagnetic interactions are believed not to affect the number and internal mass distribution of halos below virial mass $\sim 10^{10}$M$_\odot$ \citep{bullock_small-scale_2017}. When halos are gravitationally bound to a more massive halo, they are called subhalos, and their abundance and internal mass distribution are affected by tidal interactions with the host halo \citep{Du++24,Du++25}. Many subhalos above $10^{7-8}$M$_\odot$ are sufficiently massive to retain gas and stars and are therefore visible as satellites of galaxies, for example of the Milky Way \citep{Nadler++20b}.

A fundamental prediction of CDM is thus that the universe is filled with an abundance of dark halos and subhalos with mass below $\sim10^7$M$_\odot$. Gravitational lensing makes this prediction testable: confirmation would be a major triumph for CDM; falsification would close the book on CDM and accelerate the development and testing of alternative dark matter models. Statistical studies are getting close to this threshold: \citet{Nierenberg2026} proved that dark matter halos with mass below $\sim10^{8}$M$_\odot$ exist and that with 200 quadruply imaged quasars one will probe below $10^7$M$_\odot$. Note that this is the mass of the subhalos prior to infall and tidal stripping. Individual detections have reached observed mass $10^6$M$_\odot$ \citep{Powell++25}. A typical subhalo observed below redshift 1 has lost $\sim$95\% of its mass since infall \citep{Gannon:2025}.
A statistical analysis of more systems with comparable angular resolution  needs to be performed to determine whether the abundance of these individually detected subhalos is consistent with the prediction of CDM.

Cold dark matter makes specific predictions for the mass-concentration relation of dark matter halos. Intriguingly, some of the subhalos individually detected by gravitational imaging have turned out to have higher density/concentration than expected based on simple extrapolations of CDM predictions to these masses \citep[see, e.g.,][and references therein]{Tajalli:2025,Vegetti:2026}. Likewise, the number of lensed quasars with large flux ratio anomalies seems to be higher than expected based on CDM \citep{Gilman:2026c} as indicated by the preference from these measurements for higher abundances of subhalos relative to predictions from numerical simulations. This is potentially a major problem for CDM, perhaps pointing towards dark matter with non-negligible self interaction (SIDM), as discussed below. 

\subsection{Warm dark matter (WDM)}
\label{ssec:WDM}

Warm dark matter (WDM) is a broad label used by astrophysicists to describe models in which power on small scales is suppressed by a non-negligible free-streaming length. This effect introduces a turnover in the halo mass function around a characteristic mass known as the half-mode mass m$_{\rm hm}$. From the point of view of particle physics, WDM maps into a set of models, including thermal relics and sterile neutrinos with a variety of formation mechanisms \citep{Abazajian:2017}.

\begin{figure}
    \includegraphics[trim=1cm 1cm 1cm
    0.5cm,width=0.48\textwidth]{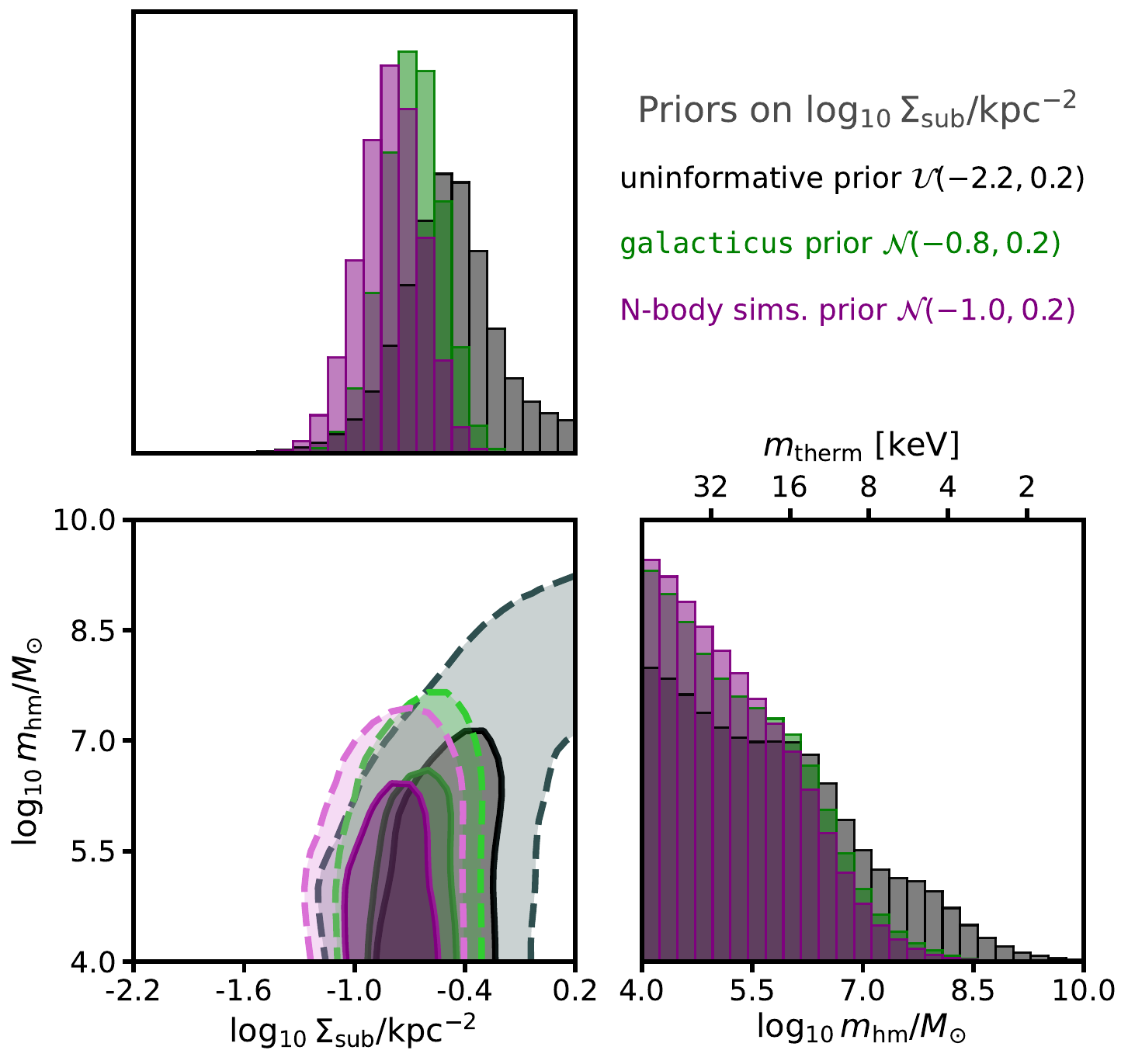}
    \caption{\label{fig:mhminference} The joint posterior probability distribution for the inferred half-mode mass ($m_{\rm{hm}}$) and normalization of the projected subhalo mass function ($\Sigma_{\rm{sub}}$) shown for three different priors on $\Sigma_{\rm{sub}}$, jointly inferred from 28 lenses. Contours show $68\%$ and $95\%$ confidence regions. The top axis of the bottom-right panel shows the corresponding mass of a thermal relic. Figure and caption from~\citet{Gilman26}.}
\end{figure}

In terms of strong lensing phenomenology, WDM models are distinguishable from CDM by a suppression of the abundance of subhalos below a characteristic mass scale and a lower density/concentration of the existing subhalos \citep{Sengul:2022b,Gilman26}.  No evidence of departure from CDM in the sense expected for WDM has been found by strong lensing so far, resulting in increasingly stringent limits on dark matter free streaming length or mass of the warm dark matter particle \citep{Vegetti:2018,Nierenberg++20,Gilman++20,Hsueh:2020,Keeley26,Gilman26,Gilman26b,Despali:2020}. A recent example is shown in Figure~\ref{fig:mhminference}, corresponding to free-streaming length less than 6-7 kpc, and thermal relic mass greater than 6-7 keV, where the ranges depend on theoretical assumptions on subhalo abundance. These limits are further strengthened by combining strong lensing with independent probes \citep{Enzi:2021,Nadler++21,Zelko++22}.

\subsection{Self-interacting dark matter (SIDM)}
\label{ssec:SIDM}

Strong lensing has figured prominently in the investigation of possible self-interaction between dark matter particles 
\citep[see, e.g.,][]{Adhikari:2025}. The so-called self-interacting dark matter models (SIDM) are often characterized, for the purpose of describing their astrophysical effects, by a self-interaction cross section, which can depend on the relative speed of the particles. The cross-section is often expressed in cm$^2$/g to account for the unknown mass of the particle.

Phenomenologically, SIDM can manifest itself in several ways that are potentially detectable via lensing. In isolated halos,  interactions initially reduce the density of the central parts of halos producing so-called isothermal cores  \citep[e.g.,][]{Rocha:2013}. Eventually, however, interactions trigger a so-called gravothermal catastrophe leading to core collapse \citep{Zeng:2025}. At any given time in cosmic history, the fraction of collapsed cores depends on the strength of the interaction. Interactions also tend to make halos more spherical than in standard CDM, especially in the central regions where densities and rates are highest \citep{Peter:2013,Despali:2022}. In colliding/non-relaxed halos, SIDM can manifest itself as offsets (or lack thereof) between the position of the dark matter and that of the stars or gas \citep{Randall:2008}.

Dark matter cores are difficult to detect via lensing in subhalos, but have been detected in massive halos at the centers of clusters in combination with stellar kinematics \citep[e.g.,][]{Newman:2013b,Odonnell:2026}. Unfortunately, the cores cannot be uniquely interpreted in terms of dark matter self-interaction, because they could also be due in part or completely to energy injection from supernovae and accretion onto supermassive black holes \citep[e.g.,][]{P+G14}. These are prominent physical processes in the center of clusters.  Modeling them with sufficient precision requires understanding and calculating complex phenomena over a daunting range of scales \citep{Martizzi:2013,Koudmani:2025,Despali:2026}.

In colliding clusters, such as the Bullet \citep{Clowe:2006,Bradac:2008,Ritarsic:2026} and Baby-Bullet \citep{Bradac:2008} clusters, the total mass distribution inferred from lensing is clearly displaced from the X-ray emitting gas, demonstrating that dark matter is non-baryonic. Furthermore, the close alignment between the distribution of total mass and that of collisionless galaxies sets upper limits to the self-interaction cross-section of the order of 1 cm$^2/g$ \citep{Randall:2008,Robertson:2017,Robertson:2019}.

The upper limits at cluster scales have stimulated interest in dark matter models with a velocity-dependent cross section, increasing towards smaller velocities \citep{Kaplinghat:2016}. Such behavior arises naturally in a number of circumstances \citep{Kaplinghat:2016} and would result in dark matter being effectively collisionless at cluster scales, while self-interaction would be strong enough to induce cores and core-collapse at velocities typical of dwarf galaxies. This signal would be detectable via galaxy-scale strong lensing \citep{Gilman++23,Gilman:2026c}.

\begin{figure}
    \includegraphics[width=0.48\textwidth]{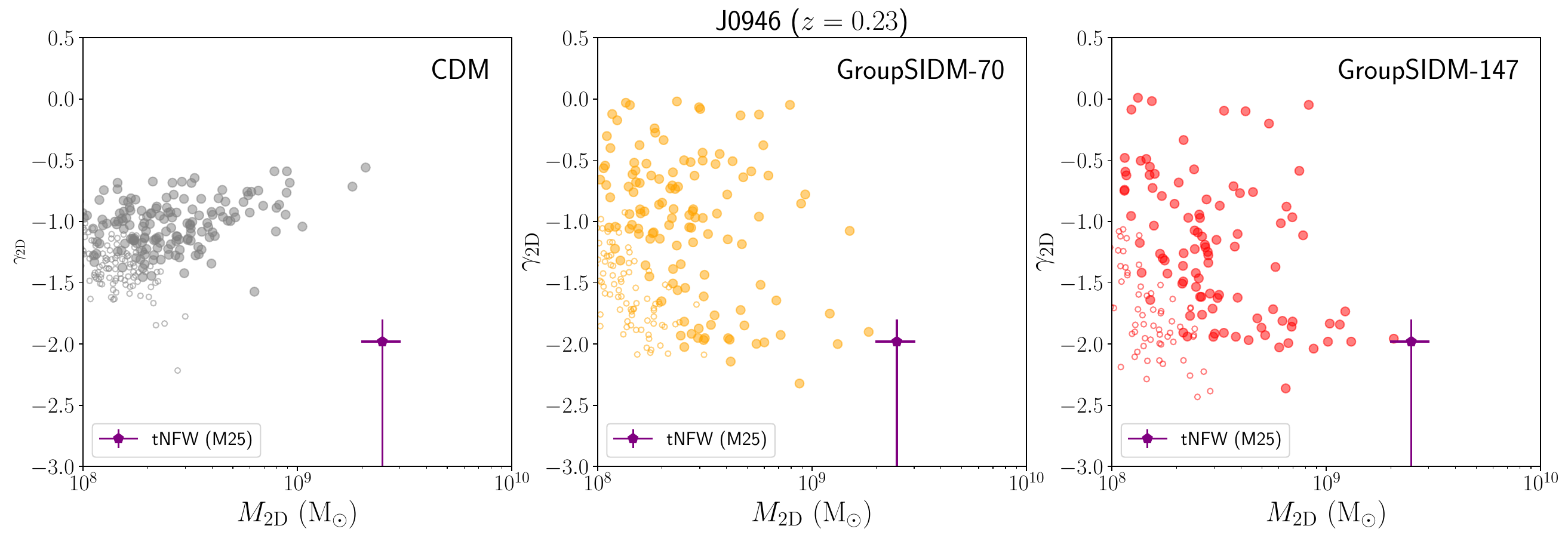}
    \caption{\label{fig:kong} Projected enclosed masses and density slopes of subhalos (R = 1 kpc) at redshift $z = 0.23$ in Group simulations
for CDM (left, gray), GroupSIDM-70 (middle, orange), and GroupSIDM-147 (right, red). Subhalos resolved with more than
2000 particles are shown as large filled markers, while those with 1000–2000 particles are shown as small open markers. For
comparison, we show the inferred enclosed mass and density slope of the J0946 perturber from \citep{Minor++21} (purple), based on a truncated NFW fit including both multipole and kinematic priors; the error bar indicates the 95\% CL. Figure and caption from~\citet{Kong:2026}.}
\end{figure}

Gravitational imaging studies have detected a handful of individual subhalos
\citep{Vegetti:2010,Vegetti++12,Minor:2021,Minor25,Vegetti:2026,Powell++25,Li:2025,Tajalli:2025}. Remarkably, they appear to be denser than one would expect based on pure CDM simulations. It has been suggested that they could be more easily explained as core-collapsed halos in a SIDM scenario \citep{Vegetti:2026}. Figure~\ref{fig:kong} by \citet{Kong:2026} illustrates the tension between the observed properties of an exemplar subhalo and those of simulated ones in CDM.

Flux ratios are also sensitive to SIDM through core collapse, which would increase the incidence of anomalies compared to CDM \citep{Gilman++21}. The recent analysis of 29 quadruply imaged quasars imaged by JWST MIRI in the mid-infrared finds that SIDM is preferred over CDM by Bayes ratios ranging from 1:6 to 1:15, depending on modeling assumptions \citep{Gilman:2026c}. This analysis includes the effects of globular clusters, deviations from ellipticity, and accounts for the possibility that outliers in the concentration-mass relation, or heavily tidally stripped subhalos, impart lensing perturbations similar to core collapsed SIDM halos. This population-level analysis strengthens significantly the evidence pointing away from simple CDM based on the overdensity of individual subhalos. However, larger samples and the corresponding reduction of statistical uncertainties are needed to increase the statistical significance from tantalizing to conclusive.

Reaching a definitive conclusion on the detection of collapsed halos and the interpretation as SIDM requires more work both in theory and in observations. On the theoretical side, computing the mass density profiles of subhalos that are perhaps 7-8 orders of magnitude less massive than their host halos requires numerical precision beyond what has been achieved to date \citep{Nadler++23}. Observationally, one needs to fully characterize the selection function and potential systematics of gravitational imaging and flux ratio anomalies alike.

\subsection{Other models}
\label{ssec:ODM}

Primordial black holes (PBH) have been suggested as a dark matter candidate \citep{Carr:1974nx}. Many mass ranges for PBH have been empirically ruled out. However, there are still mass ranges in which PBH could constitute DM in its entirety or in part \citep{F+K25}.  Strong lensing is sensitive to PBH, since they are effectively point masses. Thus, they have the most pronounced strong lensing signature at a fixed mass. The sensitivity range of strong lensing to PBH depends on source size, with the smallest sources sensitive to the smallest masses. While lensed quasars can be sensitive to PBH with masses as low as 10$^4$ M$_\odot$, lensed stars can reach significantly smaller masses \citep{Venumadhav:2017njf,Dike23}.

Strong lensing can also set some of the most stringent limits on ``fuzzy" dark matter, i.e., classes of models in which the de Broglie wavelength of the dark matter particle is macroscopic and interference patterns can form structures on astrophysically interesting scales   \citep{O+K26,Ephremidze:2026,Laroche++22,Powell:2023,Basu:2021,Chan:2020}.

Naturally, dark matter can be made of multiple particles. It is thus sensible to consider mixed dark matter models, for example a combination of cold and warm \citep{Keeley++23}. These can be tested with lensing data in the same way as single-type dark matter models. The challenge is the larger number of free parameters required. 

In this context, an interesting approach is to constrain effective quantities such as the matter power spectrum on some scale, which can then be mapped to broad classes of dark matter models \citep[e.g.,][]{Diaz-Rivero:2018,Gilman++22}.

These results are promising and showcase the potential of strong lensing as a probe of a variety of dark matter models. As the data improve, however, considerable effort will need to be devoted to improve the comparison with theory.

\subsection{The curious case of cluster lensing cross section}

One issue that deserves discussion is the difficulty of cluster simulations in reproducing the observed number of strongly lensed features, as discussed by \citet{Meneghetti:2020,Meneghetti:2023}. It appears that ray tracing through cosmological numerical simulations produces a significantly lower strong lensing cross section than observed. The tension is reduced when line-of-sight effects and selection function  closely matching simulated and observed clusters are taken into account \citep{Roche:2026}. However, the problem is not solved to date. If it survives further scrutiny, and improvements in cluster simulations and matching real and simulated data, the tension may require non-standard dark matter to be solved \citep{Dutra:2025,Natarajan:2026}.

\subsection{Systematics and open issues}
\label{ssec:sysDM}

One source of systematic error that has been the focus of much discussion in the literature since at least \citet{EW2003} is possible inadequacies of the so-called ``macro-model" used as a baseline for the detection of anomalies. The simplest example is that of a disk in the deflector that, if not properly modeled, can cause flux-ratio anomalies with respect to an elliptical mass distribution \citep{Gilman++17,Hsueh++18}. Modern-day analyses bypass the problem by excluding deflectors that show evidence for disks and by explicitly modeling and marginalizing departures from elliptical symmetry expressed as azimuthal multipoles \citep{PaugnatGilman25} and constraining the amplitude of the multipoles with the information contained in the extended arcs \citep{Stacey:2024,Gilman:2024,O+V24,Gilman26}.

Another astrophysical signal that could masquerade as dark-matter anomalies is that arising from globular clusters. Fortunately, it has been shown that globular clusters are not a significant contributor to the substructure signal \citep[e.g.,][]{Mao:1998,Gilman++17}. However, as the precision improves, it is good practice to include them in the models in order to account for them in the error budget \citep{Gilman26, Gilman:2026c}. 

One alternative approach is to bypass macro-models relying instead on local expansions of the gravitational potential, as in the curved arc formalism proposed by \citet{Birrer:2021b}. In this approach, the extended lens source is used to constrain parameters describing a series expansion of the local potential up to a certain order, which is then used to predict the flux ratios against which anomalies can be detected. This approach is by design more flexible than those based on a physically motivated macromodel. It thus produces results that are robust to the specific choice of macromodel, at the cost of increased uncertainties. Simulations show that this approach is viable \citep{Paugnat++25}, and can be used to set an upper limit to any potential bias or systematic related to the choice of macro model.

In terms of gravitational imaging, two issues remain to be addressed in order to interpret the result in terms of fundamental physics. The first one is the aforementioned selection function. The sensitivity of gravitational imaging to individual subhalos depends strongly on their density, and the selection function needs to be known to extreme accuracy in order to compare the few detections to the underlying parent population they are selected from.  The second one is collective effects \citep{Cyr:2019}. Currently, gravitational imaging signals are interpreted as being due to individual subhalos, under the assumption that sufficiently massive subhalos are rare. However, as the angular resolution and sensitivity of gravitational imaging extends to smaller masses, the collective effect of multiple subhalos will need to be modeled. 

From a theoretical point of view, an area that needs work is calculating the properties of substructure for a wide range of particle-physics-inspired dark matter models. High-precision and high-resolution cosmological numerical simulations in non-standard dark matter models are possible but expensive \citep{Nadler:2025,M+S23}. Including baryons further increases the computational cost and the number of free parameters to explore \citep{Despali:2026}. However, this may be necessary to fully characterize the impact of baryons on the subhalos. A practical way forward to explore a wide range of models is to use semi-analytic models, calibrated against detailed numerical simulations \citep{Lovell:2020,Benson20,Zeng:2025} or calculations of specific phenomena such as tidal heating and stripping \citep{G+V19,GVJ21,Du++24,Du++25}. 

\section{The future of strong lensing cosmology}
\label{sec:future}

\subsection{Mass production: opportunities}

\begin{figure}
    \includegraphics[width=\linewidth]{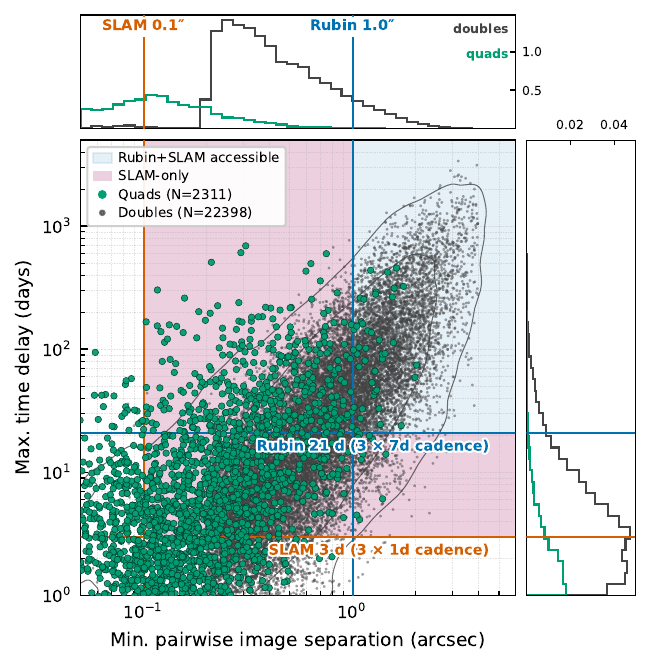}
    \caption{Distribution of maximum time delay and image separation for simulated samples of lensed quasars to be discovered by Euclid, Rubin, and Roman. The approximate sensitivity to time delays of Rubin/LSST and a dedicated space mission (SLAM) is shown as a multiple of cadence and angular resolution ($0.08"$ vs $0.8"$ for SLAM and LSST, respectively). As discussed in the main body of the text, blending limits the LSST measurements to the wider separation lenses and requires much longer monitoring campaigns than SLAM, in order to catch higher amplitude fluctuations.
    Doubles are identified by black symbols and quads by green symbols. Note that most quads are in the SLAM only region. In conclusion, SLAM can make precise time-delay measurements for a large population of lenses not otherwise accessible to Rubin. Figure courtesy of Eric Huff, based on forecasts by Paras Sharma.}
    \label{fig:sample}
\end{figure}

In the next few years, wide-area surveys from Euclid, Rubin, Roman, and SKA will enable the discovery of samples of strong gravitational lenses that are orders of magnitude larger than those currently known \citep{Wedig2025,OM10,Collett:2015,Erickson26,Lines:2025}. As an example, Figure~\ref{fig:sample} shows the forecast samples of lensed quasars to be discovered by Euclid, Rubin, and Roman based on \citet{Khadka:2026}, highlighting the distribution of image separation and time delay. The distributions for doubles and quads are shown separately. 

As far as lensed supernovae are concerned, the Legacy Survey of Space and Time (LSST) with Rubin is expected to discover $\sim 40$ SNe\,Ia and $\sim 80$ Type~II SNe \citep{Wojtak2019, Arendse2024} per year, via the detection of multiple images of the lensed transient \citep[see, also,][]{Bag:2026,Grupa:2026}.  Approximately 10 SN/year are expected to be discovered by the Roman Space Telescope High Latitude Time-Domain Survey \citep[][see also approved program 19038]{Pierel21}. Ongoing monitoring programs of clusters with JWST have been shown to be highly effective at identifying lensed supernovae suitable for cosmology. For example, the VENUS Program has already discovered six lensed supernovae \citep[e.g.,][]{Coulter:2026}. Future targeted programs may be able to provide a steady supply.

The abundance of targets will transform the field of strong lensing cosmology. For every application, statistical errors will be reduced by orders of magnitude. 

\subsubsection{H$_0$ and dark energy}

The precision on H$_0$ in flat $\Lambda$CDM will reach subpercent, comparable or better than the local distance ladder. In the immediate future, approved programs JWST-GO-9637 and JWST-GO-7184 will complete NIRSpec spectroscopy and NIRcam imaging of 18 lensed quasars with measured time delays. This is forecast to enable a measurement with 1.5\% precision on H$_0$. 

\begin{figure}[!t]
\centering\includegraphics[width=0.5\textwidth]{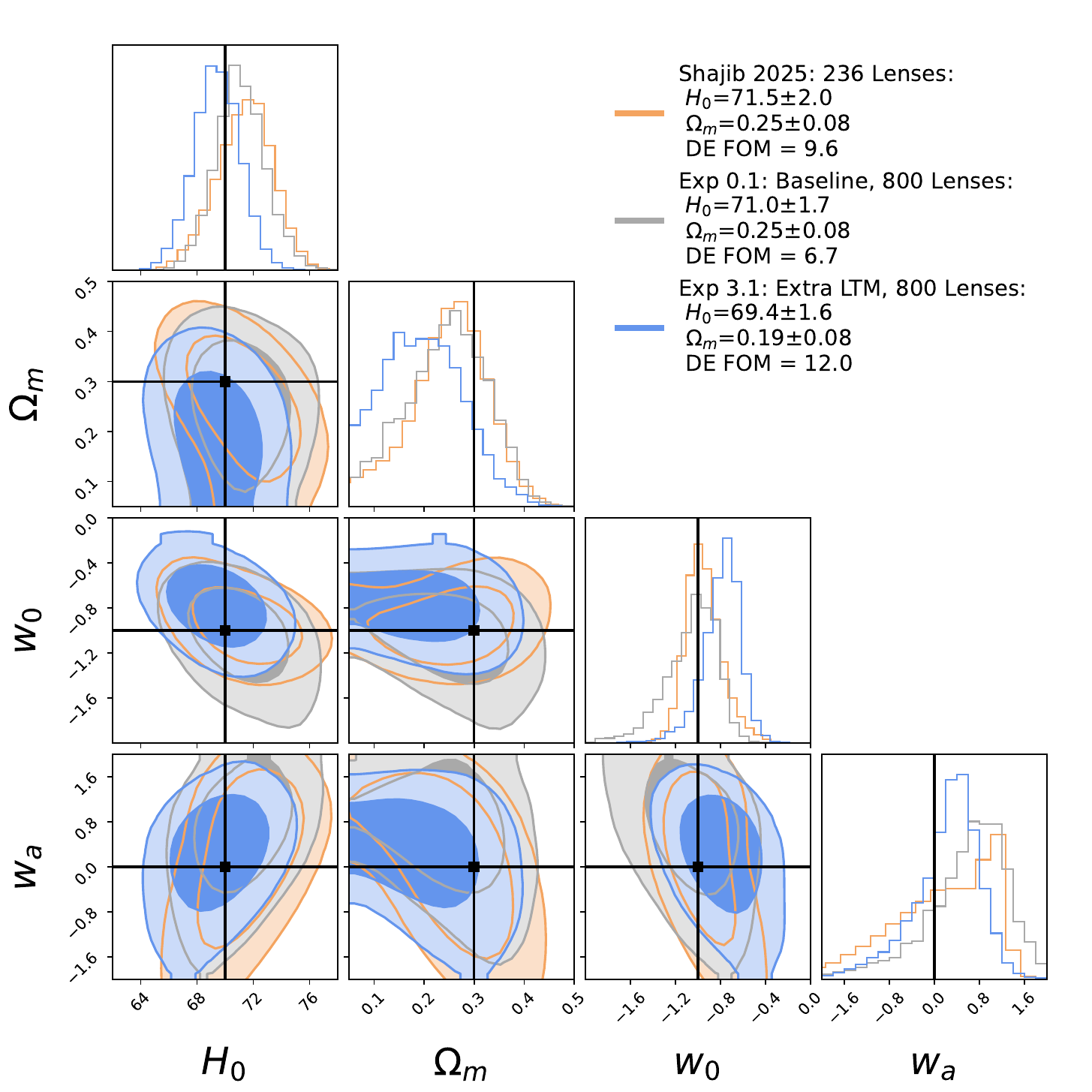}
\caption{Forecast precision on H$_0$ in $w_0w_a$ cosmology based on Rubin + ancillary data with existing facilities. The baseline experiment is based on 800 time delays from 10-years of LSST monitoring, assuming 10 systems have lens models and spatially resolved kinematics from JWST data, 40 with VLT kinematics and HST-images, 150 have unresolved kinematics and HST imaging, 300 have unresolved kinematics and LSST imaging, and 300 have no kinematics and LSST imaging. The LTM case shows the improvement resulting from assuming an additional 60 lenses in the 'unresolved kinematics + HST imaging' get dedicated long-term monitoring. The forecast by~\citep{Shajib++25} based on alternative assumptions is shown for comparison. Figure~from \citet{Erickson26}.} 
\label{fig:NOSLAMGENERIC}
\end{figure}

Beyond this horizon, the pace of progress will depend on the resources applied for follow-up and analysis of newly discovered samples. In flat $\Lambda$CDM breaking the 1\% precision barrier is within reach. If data quality comparable to that achieved with the current JWST program can be achieved, samples comprising $\sim$40 time delay lenses should be sufficient \citep{B+T20}. Considering cosmological models with dynamical dark energy, for example in the $w_0$--$w_a$ parametrization, two possible scenarios are shown in Figure~\ref{fig:NOSLAMGENERIC}. \citet{Erickson26} considers a  sample of 800 time delays measured from 10-years of monitoring with Rubin, supplemented by a layer cake strategy of follow-up: a small subset of systems will have the highest quality imaging and spectroscopy, and increasingly larger subsets will have coarser ancillary data (see caption for detail). The same figure compares the forecast with those by \citet{Shajib:2025review}, based on 10-year monitoring with LSST and somewhat different assumptions. \citet{Shajib:2025review} also investigates the combination of four probes: i) time-delay cosmography with lensed SNIa \citep{Birrer:2022}; ii) single-plane lenses with kinematics, a probe not discussed in this review \citep[see][for discussion]{Li:2024,GLB08}; iii) 236 lensed quasars with measured time delays, including 40 with spatially resolved kinematics measured with the JWST  NIRSpec integral field spectrograph; iv) 87 lenses with double-source planes\footnote{Further improvement can be obtained with the recently suggested Pseudo-Double Source Plane technique, not included in the \citet{Shajib:2025review} forecast.} \citep{SCL23}.  

\begin{figure}[!t]
\centering\includegraphics[width=0.5\textwidth]{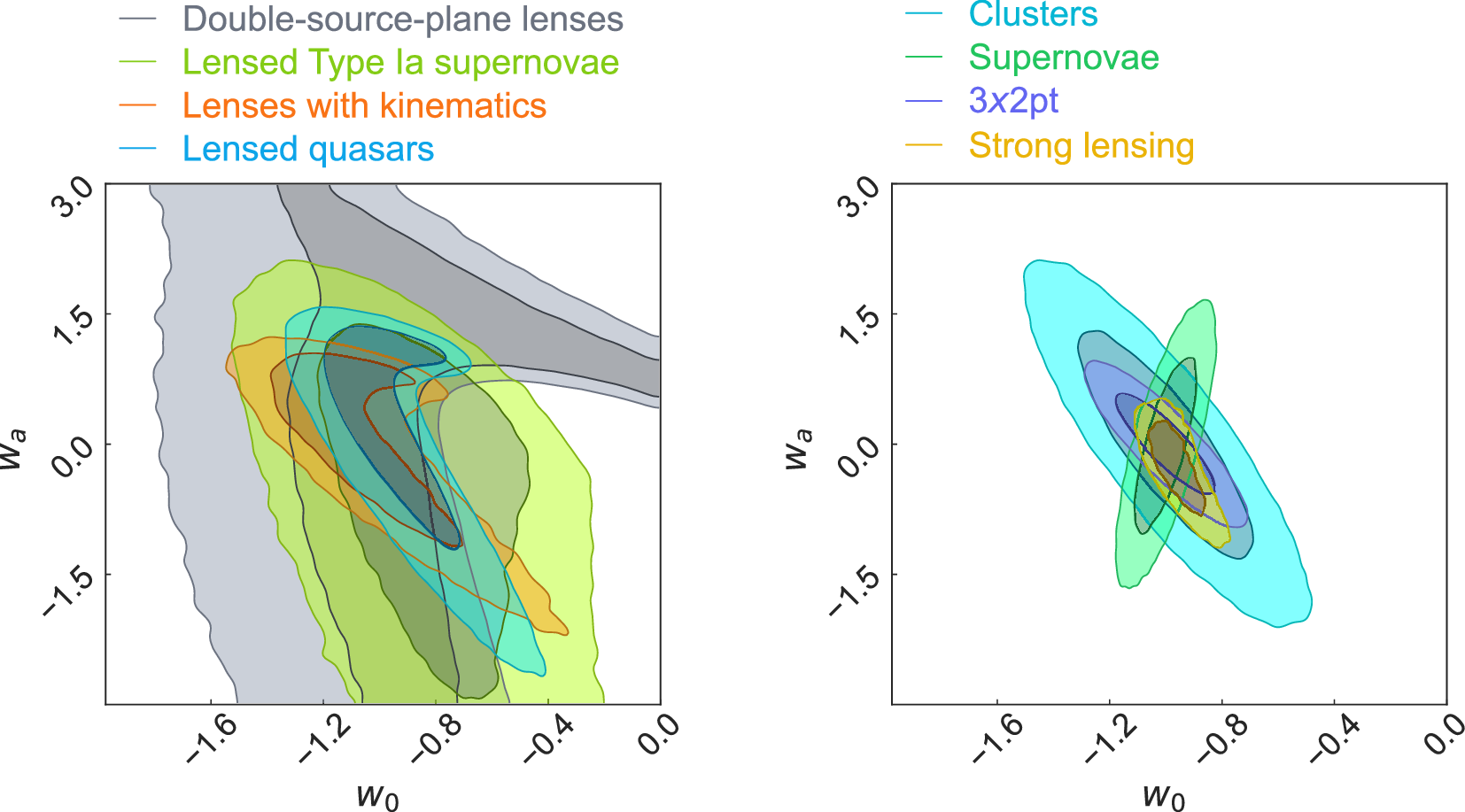}
\caption{Predicted constraints in the $w_0$--$w_a$ plane from LSST 10-year dataset. Left-hand panel: Individual constraints from four strong-lensing probes (see text for details). The particular shape of degeneracy for a given strong-lensing probe originates from using the particular ratio(s) of angular diameter distances to constrain the cosmological parameters. Right-hand panel: joint constraints from combining the four strong-lensing probes (yellow), in comparison with three other cosmological probes of Rubin: clusters (cyan), supernovae (green), and 3x2 pt correlations (blue). The figures of merit \citep{DETFreport} for these probes are 11, 32 and 48, respectively. The figure of merit for the combined strong-lensing probes is 69, making it the most constraining probe of the dark energy from the Rubin LSST. Figure and caption (adapted) from \citet{Shajib:2025review}.} 
\label{fig:NOSLAMGENERIC2}
\end{figure}

Even with rather conservative assumptions regarding follow-up, it seems clear that strong lensing can measure H$_0$ almost to 2\% precision in any arbitrary cosmology, and it can contribute in a significant manner to the determination of the equation of state of dark energy.  Figure~\ref{fig:NOSLAMGENERIC2} compares the predicted precision of the strong lensing probes under these assumptions with the predicted precision for other probes. It shows that strong lensing is the most powerful probe of dark energy if sufficient resources are devoted to follow-up and analysis. 

\begin{figure}[!t]
\centering\includegraphics[width=0.45\textwidth]{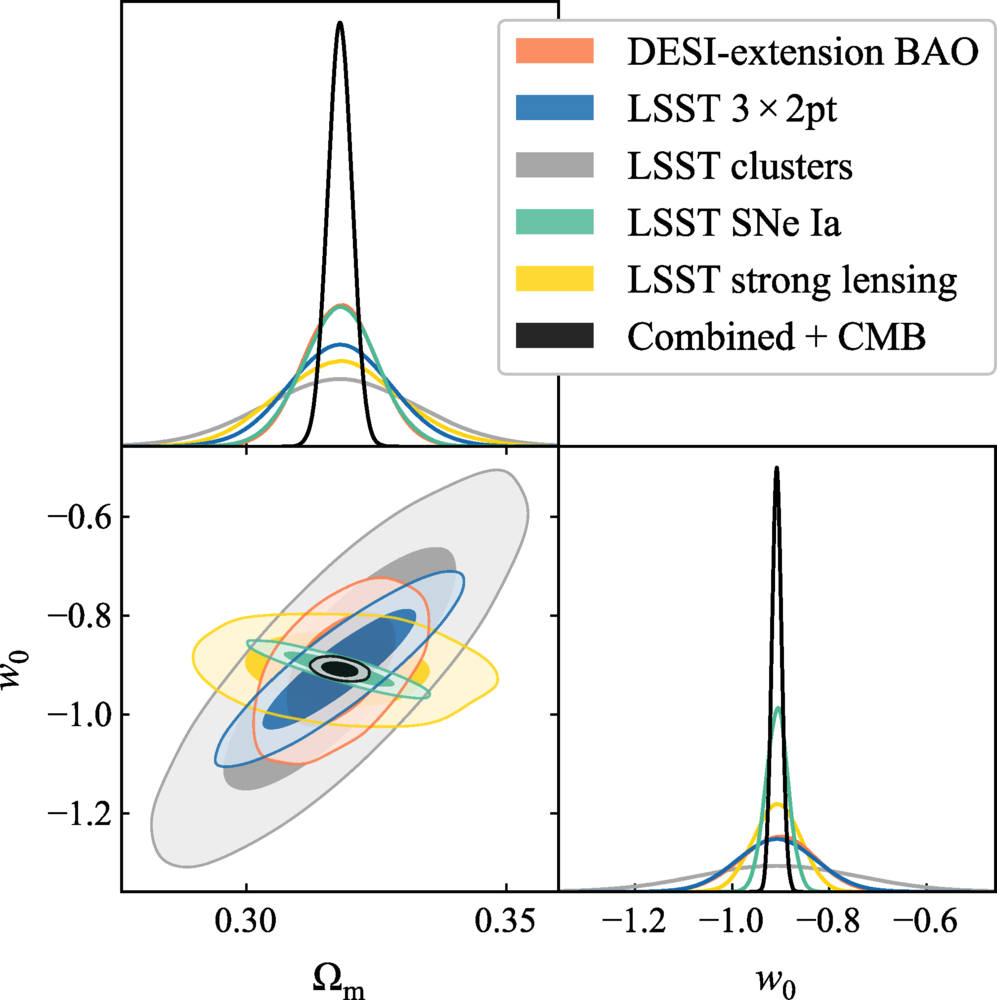}
\caption{Forecast constraints on $\Omega_{\rm m}$ and $w$ from DESI-extended BAO (orange) and Rubin LSST probes of cosmology after year 10: 3x2pt correlations (blue), clusters (gray), SNe Ia (emerald), and strong lensing (yellow). The fiducial values are set at $\Omega_{\rm m}=0.318$ and $w =-0.908$ \citep[see][for details]{S+F25}. The Combined+CMB case (black) corresponds to the combination of all the future datasets and the current CMB constraints (from Planck+ACT). The displayed posteriors have been adjusted to remove random fluctuations of the means, which are consistent with the statistical uncertainty levels, allowing for clearer visualization of each probe’s relative constraining power. Figure and caption from~\citet{S+F25}.} 
\label{fig:NOSLAMwphi}
\end{figure}

The same result is obtained considering an alternative parametrization of dark energy $w_{\phi}$CDM, which is generally well-behaved and physically motivated \citep{S+F25}. As summarized in Figure~\ref{fig:NOSLAMwphi}, the authors show that strong lensing is complementary with and comparable in precision with other probes. The combination of probes can distinguish the current best fit $w_{\phi}$CDM from $\Lambda$CDM at more than 9$\sigma$.

\begin{figure}[!t]
\centering\includegraphics[width=0.5\textwidth]{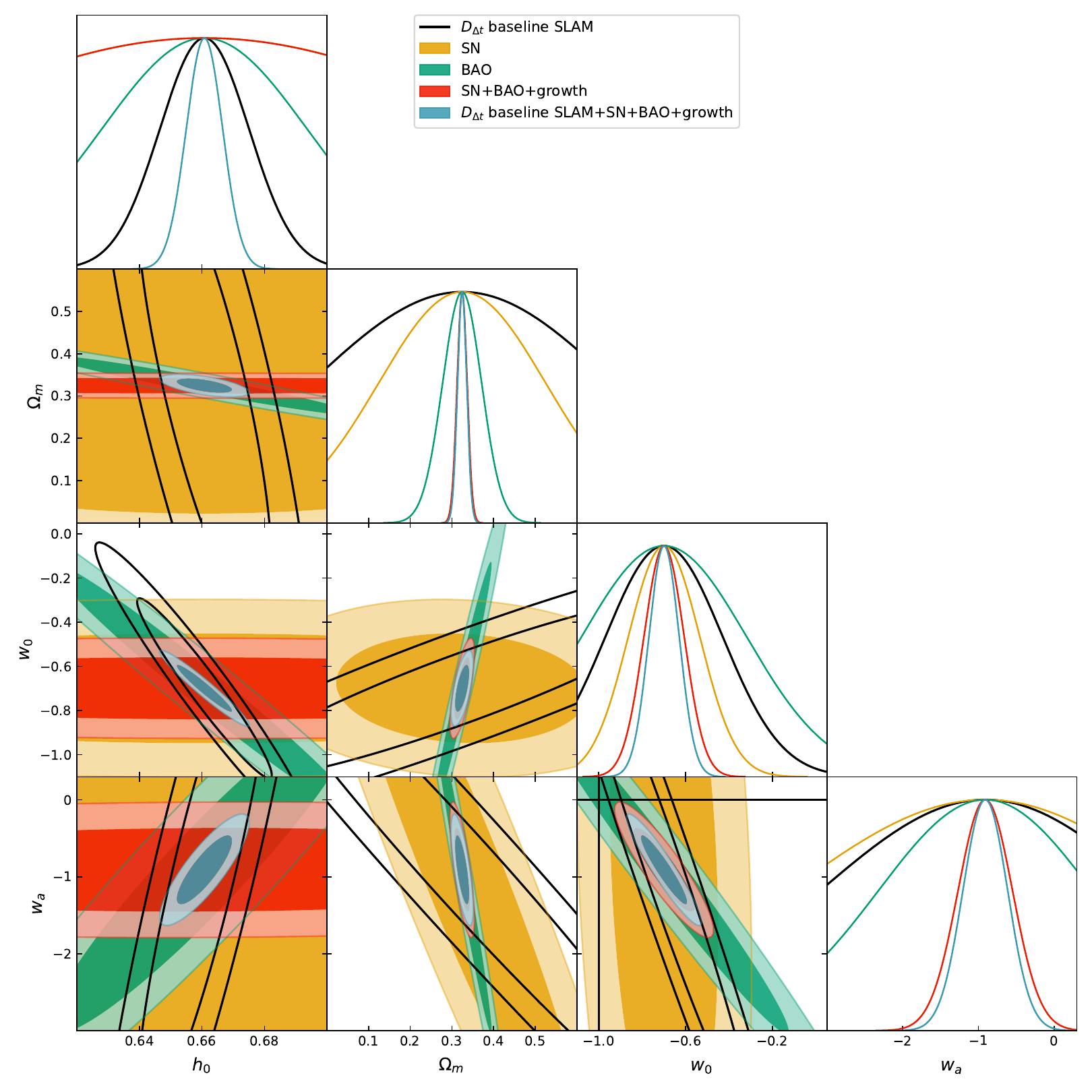}
\caption{Forecast precision on H$_0$ in $w_0$--$w_a$ cosmology based on 400 time delay lenses measured with 2\% precision ($\Ddt$ baseline SLAM). The expected combined outcome of all Stage IV SN BAO and growth probes is shown in red,  while SLAM baseline + Stage IV is shown in blue. Figure courtesy of Ryan Keeley.} 
\label{fig:SLAMBASELINEw0wa}
\end{figure}

It is interesting to consider a more optimistic forecast, in which additional resources are available for follow-up. Using the assumptions by \citet{Erickson26} as a baseline, LSST will support the measurement of some 800 time delays, with an average precision of 7\%, leading to the cosmological constraints in the figures above. Let me assume, instead, that 400 time delays can be measured with an average precision of 2\%, including both quasars and supernovae lensed by galaxy-scale deflectors. Let me also assume that integral field spectroscopy is available for all of them. As I discuss in Section~\ref{sssec:timedelays} and Section~\ref{sssec:spec}, meeting these goals will require dedicated monitoring from space (i.e. from a dedicated Strong Lensing Astrophysics Mission, referred to hereafter as SLAM) and spatially resolved spectroscopy from JWST-NIRSpec or from planned large and extremely large ground based telescopes with adaptive optics. For this forecast, I assume that one can meet the precision  goals listed in column 4 of Table~\ref{tab:bias}. The result is shown in Figure~\ref{fig:SLAMBASELINEw0wa}, in comparison with the forecast for the other Stage IV probes. Strong lensing is clearly complementary to the other probes. When combined with the other probes, it improves the figure of merit by almost one order of magnitude. 

\begin{figure}
    \centering
    \includegraphics[width=\linewidth]{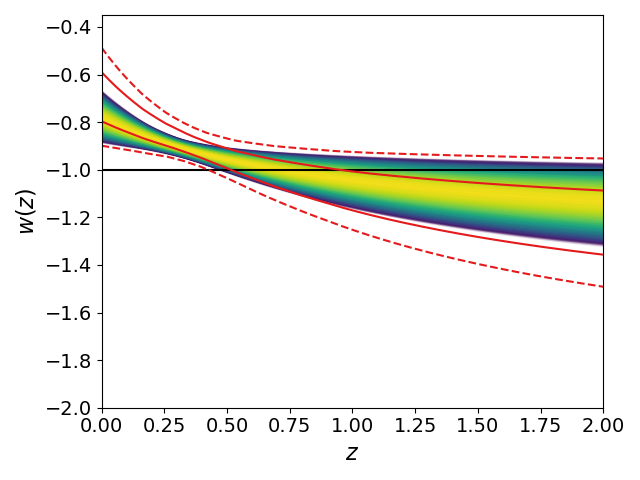}
    \caption{Reconstruction of $w(z)$ from a Gaussian process regression using SLAM plus Stage-IV cosmology probes. Each sample in the reconstruction is color-coded by its likelihood, with yellow being the maximum likelihood and blue being 2-$\sigma$ worse. Red lines bracket the 1 (solid) and 2 (dashed)-$\sigma$ reconstructions from the Stage-IV cosmology probes only. Figure courtesy of Ryan Keeley.}
    \label{fig:recon}
\end{figure}

This precision would be sufficient to distinguish $\Lambda$CDM from dynamical dark energy described by the parameters suggested by \citet{DESI-DR2} by more than 10-$\sigma$. More interestingly, if dark energy is indeed confirmed to be not the cosmological constant, one will be able to reconstruct $w(z)$, moving beyond simple parametrizations and gaining discriminatory power between competing dark energy models.  This is illustrated via a ``non-parametric'' reconstruction of $w(z)$, shown in Figure~\ref{fig:recon}. In this figure, Gaussian process regression is used to reconstruct $w(z)$ following \citet{Keeley_GP1,Keeley_GP2,Keeley_GP3}.  It is clear that having 400 high-precision time-delay lenses adds significant discriminatory power to this reconstruction.

\subsubsection{Dark matter}

\begin{figure}
    \includegraphics[trim=1cm 1cm 1cm
    0.5cm,width=0.48\textwidth]{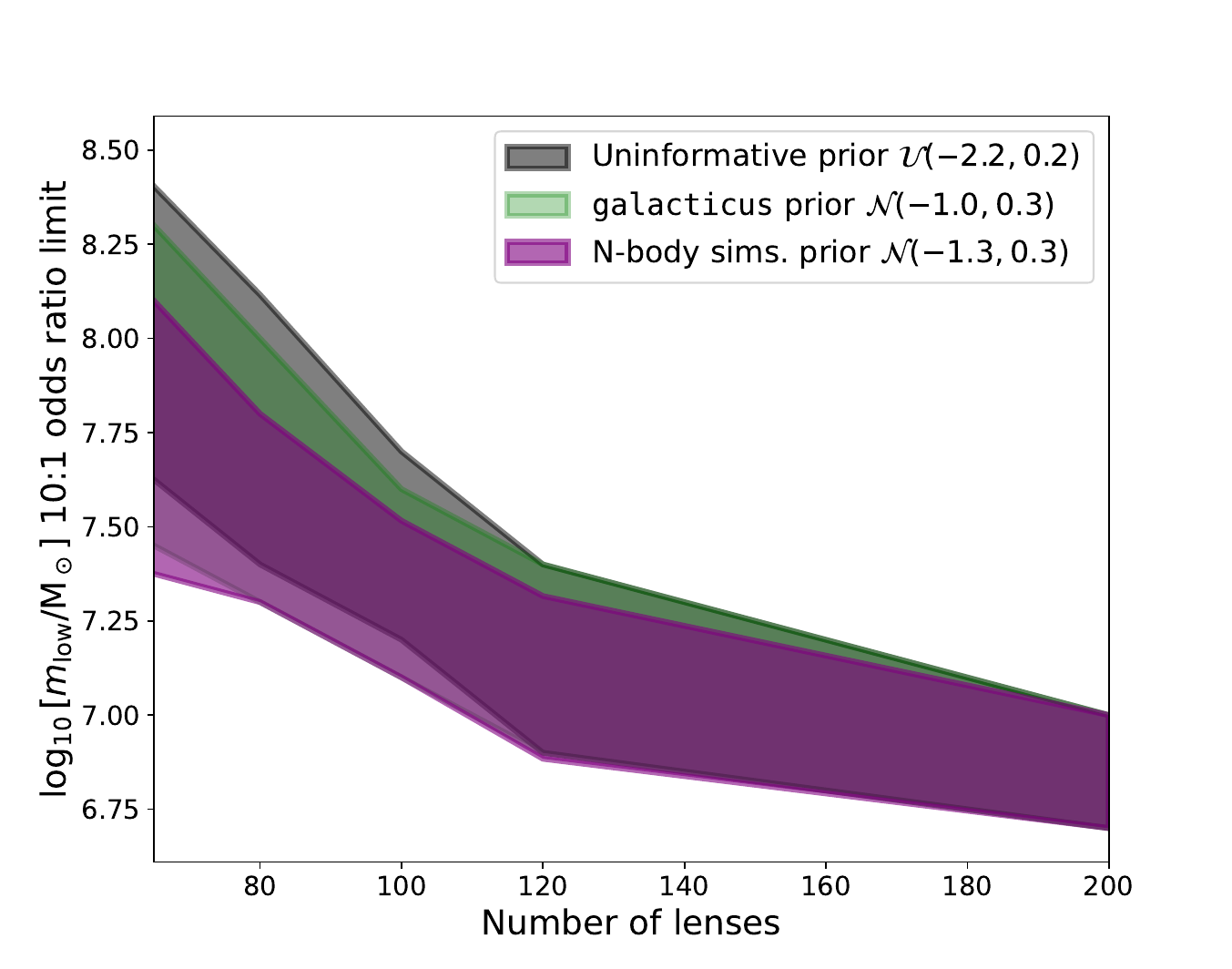}
    \caption{\label{fig:mlow_forecast} 
    Forecast of the sensitivity to minimum dark matter halo mass m$_{\rm low}$ as a function of number of quads with measured flux ratios and high resolution imaging. The current sample of 28 shows the existence of halos as small as 10$^8$ M$_\odot$ but reaching $<10^7$, where halos are expected to be completely dark, requires 200 quads. With 200 quads, the constraint becomes independent of theoretical priors on the tidal stripping and disruption of subhalos, because line-of-sight halos break the degeneracy. Figure from \citet{Nierenberg2026}.}
\end{figure}

The increase in sample size will also benefit strong lensing probes of dark matter. Flux ratio anomalies and gravitational imaging are currently limited by small number statistics. The number of quasar lenses usable for flux ratio anomalies will increase, as shown in Figure~\ref{fig:sample}. The number of galaxy-galaxy lenses amenable for gravitational imaging will increase even more. In the near future, the limiting factor will be the amount of follow-up: flux ratio anomalies will require integral field spectrographs \citep{ZNT24,Paugnat:2026} or space-based mid-infrared imaging \citep{Nierenberg++24}, while gravitational imaging will require imaging at high angular resolution, which is crucial to probe subhalos with mass 10$^6 M_\odot$ that are more informative for dark matter models \citep{Vegetti++24}. The challenges of follow-up are discussed in the next subsection. In this section, I illustrate the information gain that is achievable on dark matter using samples of lenses for which follow-up is feasible. One example is shown in Figure~\ref{fig:mlow_forecast}. Assuming that dark matter is CDM, with a sample of $\sim200$ quads with measured flux ratios one will be able to prove beyond doubt the existence of dark halos of mass 10$^{6}$M$_\odot$.

\begin{figure}
\includegraphics[width=0.95\linewidth]{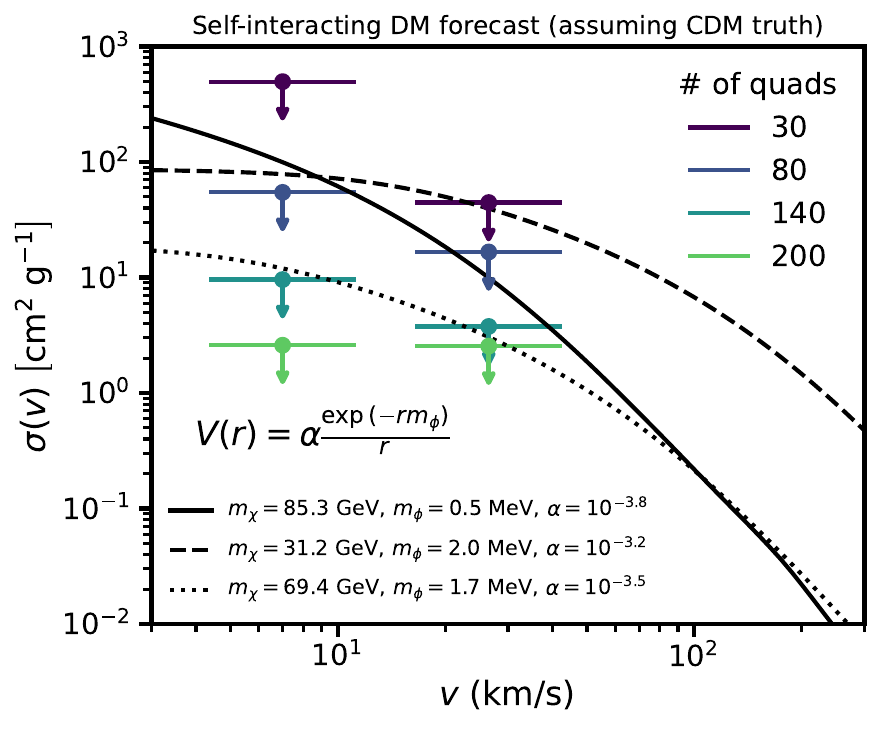}   \includegraphics[width=0.95\linewidth]{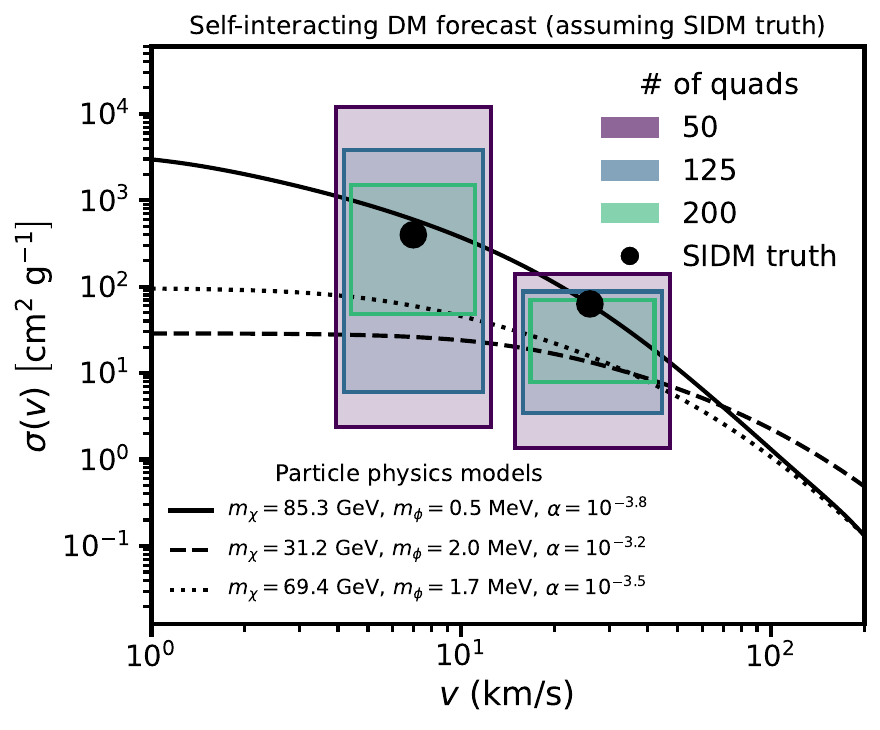}
\caption{\label{fig:sidmforecast}Forecasts for constraints on self-interacting dark matter for 200 quads. {\bf{Top:}} Upper limits (95$\%$ CI) on the self-interaction cross section as a function of the number of quad lenses in the sample, assuming a CDM ground truth. The halo masses probed by lenses $10^6 - 10^{10} M_{\odot}$ constrains the cross section across velocities $1 - 35 \ \rm{km} \ \rm{s^{-1}}$. The black curves in the figure show three representative examples of the cross section strength per dark matter particle mass $m_{\chi}$ for SIDM models with a light mediator $\phi$ and Yukawa coupling $\alpha$. {\bf{Bottom:}} A forecast assuming an SIDM ground truth. The solid black points mark the true cross section at $v=5 \ \rm{km} \ \rm{s^{-1}}$ and $v=25 \ \rm{km} \ \rm{s^{-1}}$. The shaded boxes show the $95 \%$ confidence interval on the cross section at each $v$. Figure courtesy of Daniel Gilman.}
\end{figure}

Another example of what will be possible with larger lens samples is shown in Figure~\ref{fig:sidmforecast}, showing a SIDM forecast. 
The top panel demonstrates that 200 quads would enable stringent upper limits on the self-interaction cross section strength, assuming a CDM ground truth. The constraining power derives from the abundance of core-collapsed halos, which impart stronger lensing perturbations than the more diffuse NFW density profiles predicted by CDM. As quad lenses probe halo masses below $10^7 M_{\odot}$, which corresponds to velocity scales below $10 \ \rm{km} \ \rm{s^{-1}}$, quads can access a regime below the velocity scales one can probe with classical dwarf galaxies and other studies of more massive halos. As shown in the figure, SIDM models with light mediators have velocity-dependent cross sections that can exceed $200 \ \rm{cm^2} \ \rm{g^{-1}}$ on these scales. The bottom panel of the same figure
 shows a forecast for 200 lenses assuming an SIDM ground truth, demonstrating that 200 quads would be sufficient to distinguish between different velocity-dependent cross sections, and hence dark matter properties.

\subsection{Mass production: challenges}

What are the challenges that need to be overcome to use strong lensing to achieve the cosmological breakthroughs described above? In this section, I discuss the main ones: i) identification of these rare events in the huge datasets (\S~\ref{ssec:disc}); ii) measurements of spectroscopic redshifts, supernova typing, and stellar velocity dispersions (\S~\ref{sssec:spec}); iii) obtaining high-resolution imaging data for modeling (\S~\ref{ssec:highres}); iv) measurement of time delays (\S~\ref{sssec:timedelays}); v) investigator and computation time required for modeling (\S~\ref{ssec:future-analysis}). I conclude with a discussion of systematic errors in \S~\ref{ssec:fsys}.

\subsubsection{Discovery}
\label{ssec:disc}

The first problem to be solved will be how to find lensed quasars and supernovae in the survey data. This is not a new problem, and many techniques have been developed over the years to deal with it \citep[see][for a review, and references therein]{Lemon2024}. Lens candidate searches have relied so far on techniques including automated modeling \citep{Marshall:2009,Oguri:2006,Chan:2015,Schechter:2025}, visual inspection \citep{Marshall:2016}, variability \citep{Kochanek:2006,Bag:2021,Bag:2022,Shu:2021,Lacki:2009}, spectroscopy \citep{Bolton:2006,Hsu:2026}, machine learning \citep{Agnello:2015,Jacobs:2019,Akhazhanov22}, and combinations thereof \citep{Sonnenfeld:2020,Holloway:2024,Gonzalez:2026}. The main difficulty is that strong lensing is a rare event \citep{OM10}, and candidates usually require expensive follow-up for confirmation, usually spectroscopy or high-resolution imaging, or both \citep[e.g.,][]{Treu2018}. Therefore, algorithms that are good at suppressing false positives have been preferred so far, to avoid wasting follow-up resources.  The downside of this approach is incompleteness, potentially missing out on many viable gravitational lenses. Fortunately, in the era of target abundance, this will be less of a problem. However, characterizing the selection function will remain a priority for statistical studies.

\subsubsection{Observational Follow-up: spectroscopy}
\label{sssec:spec}

Spectroscopic redshifts are a crucial ingredient of strong lensing cosmology. Knowing the redshift of the source and deflector greatly enhances the information content of each system, compared to relying on photometric redshifts. In the early days, redshifts had to be obtained one by one using 8-10m-class telescopes \citep[e.g.][]{Fassnacht:1996,T+K03}. Integral field spectroscopy from the ground and from space has been a game-changer in this respect  \citep[e.g.,][]{Scialpi:2024,Scialpi:2026,Dux:2025b,Bergamini:2023}. However, targeted follow-up will remain expensive for the large samples of lenses that are about to be discovered. Therefore, it is likely that the path forward will depend on redshifts measured as part of spectroscopic surveys \citep{Collett:2023}, photometric redshifts, and targeted efforts for the most valuable systems. For lensed supernovae, the problem of determining the redshift is exacerbated by the short time frame available after discovery. JWST is clearly unparalleled for this kind of effort, but the number of target-of-opportunity visits is limited by policy and programmatic considerations. Progress will require a way to deal with such time-critical events. For multiplane lenses, integral field spectroscopy on 8-10m-class telescopes or highly multiplexed spectroscopy from JWST makes it eminently feasible to complete  samples of order 100 systems \citep{S+L22}.

Beyond redshifts, deflector stellar velocity dispersion is the most important piece of information that requires spectroscopy. Spatially unresolved stellar velocity dispersions are feasible with ground-based telescopes \citep{Knabel:2025}
, and ongoing spectroscopic surveys should deliver large numbers of them \citep{Collett:2023}. The challenge will be to ensure that the data quality and the analysis methods are sufficient to control systematic errors to the level needed for cosmography \citep{Knabel:2025,Mozumdar:2025}. In contrast, measuring spatially resolved stellar velocity dispersion requires 2-10 hours per lens on 8-10m-class telescopes or in space \citep{Shajib:2023,Sheu:2026,Knabel:2026b,Mozumdar:2026,Shajib:2026}. Things will speed up a bit with improvements in adaptive optics at optical and IR wavelengths \citep[e.g.,][]{Scialpi:2024,LIGER24}, but the real game-changer will be the deployment of 30m-class extremely large telescopes. Those should enable spatially resolved kinematics of lens galaxies at superior angular resolution than JWST or 8-10m-class telescopes with exposure times under one hour (depending on brightness of the target). With the ELTs, samples of 400  time delay lenses as discussed above are within reach.

Measuring large numbers of narrow line flux ratios will likely require dedicated follow-up, as medium and wide tier survey data with Roman will be too shallow for the vast majority of targets. This can be done with pointed observations with HST \citep{Nierenberg++17}, JWST \citep{Paugnat:2026}, or with adaptive-optics assisted ground-based observations \citep{Nierenberg++14,ZNT24}. The proposed space telescope Lazuli \citep{Lazuli1} has an integral field spectrograph that should be sensitive enough for flux ratios. Mid-IR observations with JWST remain extremely powerful \citep{Nierenberg++24} and it is imperative to obtain them for large samples of lensed quasars during the course of the JWST mission.

\subsubsection{Observational Follow-up: high-resolution imaging}
\label{ssec:highres}

One piece of observational follow-up not to be underestimated is high-resolution imaging. Survey data from Euclid and Roman will provide some, but the signal-to-noise ratio and angular resolution will not always be sufficient for cosmography \citep{Meng:2015}. Pointed observations with Roman, HST, or JWST will do the job, but it will be expensive for large samples of lenses. A dedicated space mission such as SLAM could deliver the needed imaging data by stacking the high-resolution images needed for monitoring. Alternatively, adaptive-optics imaging can also be used, especially for sources that can be observed in the near infrared \citep[e.g.,][]{Chen:2019,Chen:2022}. The high angular resolution of ELTs will be revolutionary for applications such as gravitational imaging \citep{Vegetti2024}.

\subsubsection{Observational Follow-up: time delays}
\label{sssec:timedelays}

The single most important limiting factor of the aforementioned strong lensing cosmology forecasts is the time delays. For lensed quasars, the precision of LSST adopted in the forecasts shown in Figures~\ref{fig:NOSLAMGENERIC}~and~\ref{fig:NOSLAMGENERIC2} is set by the cadence of the survey and by deblending errors. Even with single-epoch high-resolution images to guide forced photometry, measuring time delays for images with separation smaller than the PSF leads to deblending errors that introduce an effective noise floor.   In order to quantify this statement, Venkatraman et al. (2026, in prep; hereafter V26) simulated the light curves that would be obtained with  Rubin if it could be dedicated entirely to lens monitoring after the end of LSST. With a dedicated Rubin Observatory, one could increase the monitoring cadence and exposure time. V26 ran a grid of simulations assuming 30-minute exposure time per epoch and daily cadence, as a function of campaign duration. In order to account for the noise associated with deblending the point sources from each other, from the deflector, and the host galaxy, V26 relied on data from \citet{Dux:2025}. Their data show that excess variance over the noise expected for an isolated point source depends as expected on two factors: image separation and brightness of the point sources.  The result is that even a dedicated campaign with the Rubin Observatory has a relative noise floor of a few percent at image separations comparable or smaller than the PSF. The noise floor makes it difficult to capture fluctuations on small timescales and thus would require Rubin to observe for many years, in order to catch higher-amplitude fluctuations. Furthermore, the LSST time-delay measurements made prior to this new era will have been limited to the wider separation image systems, missing out on the bulk of the population of strong lenses shown in Figure~\ref{fig:sample}. Note that the image separation shown in Figure~\ref{fig:sample} is the largest one for each system. Ninenty per cent of the LSST lensed quasars with time delays will consist of doubles. For quads, even in the cases for which Rubin will be able to measure the delay associated with the widest separation images, measuring the other two independent time delays will require space based resolution. 

Of the SNe\,Ia, only $\sim10$/yr will be suitable for follow-up, and at least half of these will require space-based follow-up, particularly to achieve a meaningful time-delay precision \citep[e.g., a time-delay uncertainty $\leq1$~day,][]{Hayes2026}. Additionally, $\sim 20$ SNe\,Ia per year will be identified via association with known galaxy-galaxy lenses, and hence will not require observation of multiple images of the transient \citep{SainzdeMurieta2024}. This will enable much earlier detection, allowing for precision time-delay measurements that would otherwise be impossible \citep{HS24}. However, detecting these supernovae comes at the cost of a high false positive rate, and they will require rapid follow-up with space-based imaging for verification and photometric monitoring \citep{SainzdeMurieta2024}. This is a strong argument for launching SLAM before the end of LSST.

Some space-based imaging capability is available from existing missions such as HST, Roman, and JWST. All three of these telescopes have the sensitivity and angular resolution to follow-up lensed quasars and supernovae. However, they are not optimized for monitoring and thus overhead and schedule disruption for time critical observation would be significant. A dedicated space mission, optimized for monitoring with low overhead and flexible scheduling, could be transformative. SLAM is a concept of such a mission, based on a diffraction-limited 1m ultraviolet and optical telescope. Monitoring of lensed supernovae is also among the science drivers of the proposed 3m class space telescope Lazuli \citep{Lazuli1}. Lazuli is optimized for time domain astronomy \citep{Lazuli2}. As a 3m space telescope, Lazuli should have the angular resolution and sensitivity to carry out photometric monitoring of lensed quasars as well, provided sufficient time is allocated.

\subsubsection{Investigator time, computing time, and artificial intelligence}
\label{ssec:future-analysis}

One issue of concern is the time that will be needed to model and analyze all the data that are going to be acquired in the next decade. At present, lens modeling is still expensive in terms of investigator and compute time. Progress has been made in speeding things up \citep{Shajib:2019,Schmidt:2023,Ertl:2023,Sheu:2026b}, but cosmography-grade analysis still requires months of investigator time and 10$^5$-10$^6$ CPU hours per system \citep{Williams:2026}. Further progress has been achieved by adopting new computing methods that can now carry out detailed lens models in a few hours on GPU clusters \citep{Lombardi:2024,Galan:2022,krizic2026,Li:2026,GIGALENS,Nightingale:2021}. However, the field is not yet in a situation where fully automated models are reliable. The main problem is that parameter space often spans 50 or more dimensions and the likelihood is simultaneously  informative and degenerate. Therefore, expert investigators are needed to guide the models towards physical solutions. Finding a principled way to properly weigh the information content of a space-based image against that of a spectrum is non-trivial and not a solved problem. For example, it is not desirable to try to fit every small feature in the surface brightness of a lensed source at the expense of the kinematic data that break the mass-sheet degeneracy. However, a brute-force approach facing millions of imaging pixels and 100 spectral pixels may lean in that direction. Artificial intelligence may help reduce the investigator and compute cost per system, allowing the analysis of large numbers of systems, as already demonstrated in some cases \citep{Erickson2025,Shajib_dolphin,PADMA:2025}.

\subsubsection{Systematic errors}
\label{ssec:fsys}

If sufficient resources are invested in follow up, cosmological applications will not be limited by statistical errors as they are now, but by systematic errors. 
Sections~\ref{sec:sys} and~\ref{ssec:sysDM} discuss at length the main sources of systematic errors for the two primary applications of strong lensing considered in this review. It is clear that identifying and mitigating those systematics will require considerable work, involving theory, computation, comparison of multiple datasets, and data challenges. This will be a non-negligible effort, possibly larger in terms of investigator time than that required to make the measurements. This is a positive development, another indication that strong lensing as a cosmological probe has moved on from youthful exuberance to full maturity.

\section{Summary}
\label{sec:summary}

Strong gravitational lensing is a purely gravitational cosmological probe, independent of and complementary to traditional methods (distance ladder, CMB, BAO, supernovae). Its power rests on two facts: (1) lensing observables depend on redshifts and cosmological distances, enabling measurements of the expansion history, H$_0$, and dark energy; (2) lensing is sensitive to matter on sub-galactic scales regardless of whether that matter emits light, making it a unique probe of dark matter's nature. The field is entering a pivotal, data-rich era: current samples of tens to hundreds of systems will grow by orders of magnitude with Euclid, Roman, Rubin, and the Square Kilometer Array.

The main takeaway messages from this article are:

\begin{enumerate}
    \item Time delay cosmography has come of age as a cosmological probe. The measurement H$_0=71.1^{+2.4}_{-2.3}$ \hunit, based on just 11 quasars lensed by galaxies, has reached 3\% precision in flat $\Lambda$CDM \citep{TDCOSMO26,TDCOSMO26L}, second only to the local distance ladder in terms of late universe probes. Ongoing programs should soon reach 1.5\%, sufficient to settle the tension.
    \item Multiply imaged supernovae are being discovered at an accelerating pace. The four published measurements of H$_0$ \citep{Kelly:2023,Grillo:2024,Pascale:2025,Pierel:2025,Suyu:2026,Bazzanini:2026,Winny5} are consistent with each other and with the value from lensed quasars within the uncertainties.
    \item Time delays and compound lenses are already providing interesting constraints on dark energy, highly complementary to those obtained with traditional techniques such as BAO or SN. For example, the combination of \citet{TDCOSMO26} with DESI-BAO \citep{DESI-DR2} and the Dovekie SN improves the figure of merit by 20\% and is in tension with $w=-1$ at the 2.3-2.5$\sigma$ level.
    \item Strong lensing studies set the most stringent limits on warm dark matter free streaming length $<6-7$kpc, corresponding to a thermal relic with mass $>6-7$ keV, where the ranges depend on theoretical assumptions \citep{Gilman++25}.
    \item Individual detections via gravitational imaging and population studies via flux ratio anomalies find that subhalos appear to be denser than expected in standard cold dark matter. Core-collapsed subhalos, as expected in self-interacting dark matter models seem to be preferred by the data by 2-$\sigma$ \citep{Gilman:2026c,Vegetti:2026}. The evidence is not conclusive. Verification requires better statistics and a better theoretical understanding of core collapse in SIDM.
    \item Forthcoming large samples of lenses have the potential to: improve significantly stage-IV measurement of dark energy parameters; detect dark halos down to 10$^{6.5}$ M$_\odot$, testing a key prediction of CDM; measure the self-interaction cross-section of dark matter, if it is non-zero, as a function of velocity.
    \item Fully exploiting the cosmological information in the forthcoming samples of strong lenses will require considerable investment in follow-up. In terms of facilities, the key priorities are: i) a dedicated space mission (SLAM) for photometric monitoring of lenses; ii) ground-based extremely large telescopes for spectroscopic follow-up. In terms of analysis, key priorities are the development of fast modeling methods and efforts to uncover and correct systematic errors. 
\end{enumerate}

\acknowledgments

I am grateful to many friends and colleagues for illuminating conversations on strong lensing over the past 30 years. I am particularly grateful to Andrew Benson, Simon Birrer, Sydney Erickson, Richard Ellis, Josh Frieman, Daniel Gilman, Eric Huff, Ryan Keeley, Shawn Knabel, Alexander Kusenko, Phil Marshall, Martin Millon,  Pritom Mozumdar, Anna Nierenberg, Massimo Pascale, Hadrien Paugnat, Piero Rosati, Anowar Shajib, Paras Sharma, William Sheu, Massimo Stiavelli, Sherry Suyu, Michael Turner, Padma Venkatraman, and Devon Williams for many contributions and insightful comments on this manuscript. 

I thank the National Science Foundation, the National Aeronautics and Space Administration, the Moore Foundation, the Sloan Foundation, and the Packard Foundation, for support over the years.

I asked Claude to help me identify typos, grammar, syntax, and style errors in an almost final draft of this article. It nearly drove me crazy with the hyphenation rules. Any residual mistakes in hyphenation are Claude's fault. Mistakes anywhere else are my responsibility. I was assisted by Claude in writing the jupyter notebooks that produced Figures~\ref{fig:timeline}, \ref{fig:tension_history}, and~\ref{fig:tension26}.

%apsrmp4-2.bst 2018-12-27 (MD) hand-edited version of apsrmp4-1.bst
%Control: key (0)
%Control: author (3) reversed first dotless
%Control: editor formatted (0) differently from author
%Control: production of article title (0) allowed
%Control: page (1) range
%Control: year (0) verbatim
%Control: production of eprint (0) enabled
%

\end{document}